\documentclass[a4paper,11pt]{article}
\pdfoutput=1 

\usepackage{jheppub} 

\usepackage[T1]{fontenc} 
\usepackage{amssymb}
\usepackage{amsmath}
\usepackage{color}
\usepackage{xspace}
\usepackage{appendix}
\usepackage{float}
\usepackage{hyperref}
\usepackage{diagbox}
\usepackage[utf8x]{inputenc}
\usepackage{soul}
\usepackage{float}
\usepackage{subfig}
\usepackage{multirow}

\newcommand{\bea}{\begin{align}}
	\newcommand{\eea}{\end{align}}
\newcommand{\beq}{\begin{equation}}
	\newcommand{\eeq}{\end{equation}}
\newcommand{\bqa}{\begin{eqnarray}}
	\newcommand{\eqa}{\end{eqnarray}}

\newcommand{\M}{\ensuremath \text{M}}

\newcommand{\ep}{\epsilon}

\allowdisplaybreaks[4]

\title{Analytic two-loop master integrals for $tW$ production at hadron colliders. Part \uppercase\expandafter{\romannumeral2}}
\author[a]{Jian Wang,}
\author[a,1]{Yefan Wang\note{Corresponding author.},}

\affiliation[a]{School of Physics, Shandong University, Jinan, Shandong 250100, China}

\emailAdd{j.wang@sdu.edu.cn}
\emailAdd{wangyefan@sdu.edu.cn}

\abstract{ 
We present analytic results of the two-loop master integrals for hadronic $tW$ production that contain two massive propagators.
For the planar integral family, we succeed in constructing the canonical basis, so the results are written in terms of multiple polylogarithms.
The master integrals in the non-planar integral families have been calculated by Taylor series expansion around $m_W^2=0$.
Even with this simplification, there are multiple square roots involved in the differential equations, which can not be rationalized simultaneously. 
We find that a proper linear combination of the integral basis alleviates the problem so that the results are expressed by multiple polylogarithms or single integrals over multiple polylogarithms up to weight four.
These single integrals can be evaluated using the classic Newton-Cotes formulas.
Our analytic results of master integrals would be used in the computation of the two-loop amplitudes for $tW$ production.
}

\begin{document}
	\maketitle
	\flushbottom
\section{Introduction}

The top quark has a mass at the electroweak scale and possesses
a large coupling with the Higgs field,
and therefore it is closely related to the electroweak symmetry breaking.
Top quarks have been copiously produced at the large hadron collider  (LHC)
and the top-quark properties have been investigated
by both the ATLAS and CMS collaborations.
Though the top-quark pair production via strong interaction has a larger cross section, the single  top-quark production via electroweak interaction depends on the electroweak coupling and thus can be used to measure the Cabibbo-Kobayashi-Maskawa matrix element $V_{tb}$.
	
There are three different modes for single top-quark productions, i.e., the s-channel, t-channel, and associated $tW$ production.
They involve $W$ bosons of different offshellness,
and thus can be used to probe different new physics beyond the standard model (SM).
The  s-channel and t-channel processes would get contributions from four-fermion operators while the $tW$ production is affected only by the modified $Wtb$ vertex.
	
At the LHC, the associated $tW$ production has the second largest
cross section after the t-channel, 
making it accessible to experimental measurements \cite{Aad:2012xca,Aad:2015eto,Aaboud:2016lpj,Aaboud:2017qyi,ATLAS:2020cwj,Chatrchyan:2012zca,Chatrchyan:2014tua,Sirunyan:2018lcp,CMS:2021vqm}.
The cross section has  been measured 
with an uncertainty of $10\%$ using the $138 ~{\rm fb}^{-1}$ data of the proton-proton collisions at 13 TeV  \cite{CMS:2022ytw}.

\begin{figure}[H]
	\centering
	\begin{minipage}{0.2\linewidth}
		\centering
		\includegraphics[width=0.8\linewidth]{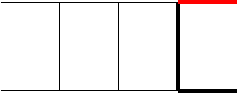}
		\caption*{P1}
	\end{minipage}
	\begin{minipage}{0.2\linewidth}
		\centering
		\includegraphics[width=0.8\linewidth]{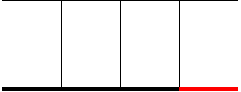}
		\caption*{P2}
	\end{minipage}
	\begin{minipage}{0.2\linewidth}
		\centering
		\includegraphics[width=0.8\linewidth]{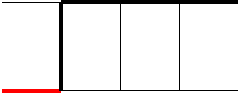}
		\caption*{P3}
	\end{minipage}
	\begin{minipage}{0.2\linewidth}
		\centering
		\includegraphics[width=0.8\linewidth]{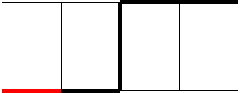}
		\caption*{P4}
	\end{minipage}
	\begin{minipage}{0.2\linewidth}
		\centering
		\includegraphics[width=0.8\linewidth]{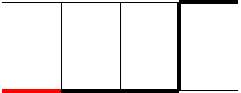}
		\caption*{P5}
	\end{minipage}
	\begin{minipage}{0.2\linewidth}
		\centering
		\includegraphics[width=0.8\linewidth]{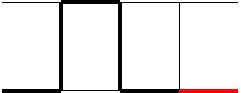}
		\caption*{P6}
	\end{minipage}
	\begin{minipage}{0.2\linewidth}
	\centering
	\includegraphics[width=0.8\linewidth]{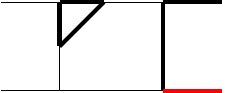}
	\caption*{P7}
\end{minipage}
	\begin{minipage}{0.2\linewidth}
	\centering
	\includegraphics[width=0.8\linewidth]{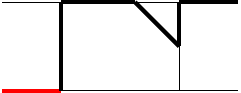}
	\caption*{P8}
\end{minipage}
	\caption{Planar integral topologies in the top sectors for $gb\to tW$.
	The thick black and red lines represent the top quark
and the $W$ boson, respectively. The other lines denote massless particles. }
	\label{fig:P}
\end{figure}

\begin{figure}[H]
	\centering
\begin{minipage}{0.2\linewidth}
		\centering
		\includegraphics[width=0.8\linewidth]{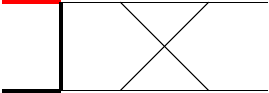}
		\caption*{NP1}
\end{minipage}
\begin{minipage}{0.2\linewidth}
	\centering
	\includegraphics[width=0.8\linewidth]{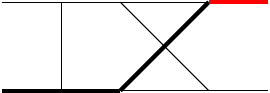}
	\caption*{NP2}
\end{minipage}
\begin{minipage}{0.2\linewidth}
	\centering
	\includegraphics[width=0.8\linewidth]{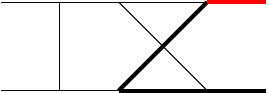}
	\caption*{NP3}
\end{minipage}
\begin{minipage}{0.2\linewidth}
	\centering
	\includegraphics[width=0.8\linewidth]{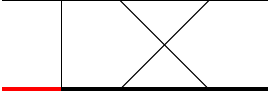}
	\caption*{NP4}
\end{minipage}
\begin{minipage}{0.2\linewidth}
	\centering
	\includegraphics[width=0.8\linewidth]{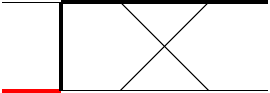}
	\caption*{NP5}
\end{minipage}
\begin{minipage}{0.2\linewidth}
	\centering
	\includegraphics[width=0.8\linewidth]{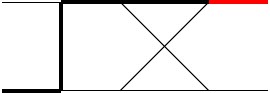}
	\caption*{NP6}
\end{minipage}
\begin{minipage}{0.2\linewidth}
	\centering
	\includegraphics[width=0.8\linewidth]{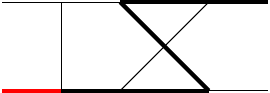}
	\caption*{NP7}
\end{minipage}
\caption{Non-planar integral topologies in the top sectors for $gb\to tW$.
	The thick black and red lines represent the top quark
and the $W$ boson, respectively. The other lines denote massless particles.}
\label{fig:NP}
\end{figure}

Precise theoretical predictions of scattering cross sections
are important in testing the  SM and searching for new physics using the data accumulated at the LHC.
Though the state-of-the-art results for many $2\to 2$ processes have included  QCD next-to-next-to leading order (NNLO)  corrections,
the $tW$ process is only known at QCD NLO \cite{Giele:1995kr,Zhu:2001hw,Cao:2008af, Kant:2014oha,Campbell:2005bb}.
Approximate higher order corrections have been obtained by expanding the threshold resummation formula \cite{Kidonakis:2006bu,Kidonakis:2010ux,Kidonakis:2016sjf,Kidonakis:2021vob,Li:2019dhg} in the strong coupling.
This approximation neglects  contributions from hard parton emissions and does not include two-loop virtual corrections. 
It is necessary to compare with the exact NNLO result to estimate the accuracy of this approximation. 
As an important ingredient of a complete NNLO correction, the two-loop virtual correction can be reduced to  a set of master integrals (MIs) and the corresponding coefficients.
The planar and non-planar  integral topologies in the top sectors are shown in figure \ref{fig:P}  and figure \ref{fig:NP}, respectively.
	
In ref. \cite{Chen:2021gjv}, we started to compute all the two-loop MIs, and presented the results for the integrals with one massive propagator, 
which correspond to the integral topologies P1 in figure \ref{fig:P}  and NP1 in figure \ref{fig:NP}.
In this paper, we provide analytic results for the integrals with two massive propagators,
which include both the planar P2 in figure \ref{fig:P}  and the non-planar NP2, NP3, NP4 topologies  in figure \ref{fig:NP}.
With these results at hand,
it is feasible to obtain the leading color contribution to the two-loop amplitudes \cite{Chen:2022yni}.
We notice that the integrals in the planar P2 topology have been calculated analytically in \cite{Long:2021vse}.
However,  the analytic results of non-planar NP2, NP3 and NP4 topologies are still unknown.

We find that the systems of differential equations in the 
NP2, NP3 and NP4 topologies involve multiple square roots which can not be rationalized simultaneously. 
To simplify the calculation, 
we will make use of the method of  expansion in $m_W^2$, 
which provides a good approximation to the exact result,
as we found in the calculation of one-loop squared amplitudes \cite{Chen:2022ntw}.
In particular, the accuracy can be improved systematically, since the  expansion to higher orders in $m_W^2$ does not require a calculation of new MIs.

This paper is organized as follows.
In section \ref{sec:planar}, we present the result for the integral family in the planar P2 topology.
In section \ref{sec:nonplanar}, we investigate the non-planar MIs in the NP2, NP3 and NP4 topologies under the condition $m_W^2= 0$.
The full results can be considered as a series of $m_W^2$, in which all orders share the same MIs.
The differential equations in the NP3 topology involve multiple square roots that can not be rationalized simultaneously. 
However, we find that a proper linear combination of the basis is able to bypass this difficulty. 

We conclude in section \ref{sec:conclusions}.


\section{Planar integral family with two massive propagators}
\label{sec:planar}

\subsection{Kinematics and notations}

We consider the two-loop integrals for the process 
\begin{align}
g(k_1) +b(k_2) \longrightarrow W(k_3)+t(k_4).
\end{align}
In this section, we focus on the P2 integral topology shown in figure \ref{fig:P}.
The integral family 
in this topology
is defined by  
\begin{align}
	I^{P2}_{n_1,n_2,\ldots,n_{9}}=\int{\mathcal D}^D q_1~{\mathcal D}^D q_2\frac{D_8^{-n_8}~D_9^{-n_9}}{D_1^{n_1}~D_2^{n_2}~D_3^{n_3}~D_4^{n_4}~D_5^{n_5}~D_6^{n_6}~D_7^{n_7}},
\end{align}
with
\begin{align}
	{\mathcal D}^D q_i = \frac{\left(m_t^2 \right)^\epsilon}{i \pi^{D/2}\Gamma(1+\epsilon)}  d^D q_i \ ,\quad D=4-2\epsilon \,,
\end{align}
where all $n_i\ge 0,i=1,\cdots, 7$ and $n_i\le 0,i=8,9$. The denominators $D_i$ read
\begin{align}
	D_1 &= q_1^2,&
	D_2 &= q_2^2,&
	D_3 &= (q_1-k_2)^2,\nonumber\\
	D_4 &= (q_1-k_3)^2-m_t^2,&
	D_5 &= (q_1 + q_2-k_2)^2,&
	D_6 &= (q_2+k_1)^2,\nonumber\\
	D_7 &= (q_2-k_2+k_3)^2-m_t^2,&
	D_8 &=(q_1-k_1-k_2)^2,&
	D_9 &=(q_2+k_1-k_2)^2,
\end{align}
where a conventional Feynman prescription $+i\varepsilon$ has been suppressed.
The Mandelstam variables are defined as
\begin{align}
	s=(k_1+k_2)^2\,, \qquad t=(k_1-k_3)^2\,, \qquad u=(k_2-k_3)^2 \,
\end{align}
with $s+t+u=m_W^2+m_t^2$. 

\subsection{Canonical basis}
All  the scalar integrals in each family can be reduced to MIs
using  integration-by-parts (IBP) identities \cite{Tkachov:1981wb,Chetyrkin:1981qh,Laporta:2000dsw}. 
In practice, we have used the  {\tt FIRE}  \cite{Smirnov:2019qkx} package  to perform the reduction.
Our purpose is to calculate the MIs in this integral family.
We adopt the method of differential equations \cite{Kotikov:1990kg,Kotikov:1991pm}.
Once a good basis that turns the differential equations to the canonical form is found,
the solution can be obtained recursively to higher orders in $\ep$ \cite{Henn:2013pwa}.
There are two prerequisites in this procedure, i.e.,  the boundary conditions have been determined and 
no square roots appear in the differential equations.	
We have the freedom to choose special boundaries where direct computation of the integral is simple or one integral can be related to others.
The square roots in the differential equation can be converted to polynomials once a judicious redefinition of integration variables can be found.
This is not always possible
when there are multiple scales or elliptic curves are involved in the differential equations.

We find  38 MIs in the P2 family after considering the  symmetries between integrals.
To construct the canonical basis, we first select  the MIs whose differential equations are  linear in $\epsilon$. 
\begin{align}
	\M^\text{P2}_{1}&=\epsilon^2I^\text{P2}_{0,0,0,2,0,0,2,0,0},\quad&
	\M^\text{P2}_{2}&=\epsilon^2I^\text{P2}_{0,0,0,2,2,1,0,0,0},\quad\nonumber\\
	\M^\text{P2}_{3}&=\epsilon^2I^\text{P2}_{0,0,1,0,2,0,2,0,0}, \quad&
	\M^\text{P2}_{4}&=\epsilon^2I^\text{P2}_{0,0,2,0,2,0,1,0,0},\quad\nonumber\\
	\M^\text{P2}_{5}&=\epsilon^2I^\text{P2}_{0,0,1,2,0,0,2,0,0},\quad&
	\M^\text{P2}_{6}&=\epsilon^2I^\text{P2}_{1,0,0,0,2,0,2,0,0}, \quad\nonumber\\
	\M^\text{P2}_{7}&=\epsilon^2I^\text{P2}_{2,0,0,0,2,0,1,0,0},\quad&
	\M^\text{P2}_{8}&=\epsilon^2I^\text{P2}_{1,0,0,0,2,2,0,0,0},\quad\nonumber\\
	\M^\text{P2}_{9}&=\epsilon^2I^\text{P2}_{1,0,0,2,0,0,2,0,0}, \quad&
	\M^\text{P2}_{10}&=\epsilon^3I^\text{P2}_{0,0,1,1,1,2,0,0,0},\quad\nonumber\\
	\M^\text{P2}_{11}&=\epsilon^2I^\text{P2}_{0,1,1,2,0,0,2,0,0},\quad&
	\M^\text{P2}_{12}&=\epsilon^3I^\text{P2}_{1,0,0,0,2,1,1,0,0}, \quad\nonumber\\
	\M^\text{P2}_{13}&=\epsilon^3I^\text{P2}_{1,0,0,1,1,2,0,0,0},\quad&
	\M^\text{P2}_{14}&=\epsilon^2I^\text{P2}_{1,0,0,2,1,2,0,0,0},\quad\nonumber\\
	\M^\text{P2}_{15}&=\epsilon^3I^\text{P2}_{1,1,0,0,2,0,1,0,0}, \quad&
	\M^\text{P2}_{16}&=\epsilon^2I^\text{P2}_{1,1,0,2,0,0,2,0,0},\quad\nonumber\\
	\M^\text{P2}_{17}&=\epsilon^3I^\text{P2}_{1,1,0,1,2,0,0,0,0},\quad&
	\M^\text{P2}_{18}&=\epsilon^4I^\text{P2}_{0,0,1,1,1,1,1,0,0}, \quad\nonumber\\
	\M^\text{P2}_{19}&=\epsilon^3I^\text{P2}_{0,0,1,2,1,1,1,0,0},\quad&
	\M^\text{P2}_{20}&=\epsilon^4I^\text{P2}_{1,0,0,1,1,1,1,0,0},\quad\nonumber\\
	\M^\text{P2}_{21}&=\epsilon^3I^\text{P2}_{1,0,0,2,1,1,1,0,0}, \quad&
	\M^\text{P2}_{22}&=\epsilon^2(1+2\epsilon)I^\text{P2}_{1,0,0,1,1,2,1,0,0},\quad\nonumber\\
	\M^\text{P2}_{23}&=\epsilon^4I^\text{P2}_{1,0,1,0,1,1,1,0,0},\quad&
	\M^\text{P2}_{24}&=\epsilon^3I^\text{P2}_{1,0,1,0,1,1,2,0,0}, \quad\nonumber\\
	\M^\text{P2}_{25}&=\epsilon^3I^\text{P2}_{1,0,1,1,1,2,0,0,0},\quad&
	\M^\text{P2}_{26}&=\epsilon^3I^\text{P2}_{1,1,0,0,2,1,1,0,0},\quad\nonumber\\
	\M^\text{P2}_{27}&=\epsilon^4I^\text{P2}_{1,1,0,1,1,0,1,0,0}, \quad&
	\M^\text{P2}_{28}&=\epsilon^3I^\text{P2}_{1,1,0,1,1,0,2,0,0},\quad\nonumber\\
	\M^\text{P2}_{29}&=\epsilon^3I^\text{P2}_{1,1,0,2,1,0,1,0,0},\quad&
	\M^\text{P2}_{30}&=\epsilon^4I^\text{P2}_{1,1,0,1,1,1,0,0,0}, \quad\nonumber\\
	\M^\text{P2}_{31}&=\epsilon^3I^\text{P2}_{1,1,0,2,1,1,0,0,0},\quad&
	\M^\text{P2}_{32}&=\epsilon^4I^\text{P2}_{1,0,1,1,1,1,1,0,0},\quad\nonumber\\
	\M^\text{P2}_{33}&=\epsilon^4I^\text{P2}_{1,0,1,1,1,1,1,-1,0}, \quad&
	\M^\text{P2}_{34}&=\epsilon^4I^\text{P2}_{1,1,0,1,1,1,1,0,0},\quad\nonumber\\
	\M^\text{P2}_{35}&=\epsilon^3(2\epsilon-1)I^\text{P2}_{1,1,0,1,1,1,1,-1,0},\quad&
	\M^\text{P2}_{36}&=\epsilon^4I^\text{P2}_{1,1,0,1,1,1,1,0,-1}, \quad\nonumber\\
	\M^\text{P2}_{37}&=\epsilon^4I^\text{P2}_{1,1,1,1,1,1,1,0,0},\quad&
	\M^\text{P2}_{38}&=\epsilon^4I^\text{P2}_{1,1,1,1,1,1,1,-1,0}.
\end{align}
The corresponding topology diagrams are displayed in figure \ref{fig:P2_MIs} in the appendix.

Then the canonical basis $F^\text{P2}_{i}$ with $i=1,\dots,38$ can be constructed as the linear combinations of  $\M^\text{P2}_{i}$ using a method similar to that proposed in ref.~\cite{Argeri:2014qva}.
More details can be found in \cite{Chen:2021gjv}.
As a result, we obtain the following canonical basis of the P2 family:
\begin{align}
	F^\text{P2}_1 &= \M^\text{P2}_{1},\quad
	F^\text{P2}_2 = \M^\text{P2}_{2}m_t^2,\quad
	F^\text{P2}_3 = \M^\text{P2}_{3}u,\quad
	F^\text{P2}_4 = -2\M^\text{P2}_{3}m_t^2+\M^\text{P2}_{4}(u-m_t^2),\quad
	\nonumber\\
	F^\text{P2}_5 &= \M^\text{P2}_{5}u,\quad
	F^\text{P2}_6 =  \M^\text{P2}_{6}m_W^2 ,\quad
	F^\text{P2}_7 = -2\M^\text{P2}_{6}m_t^2+\M^\text{P2}_{7}(m_W^2-m_t^2),\quad
	\nonumber\\
	F^\text{P2}_8 &= \M^\text{P2}_{8}(-s),\quad
	F^\text{P2}_9 = \M^\text{P2}_{9}m_W^2,\quad
	F^\text{P2}_{10}=\M^\text{P2}_{10}(u-m_t^2),\quad
	F^\text{P2}_{11}=\M^\text{P2}_{11}u^2,\quad
	\nonumber\\
	F^\text{P2}_{12}&=\M^\text{P2}_{12}r^\text{P2}_1,\quad
	F^\text{P2}_{13}=\M^\text{P2}_{13}r^\text{P2}_1,\quad
	F^\text{P2}_{14}=-\frac{3}{2}\M^\text{P2}_{13}(m_t^2-m_W^2+s) -\M^\text{P2}_{14}m_t^2 s,\quad
	\nonumber\\
	F^\text{P2}_{15}&= \M^\text{P2}_{15}(u-m_W^2),\quad
	F^\text{P2}_{16}= \M^\text{P2}_{16}m_W^2u,\quad
	F^\text{P2}_{17}= \M^\text{P2}_{17}(u-m_W^2),\quad
	\nonumber\\
	F^\text{P2}_{18}&= \M^\text{P2}_{18}(u-m_t^2),\quad
	F^\text{P2}_{19}= \M^\text{P2}_{19}m_t^2(u-m_t^2),\quad
	F^\text{P2}_{20}= \M^\text{P2}_{20}r^\text{P2}_1,\quad
	F^\text{P2}_{21}= \M^\text{P2}_{21}m_t^2r^\text{P2}_1,
	\quad\nonumber\\
	F^\text{P2}_{22}&= \frac{1}{2}\M^\text{P2}_{12} (m_t^2-m_W^2+s)
	-\M^\text{P2}_{13}(m_t^2-m_W^2+s) 
	+2\M^\text{P2}_{20} (m_t^2-m_W^2+s)
	\nonumber\\&\quad
	+2\M^\text{P2}_{21} m_t^2(m_t^2-m_W^2)- \M^\text{P2}_{22}m_t^2s,\quad\nonumber\\
	F^\text{P2}_{23}&=\M^\text{P2}_{23}(m_W^2-s-u),\quad
	F^\text{P2}_{24}= \M^\text{P2}_{24}m_t^2(-s),\quad
	F^\text{P2}_{25}= \M^\text{P2}_{25}s(m_t^2-u),\quad
	\nonumber\\
	F^\text{P2}_{26}&= \M^\text{P2}_{26}s(m_t^2-u),\quad
	F^\text{P2}_{27}= \M^\text{P2}_{27}(u-m_W^2),\quad
	F^\text{P2}_{28}= \M^\text{P2}_{28}m_t^2(u-m_W^2),\quad
	\nonumber\\
	F^\text{P2}_{29}&= \M^\text{P2}_{29}m_t^2(u-m_W^2),\quad
	F^\text{P2}_{30}= \M^\text{P2}_{30}(m_t^2-s-u),\quad
	\nonumber\\
	F^\text{P2}_{31}&= \M^\text{P2}_{31}(-m_t^4+m_t^2(m_W^2+s+u)-m_W^2u),\quad
	F^\text{P2}_{32}= \M^\text{P2}_{32} r^\text{P2}_2 ,\quad
	\nonumber\\
	F^\text{P2}_{33}&= -\M^\text{P2}_{23}(u-m_W^2)+\M^\text{P2}_{32}s(m_t^2-u)+\M^\text{P2}_{33}(u-m_W^2),\quad
	F^\text{P2}_{34}= \M^\text{P2}_{34}r^\text{P2}_1(u-m_t^2),\quad\nonumber\\
	F^\text{P2}_{35}&= \frac{\M^\text{P2}_{1}m_t^2}{2 m_t^2-2 m_W^2}-\frac{2 \M^\text{P2}_{2}m_t^4}{m_t^2-m_W^2}
	+\frac{\M^\text{P2}_{3}m_t^2 (m_t^2+u) }{m_t^2-m_W^2}
	+\frac{\M^\text{P2}_{4}m_t^2 (m_t^2-u) }{2 (m_t^2-m_W^2)}-\frac{\M^\text{P2}_{5}m_t^2 u}{m_t^2-m_W^2}
	\nonumber\\&\quad
	+\frac{2 \M^\text{P2}_{20}m_t^2 (m_t^2-u)}{m_t^2-m_W^2}-\frac{2\M^\text{P2}_{27} (m_t^2+m_W^2) (m_t^2-u)}{m_t^2-m_W^2}
	+2\M^\text{P2}_{30}(u-m_t^2)
	\nonumber\\&\quad
	-\frac{ \M^\text{P2}_{34}(m_t^2-u) (m_t^4-m_t^2 (2 m_W^2+s)+m_W^2 (m_W^2-s))}{m_t^2-m_W^2}
	\nonumber\\&\quad
	+\frac{\M^\text{P2}_{35}m_W^2 (u-m_t^2)}{m_t^2-m_W^2}
	+\frac{2\M^\text{P2}_{36}m_t^2 (m_t^2-u)}{m_t^2-m_W^2},\quad\nonumber\\
	F^\text{P2}_{36}&= \M^\text{P2}_{20}(u-m_t^2)+\M^\text{P2}_{27}(m_t^2-m_W^2)-\M^\text{P2}_{30}(u-m_t^2)+\M^\text{P2}_{36}(u-m_t^2),\quad
	\nonumber\\
	F^\text{P2}_{37}&=\M^\text{P2}_{37}s(m_t^2-u)^2,\quad
	F^\text{P2}_{38}=\M^\text{P2}_{38}(u-m_t^2)(u-m_W^2),
\end{align}
where
\begin{align}
r^\text{P2}_1 &= \sqrt{m_t^4 + (m_W^2 - s)^2 - 2 m_t^2 (m_W^2 + s)},\nonumber\\
r^\text{P2}_2 &= \sqrt{s(m_t^2-u)\left(m_t^2 \left(-4 m_W^2+s+4 u\right)-su\right)}.
\end{align}
Notice that only $F^\text{P2}_{32},F^\text{P2}_{33},F^\text{P2}_{37}$ and $F^\text{P2}_{38}$ depend on $r^\text{P2}_2$. 

For the integrals whose differential equations contain only $r^\text{P2}_1$,
we change the integration variables,
\begin{align}
	m_W^2 = m_t^2 v^2,\quad s = m_t^2 \frac{(w+v)(1+wv)}{w},\quad u = m_t^2z.
\end{align}
We choose $-1<w<1$, then we have
\begin{align}
	r^\text{P2}_1 = \frac{(1+w)(1-w)v}{w}m_t^2.
\end{align}

For the integrals involving both $r^\text{P2}_1$ and $r^\text{P2}_2$, by using {\tt RationalizeRoots} \cite{Besier:2018jen,Besier:2019kco}, we find that
after transformation of the variables
\begin{align} 
	m_W^2 = \frac{b_1 b_2 m_t^2}{1-b_2},\quad s =  -\frac{(b_1+1)m_t^2}{b_2-1},\quad
	u =  \frac{b_2({b_3} ({b_1}+1) ({b_3}{b_2}-2)+4 {b_1})m_t^2}{{b_2} ({b_3} ({b_1}+1) ({b_3}{b_2}-2)-4)+4},
\end{align}
 the above two square roots can be rationalized simultaneously,
\begin{align}
	r^\text{P2}_1 =\frac{{b_1} ({b_2}-1)+{b_2}}{(1-{b_2})}m_t^2,\quad
	r^\text{P2}_2 =\frac{4({b_1}+1) (b_3 {b_2}-1) ({b_1} {b_2}+{b_2}-1)}{(1-{b_2}) \left(b_3^2 ({b_1}+1) b_2^2-2 {b_2} ({b_3} {b_1}+b_3+2)+4\right)}m_t^4.
\end{align}
Here we choose the solutions of $b_1,b_2,b_3$ in such a way that $r^\text{P2}_1$ and $r^\text{P2}_2$ are positive in the physical region
\begin{align}
s>(m_W+m_t)^2\wedge \frac{m_W^2+m_t^2+r^\text{P2}_1-s}{2}>u>\frac{m_W^2+m_t^2-r^\text{P2}_1-s}{2}\wedge m_t>m_W>0.
\end{align}

\subsection{Boundary conditions}
To analytically solve the differential equations, we still need boundary conditions for the canonical basis. In the P2 integral family, the analytical results of $F^\text{P2}_1$ and $F^\text{P2}_2$ can be directly computed,
\begin{align}
F^\text{P2}_1 &= 1,\nonumber\\
F^\text{P2}_2 &= -\frac{1}{4}-\frac{\pi ^2}{6}\epsilon^2-2 \zeta (3)\epsilon^3-\frac{8\pi ^4}{45}\epsilon^4+ {\cal O}(\epsilon^5).
\end{align}
The boundary conditions of $F^\text{P2}_4$, $F^\text{P2}_7$ and $F^\text{P2}_8$ are also easy to obtain at special kinematic points,
\begin{align}
F^\text{P2}_4|_{u=0} &=F^\text{P2}_7|_{m_W^2=0}= 1+\frac{\pi ^2}{3}\epsilon^2-2 \zeta (3)\epsilon^3+\frac{\pi ^4}{10}\epsilon^4+ {\cal O}(\epsilon^5),\nonumber\\
F^\text{P2}_8|_{s =m_t^2} &= -1-2 i \pi\epsilon+\frac{7 \pi ^2}{3}\epsilon^2 + \left(10 \zeta (3)+2 i \pi ^3\right)\epsilon^3+\left(-\frac{109 \pi ^4}{90}+20 i \pi  \zeta (3)\right)\epsilon^4+ {\cal O}(\epsilon^5).\nonumber\\
\end{align}
Besides, $F^\text{P2}_{18}$ and $F^\text{P2}_{23}$ vanish at kinematic points satisfying $u = m_t^2$ and  $(m_W^2 = 0, s = m_t^2, t = m_t^2, u = -m_t^2)$, respectively.
The boundary values of the remaining canonical basis integrals can be fixed using the regularity conditions at the kinematic point $( m_W^2 = 0, s = m_t^2, u = 0, t = 0 )$,
which corresponds to $(v = 0, z=0 )$ or $(b_1 = 0, b_2 = 0)$. For example,
\begin{align}
&F^\text{P2}_{38}|_{(m_W^2 = 0, s = m_t^2, u = 0, t = 0)}\nonumber\\
=& \frac{1}{36} (9 F^\text{P2}_1+18 F^\text{P2}_2-64 F^\text{P2}_3+10 F^\text{P2}_4-24 F^\text{P2}_5+34 F^\text{P2}_6-16 F^\text{P2}_7+21 F^\text{P2}_8+6 F^\text{P2}_9
\nonumber\\&
+24 F^\text{P2}_{10}
+12 F^\text{P2}_{11}
-32 F^\text{P2}_{14}+24 F^\text{P2}_{15}-12 F^\text{P2}_{16}-36 F^\text{P2}_{17}-24 F^\text{P2}_{18}-24 F^\text{P2}_{19}-12 F^\text{P2}_{22}
\nonumber\\&
-108 F^\text{P2}_{23}
-48 F^\text{P2}_{24}+20 F^\text{P2}_{25}
+16 F^\text{P2}_{26}+108 F^\text{P2}_{27}+24 F^\text{P2}_{28}+24 F^\text{P2}_{29}-108 F^\text{P2}_{30}+48 F^\text{P2}_{31}
\nonumber\\&
+36 F^\text{P2}_{32}
-24 F^\text{P2}_{33}-12 F^\text{P2}_{35}-36 F^\text{P2}_{36})|_{(m_W^2 = 0, s = m_t^2, u = 0, t = 0)},
\end{align}
where all the integrals on the right hand side of the above equation are assumed to be known because they exist in lower sectors.

In this work we focus on the physical region.
The analytic continuation of the results from unphysical  to physical regions is realized 
by adding to $s$ a small imaginary part $i\varepsilon~ (\varepsilon>0)$, i.e., $s\rightarrow s+i\varepsilon$ \cite{Chen:2021gjv}.

\subsection{Analytical and numerical results}
After rationalizing the square roots appearing in the differential equations and obtaining the boundary conditions, we can solve the differential equations recursively. The analytical results of the canonical basis can be written in terms of the multiple polylogarithms \cite{Goncharov:1998kja}, which are defined as $G(x)\equiv 1$ and
\bqa
	G_{a_1,a_2,\ldots,a_n}(x) &\equiv & \int_0^x \frac{\text{d} t}{t - a_1} G_{a_2,\ldots,a_n}(t)\, ,\\
	G_{\overrightarrow{0}_n}(x) & \equiv & \frac{1}{n!}\ln^n x\, .
\eqa
The dimension of the vector $(a_1,a_2,\ldots,a_n)$ is called the transcendental $weight$ of the multiple polylogarithm.
We need multiple polylogarithms of at most transcendental weight four in the calculations of two-loop amplitudes. And our
analytic results in terms of  multiple polylogarithms can be evaluated efficiently using the package {\tt PolyLogTools} \cite{Duhr:2019tlz}, which employs {\tt GiNaC} \cite{Vollinga:2004sn,Bauer:2000cp}.
They were compared with the results of numerical computation with the packages {\tt FIESTA} \cite{Smirnov:2021rhf} and {\tt AMFlow} \cite{Liu:2022chg}.
For this P2 integral family, we also examined the difference between our results and those in \cite{Long:2021vse}, and found full agreement.
Here we present the result of $F^\text{P2}_{37}$ at a physical kinematic point $(m_t^2 = 1, m_W^2 = 1/4, s = 10, u = -27/4, t = -2)$,
\begin{align}
F^\text{P2,\text{analytic}}_{37} &= 0.666667 - (6.00309 - 3.14159 i)\epsilon + (8.32922 - 
27.4366 i) \epsilon^2 
\nonumber\\
&\quad+ (56.1661+ 48.2519 i) \epsilon^3 - (113.281 - 
138.028 i) \epsilon^4+{\cal O}(\epsilon^5).
\end{align}
The numerical results of the other canonical basis integrals  are provided in an ancillary file submitted to the arXiv website along with this paper.

\section{Non-planar integral families with two massive propagators}
\label{sec:nonplanar}

The calculation of non-planar integrals is generally more complicated than that of  planar ones. 
In our case, the differential equations may be transformed to the $\ep$-form.
But the occurring  square roots can not be rationalized simultaneously.
Instead of attempting to obtain the full analytical results, we use the method of expansion in $m_W^2$,
which has been proposed and tested in the computation of one-loop squared amplitudes \cite{Chen:2022ntw}, to provide an analytic approximation to the full results. 
Since the two quarks in the $Wtb$ vertices have different masses, setting $m_W^2=0$ does not bring about any additional soft or collinear singularities and thus the full results can be Taylor expanded around $m_W^2=0$. 
After removing this $m_W$ scale, the number of MIs is reduced, as will be shown below,
and some square roots would disappear.
As a consequence, the calculation becomes simplified obviously.
The expansion can be performed to sufficiently high orders, once the MIs in the limit of $m_W \rightarrow 0$ are computed,
and therefore the accuracy can be improved systematically.

In the $m_W \rightarrow 0$ limit, the process $g(k_1) +b(k_2) \longrightarrow W(k_3)+t(k_4)$ contains only one massive final-state particle. 
For the external particles, there are on-shell conditions $k_1^2 = k_2^2  =k_3^2 = 0$ and $k_4^2 = (k_1+k_2-k_3)^2 = m_t^2$. The Mandelstam variables are defined as
\begin{align}
s = (k_1+k_2)^2,\quad t = (k_1-k_3)^2,\quad u = (k_2-k_3)^2
\end{align} 
with $s+t+u = m_t^2$. 

The non-planar integral families in the NP2, NP3 and NP4 topologies can be written in the form 
\begin{align}
	I^\text{NP}_{n_1,n_2,\ldots,n_{9}}=\int{\mathcal D}^D q_1~{\mathcal D}^D q_2\frac{D_8^{-n_8}~D_9^{-n_9}}{D_1^{n_1}~D_2^{n_2}~D_3^{n_3}~D_4^{n_4}~D_5^{n_5}~D_6^{n_6}~D_7^{n_7}},
\end{align}
where all $n_i\ge 0,i=1,\cdots, 7$ and $n_i\le 0,i=8,9$. 
The denominators of these integral families are defined in table \ref{tab:families}.
\begin{table*}[ht]
	\centering
	\begin{tabular}{|c|c|p{0.7 \linewidth}|}
		\hline \hline
		\multicolumn{2}{|c|}{\textbf{Topology}} & \multicolumn{1}{|c|}{\textbf{Definition of denominators}} \\
		\hline \hline \multirow{6}{*}{Non-Planar}
		&  NP2 &
		$
		q_{1}^{2},
		q_{2}^{2} - m_{t}^{2},
		(q_{1} + q_{2})^{2} - m_{t}^{2},
		(q_{1} + k_{1})^{2},
		(q_{1} + q_{2} + k_{4})^{2},
		$
		\newline
		$
		(q_{1} + q_{2} + k_{1} - k_{3})^{2},
		(q_{2} - k_{3})^{2},
		(q_{1} - k_{2})^{2},
		(q_{1} + q_{2} + k_{2} - k_{3})^{2}
		$
		\\
		\cline{2-3}
		& NP3 &
		$
		q_{1}^{2},
		q_{2}^{2},
		(q_{1} - k_{1})^{2},
		(q_{1} + k_{2})^{2},
		(q_{1} + q_{2} - k_{1})^{2},
		$
		\newline
		$
		(q_{1} + q_{2} +k_{2} - k_{3})^{2} - m_{t}^{2},
		(q_{2} - k_{3})^{2} - m_{t}^{2},
		(q_{1} - k_{4})^{2} - m_{t}^{2},
		(q_{2} - k_{1})^{2}
		$
		\\
		\cline{2-3}
		& NP4 &
		$
		q_{1}^{2},
		q_{2}^{2},
		(q_{1} - k_{1})^{2},
		(q_{1} - k_{3})^{2} - m_{t}^{2},
		(q_{1} + q_{2} - k_{1})^{2},
		$
		\newline
		$
		(q_{2} + k_{2})^{2},
		(q_{1} + q_{2} + k_{2} - k_{3})^{2} - m_{t}^{2},
		(q_{2} - k_{1})^{2},
		(q_{1} - k_{2})^{2}
		$
		\\
		\cline{2-3}
		\hline \hline
	\end{tabular}
	\caption{The denominators of the master integrals in NP2, NP3 and NP4 topologies.
	}
	\label{tab:families}
\end{table*}

\subsection{NP2 integral family}
\subsubsection{Canonical basis}
In the $m_W \rightarrow 0$ limit, the integrals in the NP2 family can be reduced to 49 MIs after considering the symmetries.
In contrast, the number of the MIs is 65 if $m_W \neq 0$. We first select the MIs whose differential equations have coefficients linear in $\epsilon$,
\begin{align}
	\M^\text{NP2}_1&= \epsilon^2 I^\text{NP2}_{0,2,2,0,0,0,0,0,0},\quad&
	\M^\text{NP2}_2&= \epsilon^2 I^\text{NP2}_{1,0,0,0,2,0,2,0,0}, \quad\nonumber\\
	\M^\text{NP2}_3&= \epsilon^2 I^\text{NP2}_{2,2,0,0,1,0,0,0,0}, \quad&
	\M^\text{NP2}_4&= (1-\epsilon)\epsilon I^\text{NP2}_{0,2,0,2,0,1,0,0,0},\quad\nonumber\\
	\M^\text{NP2}_5&= \epsilon^2 I^\text{NP2}_{0,2,0,2,1,0,0,0,0}, \quad&
	\M^\text{NP2}_6&= \epsilon^2 I^\text{NP2}_{0,1,0,2,2,0,0,0,0},\quad\nonumber\\
	\M^\text{NP2}_7&= \epsilon^2 I^\text{NP2}_{0,2,2,0,0,1,0,0,0}, \quad&
	\M^\text{NP2}_8&= \epsilon^2 I^\text{NP2}_{1,2,0,0,0,2,0,0,0},\quad\nonumber\\
	\M^\text{NP2}_9&= \epsilon^2 I^\text{NP2}_{2,1,0,0,0,2,0,0,0}, \quad&
	\M^\text{NP2}_{10}&=(1-2\epsilon)\epsilon^2 I^\text{NP2}_{1,0,2,0,0,1,1,0,0},\quad\nonumber\\
	\M^\text{NP2}_{11}&= \epsilon^3 I^\text{NP2}_{1,0,1,0,1,0,2,0,0}, \quad&
	\M^\text{NP2}_{12}&= \epsilon^3 I^\text{NP2}_{0,2,1,1,1,0,0,0,0},\quad\nonumber\\
	\M^\text{NP2}_{13}&= \epsilon^2 I^\text{NP2}_{0,3,1,1,1,0,0,0,0}, \quad&
	\M^\text{NP2}_{14}&= \epsilon^3 I^\text{NP2}_{1,1,0,0,2,0,1,0,0},\quad\nonumber\\
	\M^\text{NP2}_{15}&= \epsilon^3 I^\text{NP2}_{0,2,1,1,0,1,0,0,0}, \quad&
	\M^\text{NP2}_{16}&= \epsilon^2 I^\text{NP2}_{0,3,1,1,0,1,0,0,0},\quad\nonumber\\
	\M^\text{NP2}_{17}&= \epsilon^3 I^\text{NP2}_{0,1,2,1,0,0,1,0,0}, \quad&
	\M^\text{NP2}_{18}&= \epsilon^2 I^\text{NP2}_{0,1,3,1,0,0,1,0,0},\quad\nonumber\\
	\M^\text{NP2}_{19}&= \epsilon^3 I^\text{NP2}_{1,0,1,0,1,1,2,0,0}, \quad&
	\M^\text{NP2}_{20}&= \epsilon^4 I^\text{NP2}_{0,1,1,1,0,1,1,0,0},\quad\nonumber\\
	\M^\text{NP2}_{21}&= (1-2\epsilon)\epsilon^3 I^\text{NP2}_{0,1,1,1,1,1,0,0,0},\quad&
	\M^\text{NP2}_{22}&= \epsilon^3 I^\text{NP2}_{0,2,1,1,1,1,0,0,0},\quad\nonumber\\
	\M^\text{NP2}_{23}&= \epsilon^4 I^\text{NP2}_{1,0,1,1,1,0,1,0,0}, \quad&
	\M^\text{NP2}_{24}&= \epsilon^3 I^\text{NP2}_{1,0,2,1,1,0,1,0,0},\quad\nonumber\\
	\M^\text{NP2}_{25}&= \epsilon^4 I^\text{NP2}_{1,1,0,0,1,1,1,0,0}, \quad&
	\M^\text{NP2}_{26}&= \epsilon^3 I^\text{NP2}_{1,2,0,0,1,1,1,0,0},\quad\nonumber\\
	\M^\text{NP2}_{27}&= \epsilon^4 I^\text{NP2}_{1,1,0,1,1,0,1,0,0}, \quad&
	\M^\text{NP2}_{28}&= \epsilon^3 I^\text{NP2}_{1,2,0,1,1,0,1,0,0},\quad\nonumber\\
	\M^\text{NP2}_{29}&= \epsilon^4 I^\text{NP2}_{1,1,0,1,1,1,0,0,0}, \quad&
	\M^\text{NP2}_{30}&= \epsilon^3 I^\text{NP2}_{1,2,0,1,1,1,0,0,0},\quad\nonumber\\
	\M^\text{NP2}_{31}&= \epsilon^4 I^\text{NP2}_{1,1,1,0,0,1,1,0,0}, \quad&
	\M^\text{NP2}_{32}&= \epsilon^3 I^\text{NP2}_{1,2,1,0,0,1,1,0,0},\quad\nonumber\\
	\M^\text{NP2}_{33}&= \epsilon^4 I^\text{NP2}_{1,1,1,0,1,0,1,0,0}, \quad&
	\M^\text{NP2}_{34}&= \epsilon^3 I^\text{NP2}_{1,2,1,0,1,0,1,0,0},\quad\nonumber\\
	\M^\text{NP2}_{35}&= \epsilon^4 I^\text{NP2}_{0,1,1,1,1,0,1,0,0}, \quad&
	\M^\text{NP2}_{36}&= \epsilon^3 I^\text{NP2}_{0,2,1,1,1,0,1,0,0},\quad\nonumber\\
	\M^\text{NP2}_{37}&= \epsilon^3 I^\text{NP2}_{0,1,2,1,1,0,1,0,0}, \quad&
	\M^\text{NP2}_{38}&= \epsilon^3 I^\text{NP2}_{0,1,1,2,1,0,1,0,0},\quad\nonumber\\
	\M^\text{NP2}_{39}&= \epsilon^4 I^\text{NP2}_{1,1,1,1,0,1,1,0,0}, \quad&
	\M^\text{NP2}_{40}&= \epsilon^4 I^\text{NP2}_{1,1,0,1,1,1,1,0,0},\quad\nonumber\\
	\M^\text{NP2}_{41}&= \epsilon^4 I^\text{NP2}_{1,1,1,0,1,1,1,0,0}, \quad&
	\M^\text{NP2}_{42}&= \epsilon^4 I^\text{NP2}_{1,1,1,0,1,1,1,0,-1},\quad\nonumber\\
	\M^\text{NP2}_{43}&= \epsilon^4 I^\text{NP2}_{1,1,1,1,1,0,1,0,0}, \quad&
	\M^\text{NP2}_{44}&= \epsilon^4 I^\text{NP2}_{1,1,1,1,1,-1,1,0,0},\quad\nonumber\\
	\M^\text{NP2}_{45}&= \epsilon^4 I^\text{NP2}_{1,1,1,1,1,0,1,-1,0}, \quad&
	\M^\text{NP2}_{46}&= \epsilon^4 I^\text{NP2}_{1,1,1,1,1,0,1,0,-1},\quad\nonumber\\
	\M^\text{NP2}_{47}&= \epsilon^4 I^\text{NP2}_{1,1,1,1,1,1,1,0,0}, \quad&
	\M^\text{NP2}_{48}&= \epsilon^4 I^\text{NP2}_{1,1,1,1,1,1,1,-1,0},\quad\nonumber\\
	\M^\text{NP2}_{49}&= \epsilon^4 I^\text{NP2}_{1,1,1,1,1,1,1,0,-1}.
\end{align}
The corresponding topology diagrams are displayed in figure \ref{fig:NP2_MIs} in the  appendix.

The canonical basis integrals $F^\text{NP2}_{i},~i=1,\dots,49,$ can be constructed as the linear combinations of  $\M^\text{NP2}_{i}$,
\begin{align}
	F^\text{NP2}_1&=\M^\text{NP2}_1,\quad F^\text{NP2}_2=\M^\text{NP2}_2 (-s),\quad F^\text{NP2}_3=\M^\text{NP2}_3 m_t^2,\quad F^\text{NP2}_4=\M^\text{NP2}_4 m_t^2,\quad
	\nonumber\\
	F^\text{NP2}_5&=\M^\text{NP2}_5 u,\quad
	F^\text{NP2}_6=\M^\text{NP2}_6 \left(u-m_t^2\right)-2 \M^\text{NP2}_5 m_t^2,\quad 
	F^\text{NP2}_7=\M^\text{NP2}_7 \left(m_t^2-s-u\right),\quad 
	\nonumber\\
	F^\text{NP2}_8&=\M^\text{NP2}_8 \left(m_t^2-s-u\right),\quad 
	F^\text{NP2}_9=-\M^\text{NP2}_9 (s+u)-2 \M^\text{NP2}_8 m_t^2,\quad
	\nonumber\\
	F^\text{NP2}_{10}&=\M^\text{NP2}_{10} \left(m_t^2-s-u\right),\quad 
	F^\text{NP2}_{11}=\M^\text{NP2}_{11} \left(m_t^2-s\right),\quad 
	F^\text{NP2}_{12}=\M^\text{NP2}_{12} \left(u-m_t^2\right),\quad
	\nonumber\\
	F^\text{NP2}_{13}&=-\M^\text{NP2}_{13} m_t^2 \left(m_t^2-u\right),\quad F^\text{NP2}_{14}=\M^\text{NP2}_{14} \left(m_t^2-s\right),\quad
	F^\text{NP2}_{15}=\M^\text{NP2}_{15} \left(m_t^2-s-u\right),\quad
	\nonumber\\
	F^\text{NP2}_{16}&=\M^\text{NP2}_{16} m_t^2 \left(m_t^2-s-u\right),\quad F^\text{NP2}_{17}=\M^\text{NP2}_{17} \left(m_t^2-s-u\right),\quad
	\nonumber\\
	F^\text{NP2}_{18}&=\M^\text{NP2}_{18} m_t^2 \left(m_t^2-s-u\right),\quad 
	F^\text{NP2}_{19}=\M^\text{NP2}_{19} s (s+u),\quad 
	F^\text{NP2}_{20}=\M^\text{NP2}_{20} \left(m_t^2-s-u\right),\quad
	\nonumber\\ 
	F^\text{NP2}_{21}&=\M^\text{NP2}_{22} m_t^2 (s+u)+\M^\text{NP2}_{21} (-s-u),\quad 
	F^\text{NP2}_{22}=\M^\text{NP2}_{22} (-u) (s+u),\quad 
	\nonumber\\
	F^\text{NP2}_{23}&=\M^\text{NP2}_{23} \left(u-m_t^2\right),\quad 
	F^\text{NP2}_{24}=\M^\text{NP2}_{24} (-s) m_t^2,\quad
	F^\text{NP2}_{25}=\M^\text{NP2}_{25} u,\quad
	F^\text{NP2}_{26}=\M^\text{NP2}_{26} u m_t^2,\quad
	\nonumber\\
	F^\text{NP2}_{27}&=\M^\text{NP2}_{27} \left(m_t^2-s-u\right),\quad
	F^\text{NP2}_{28}=\M^\text{NP2}_{28} m_t^2 \left(m_t^2-s-u\right),\quad F^\text{NP2}_{29}=\M^\text{NP2}_{29} (-s),\quad
	\nonumber\\
	F^\text{NP2}_{30}&=\M^\text{NP2}_{30} (s+u) \left(m_t^2-u\right),\quad
	F^\text{NP2}_{31}=\M^\text{NP2}_{31} \left(m_t^2-s-u\right),\quad
	\nonumber\\
	F^\text{NP2}_{32}&=\M^\text{NP2}_{32} m_t^2 \left(m_t^2-s-u\right),\quad 
	F^\text{NP2}_{33}=\M^\text{NP2}_{33} \left(m_t^2-s\right),\quad 
	F^\text{NP2}_{34}=\M^\text{NP2}_{34} m_t^2 \left(m_t^2-s\right),\quad
	\nonumber\\ 
	F^\text{NP2}_{35}&=\M^\text{NP2}_{35} \left(m_t^2-s\right),\quad 
	F^\text{NP2}_{36}=\M^\text{NP2}_{36} m_t^2 \left(m_t^2-s\right),\quad 
	F^\text{NP2}_{37}=\M^\text{NP2}_{37} u m_t^2,\quad
	\nonumber\\ 
	F^\text{NP2}_{38}&=\M^\text{NP2}_{38} (s+u) \left(m_t^2-u\right),\quad F^\text{NP2}_{39}=\M^\text{NP2}_{39} \left(-m_t^2+s+u\right)^2,\quad
	F^\text{NP2}_{40}=\M^\text{NP2}_{40} s m_t^2,\quad
	\nonumber\\ 
	F^\text{NP2}_{41}&=\M^\text{NP2}_{41} (s+u) \left(m_t^2-s\right),\quad
	\nonumber\\
	F^\text{NP2}_{42}&=-\M^\text{NP2}_{31} m_t^2+\M^\text{NP2}_{33} (s+u)+\M^\text{NP2}_{42} (s+u)-\M^\text{NP2}_{25} s,\quad 
	F^\text{NP2}_{43}=\M^\text{NP2}_{43} r^{\text{NP2}}_1,\quad 
	\nonumber\\
	F^\text{NP2}_{44}&=-\frac{\M^\text{NP2}_4 m_t^2 \left(m_t^2-u\right)}{4 \left(m_t^2-s-u\right)}-\frac{\M^\text{NP2}_8 \left(m_t^2-u\right) \left(2 m_t^2-s-u\right)}{2 \left(m_t^2-s-u\right)}
	\nonumber\\&\quad
	-\frac{\M^\text{NP2}_9 (s+u) \left(m_t^2-u\right)}{4 \left(m_t^2-s-u\right)}+\M^\text{NP2}_{17} \left(u-m_t^2\right)
	-\M^\text{NP2}_{18} m_t^2 \left(m_t^2-u\right)
	\nonumber\\&\quad
	+\M^\text{NP2}_{27} \left(m_t^2-u\right)+\M^\text{NP2}_{43} s \left(u-m_t^2\right)+\M^\text{NP2}_{44} \left(u-m_t^2\right)+\M^\text{NP2}_{46} \left(u-m_t^2\right),
	\nonumber\\
	F^\text{NP2}_{45}&=
	\frac{\M^\text{NP2}_3 m_t^2 \left(-m_t^2+s+u\right)}{m_t^2-u}
	-\frac{1}{4} \M^\text{NP2}_4 m_t^2	+\frac{\M^\text{NP2}_5 \left(m_t^2+u\right) \left(m_t^2-s-u\right)}{2 \left(m_t^2-u\right)}
	\nonumber\\&\quad
	+\frac{1}{4} \M^\text{NP2}_6 \left(m_t^2-s-u\right)+\frac{1}{2} 
	\M^\text{NP2}_8 \left(-2 m_t^2+s+u\right)
	-\frac{1}{4} \M^\text{NP2}_9 (s+u)
	\nonumber\\&\quad
	+\M^\text{NP2}_{12} \left(-m_t^2+s+u\right)	+\M^\text{NP2}_{13} m_t^2 \left(-m_t^2+s+u\right)+\M^\text{NP2}_{17} \left(-m_t^2+s+u\right)
	\nonumber\\&\quad
	+\M^\text{NP2}_{18} m_t^2 \left(-m_t^2+s+u\right)+\M^\text{NP2}_{23} \left(-m_t^2+s+u\right)+\M^\text{NP2}_{27} \left(m_t^2-s-u\right)
	\nonumber\\&\quad
	+\M^\text{NP2}_{33} \left(-m_t^2+s+u\right)+\M^\text{NP2}_{43} s \left(-m_t^2+s+u\right)+\M^\text{NP2}_{44} \left(-m_t^2+s+u\right)
	\nonumber\\&\quad
	+\M^\text{NP2}_{45} \left(m_t^2-s-u\right)+\M^\text{NP2}_{46} \left(-m_t^2+s+u\right),\nonumber\\
	F^\text{NP2}_{46}&=\frac{\M^\text{NP2}_4 m_t^2 \left(m_t^2-s\right)}{4 \left(m_t^2-s-u\right)}+\frac{\M^\text{NP2}_8 \left(m_t^2-s\right) \left(2 m_t^2-s-u\right)}{2 \left(m_t^2-s-u\right)}+\frac{\M^\text{NP2}_9 (s+u) \left(m_t^2-s\right)}{4 \left(m_t^2-s-u\right)}
	\nonumber\\&\quad
+\M^\text{NP2}_{17} \left(m_t^2-s\right)+\M^\text{NP2}_{18} m_t^2 \left(m_t^2-s\right)	+\M^\text{NP2}_{43} \left(m_t^2-s\right) \left(m_t^2-u\right)+\M^\text{NP2}_{46} \left(m_t^2-s\right),
	\nonumber\\
	F^\text{NP2}_{47}&=\M^\text{NP2}_{39} s (s+u)+\M^\text{NP2}_{40} s \left(-m_t^2+s+u\right)-\M^\text{NP2}_{47} s (s+u) \left(m_t^2-u\right)-\M^\text{NP2}_{49} s (s+u),
	\nonumber\\
	F^\text{NP2}_{48}&=\frac{\M^\text{NP2}_1 \left(-m_t^2+s+u\right)}{4 \left(m_t^2-u\right)}-\frac{\M^\text{NP2}_4 m_t^4 \left(m_t^2-s-u\right)}{4 u \left(m_t^2-u\right)}
	-\frac{\M^\text{NP2}_5 \left(m_t^2+2 u\right) \left(m_t^2-s-u\right)}{2 u}
	\nonumber\\&\quad
	-\frac{\M^\text{NP2}_6 \left(m_t^2-u\right) \left(m_t^2-s-u\right)}{4 u}	
	+\frac{\M^\text{NP2}_7 \left(-m_t^2+s+u\right)^2}{m_t^2-u}
	\nonumber\\&\quad
	+\frac{\M^\text{NP2}_{15} \left(2 m_t^2+u\right) \left(-m_t^2+s+u\right)^2}{2 u \left(m_t^2-u\right)}
    +\frac{\M^\text{NP2}_{16} m_t^2 \left(-m_t^2+s+u\right)^2}{m_t^2-u}	
	\nonumber\\&\quad    
    +\frac{\M^\text{NP2}_{17} \left(-m_t^2+s+u\right)^2}{u} 	
	+\frac{\M^\text{NP2}_{20} \left(-m_t^2+s+u\right)^2}{u}
    -\frac{3 \M^\text{NP2}_{21} (s+u) \left(m_t^2-s-u\right)}{2 \left(m_t^2-u\right)}	
	\nonumber\\&\quad
	+\frac{\M^\text{NP2}_{22} (s+u) \left(5 m_t^2-u\right) \left(m_t^2-s-u\right)}{2 \left(m_t^2-u\right)}-\frac{1}{2} \M^\text{NP2}_{30} (s+u) \left(-m_t^2+s+u\right)
	\nonumber\\&\quad
	+\frac{3 \M^\text{NP2}_{35} \left(m_t^2-s\right) \left(m_t^2-s-u\right)}{u}+\frac{\M^\text{NP2}_{36} m_t^2 \left(m_t^2-s\right) \left(m_t^2-s-u\right)}{u}
	\nonumber\\&\quad	
    +2 \M^\text{NP2}_{37} m_t^2 \left(m_t^2-s-u\right)	
	-\M^\text{NP2}_{39} (s+u) \left(-m_t^2+s+u\right)
	\nonumber\\&\quad
	-\M^\text{NP2}_{40} s \left(-m_t^2+s+u\right)+\M^\text{NP2}_{41} (s+u) \left(-m_t^2+s+u\right)
	\nonumber\\&\quad
	+\M^\text{NP2}_{43} (s+u) \left(-m_t^2+s+u\right)+\M^\text{NP2}_{47} s (s+u) \left(-m_t^2+s+u\right)
	\nonumber\\&\quad
	-\M^\text{NP2}_{48} (s+u) \left(-m_t^2+s+u\right)+\M^\text{NP2}_{49} (s+u) \left(-m_t^2+s+u\right),
	\nonumber\\
	F^\text{NP2}_{49}&=\M^\text{NP2}_{39} m_t^2 \left(-m_t^2+s+u\right)+\M^\text{NP2}_{40} s \left(-m_t^2+s+u\right)+\M^\text{NP2}_{43} (s+u) \left(m_t^2-u\right)
	\nonumber\\&\quad
	-\M^\text{NP2}_{49} (s+u) \left(-m_t^2+s+u\right),
\end{align}
where 
\begin{align}
	r^\text{NP2}_1 = \sqrt{m_t^2 \left(m_t^2-u\right) \left(-m_t^2 (4 s+u)+m_t^4+4 s (s+u)\right)}
\end{align}
appears in the basis integrals $F^\text{NP2}_{i}$ with $i\ge 43$.
For the calculations of the basis integrals independent of $r^\text{NP2}_1$, we define dimensionless variables $y$ and $z$ via
\begin{align}
    s = m_t^2(1-y-z),\quad t = m_t^2y,\quad u = m_t^2z.
\end{align}
For the integrals that rely on 	$r^\text{NP2}_1$, 
 we define dimensionless variables $y_1$ and $z_1$ by
\begin{align}
	s &= \frac{y_1}{\left(y_1+1\right) \left(z_1+1\right)}m_t^2,\nonumber\\
	u &= \frac{z_1 \left(y_1-z_1-1\right)}{\left(z_1+1\right) \left(y_1-z_1\right)}m_t^2 ,
\end{align}
so that $r^\text{NP2}_1$ is rationalized to
\begin{align}
	r^\text{NP2}_1 = \frac{\left(y_1-1\right) y_1}{\left(y_1+1\right) \left(z_1+1\right) \left(y_1-z_1\right)}m_t^4.
\end{align}
Below we will utilize the boundary condition at $u=0$,
which is reproduced by $z_1 = 0$.

\subsubsection{Boundary conditions and numerical check}
The analytical results of the single scale integrals $F^\text{NP2}_1$, $F^\text{NP2}_3$ and $F^\text{NP2}_4$ can be directly computed,
\begin{align}
	F^\text{NP2}_1 &= 1,\nonumber\\
	F^\text{NP2}_3 &= -\frac{1}{4}-\frac{\pi ^2}{6}\epsilon^2-2 \zeta (3)\epsilon^3-\frac{8\pi ^4}{45}\epsilon^4+ {\cal O}(\epsilon^5),\nonumber\\
	F^\text{NP2}_4 &= 1+\frac{\pi ^2}{3}\epsilon^2-2 \zeta (3)\epsilon^3+\frac{\pi ^4}{10}\epsilon^4+ {\cal O}(\epsilon^5).
\end{align}
And the boundary conditions of $F^\text{NP2}_2$, $F^\text{NP2}_6$ and $F^\text{NP2}_9$ can be obtained at special kinematic points,
\begin{align}
	F^\text{NP2}_2|_{s=m_t^2} &= -1-2 i \pi\epsilon+\frac{7 \pi ^2}{3}\epsilon^2 + \left(10 \zeta (3)+2 i \pi ^3\right)\epsilon^3+\left(-\frac{109 \pi ^4}{90}+20 i \pi  \zeta (3)\right)\epsilon^4+ {\cal O}(\epsilon^5),\nonumber\\
	F^\text{NP2}_6|_{u=0}& = F^\text{NP2}_9|_{t=0} = 1+\frac{\pi ^2}{3}\epsilon^2-2\zeta (3)\epsilon^3+\frac{\pi^4}{10}\epsilon^4+ {\cal O}(\epsilon^5).
\end{align}
The boundary conditions of $F^\text{NP2}_{12}$, $F^\text{NP2}_{13}$ and $F^\text{NP2}_{23}$ can be found in $u = m_t^2$,
\begin{align}
	F^\text{NP2}_{12}|_{u=m_t^2} = F^\text{NP2}_{13}|_{u=m_t^2} = F^\text{NP2}_{23}|_{u=m_t^2} = 0.
\end{align}
The boundary conditions of the other canonical basis integrals can be found at the regular point 
$ (s = m_t^2, u = 0, t = 0) $,
where the differential equations should not develop any poles.
For example,
\begin{align}
&F^\text{NP2}_{49}\Big|_{(s = m_t^2, u = 0, t = 0)} \nonumber\\
=& \frac{1}{4}\Big(F^\text{NP2}_2-4 F^\text{NP2}_3+2 F^\text{NP2}_4-F^\text{NP2}_6+4 F^\text{NP2}_8-2 F^\text{NP2}_9-2 F^\text{NP2}_{11}+2 F^\text{NP2}_{12}
+12 F^\text{NP2}_{17}
\nonumber\\&
+8 F^\text{NP2}_{18}-4 F^\text{NP2}_{23}-4 F^\text{NP2}_{24}+4 F^\text{NP2}_{27}-4 F^\text{NP2}_{28}-12 F^\text{NP2}_{33}-4 F^\text{NP2}_{34}+16 F^\text{NP2}_{35}+4 F^\text{NP2}_{36}
\nonumber\\&
+4 F^\text{NP2}_{37}-2 F^\text{NP2}_{38}-8 F^\text{NP2}_{43}-8 F^\text{NP2}_{44}-8 F^\text{NP2}_{45}-8 F^\text{NP2}_{46}\Big)\Big|_{(s = m_t^2, u = 0, t = 0)} \,.
\end{align}

Then it is straightforward to solve the differential equations in terms of multiple polylogarithms.
All the analytical results have been checked with the numerical packages {\tt FIESTA} \cite{Smirnov:2021rhf} and {\tt AMFlow} \cite{Liu:2022chg}. Here we present the result of $F^\text{NP2}_{47}$ at a physical kinematic point $(m_t^2 = 1, s = 10, u = -9/4, t = -27/4)$,
\begin{align}
	F^\text{NP2,\text{analytic}}_{47} &= -0.458333 + (3.40540 - 1.57080 i)\epsilon + (2.82240 + 13.8387 i) \epsilon^2 
	\nonumber\\
	&\quad+ (-24.8447 + 5.25538 i) \epsilon^3 + (12.0588 - 15.0124 i) \epsilon^4+{\cal O}(\epsilon^5).
\end{align}
The results of the other basis integrals are provided in the ancillary file.

\subsection{NP3 integral family}
\subsubsection{Canonical basis}
In the $m_W \rightarrow 0$ limit,  the non-planar NP3 integral family contains 60 MIs after considering the integral symmetries. 
In contrast, the number of MIs is 76 when keeping finite $m_W$.
In our method to find the canonical basis, we first select the MIs whose differential equations have coefficients linear in $\epsilon$,
\begin{align}
\M^\text{NP3}_{1}  &= \epsilon^2 I^\text{NP3}_{0,0,0,0,0,2,2,0,0},\quad&
\M^\text{NP3}_{2}  &= \epsilon^2 I^\text{NP3}_{0,1,0,2,2,0,0,0,0},\quad\nonumber\\
\M^\text{NP3}_{3}  &= \epsilon^2 I^\text{NP3}_{0,1,2,0,0,2,0,0,0},\quad&
\M^\text{NP3}_{4}  &=\epsilon^2 I^\text{NP3}_{0,0,1,2,0,2,0,0,0},\quad\nonumber\\
\M^\text{NP3}_{5}  &=\epsilon^2 I^\text{NP3}_{0,0,1,0,0,2,2,0,0},\quad&
\M^\text{NP3}_{6}  &=\epsilon^2 I^\text{NP3}_{0,0,2,0,0,2,1,0,0},\quad\nonumber\\
\M^\text{NP3}_{7}  &=(1-\epsilon)\epsilon I^\text{NP3}_{0,2,0,1,0,2,0,0,0},\quad&
\M^\text{NP3}_{8}  &=\epsilon^2 I^\text{NP3}_{1,0,0,0,2,0,2,0,0},\quad\nonumber\\
\M^\text{NP3}_{9}  &=\epsilon^2 I^\text{NP3}_{2,0,0,0,2,0,1,0,0},\quad&
\M^\text{NP3}_{10}  &=\epsilon^2 I^\text{NP3}_{1,2,0,0,0,2,0,0,0},\quad\nonumber\\
\M^\text{NP3}_{11}  &=\epsilon^2 I^\text{NP3}_{2,2,0,0,0,1,0,0,0},\quad&
\M^\text{NP3}_{12}  &=\epsilon^3 I^\text{NP3}_{0,2,0,1,1,1,0,0,0},\quad\nonumber\\
\M^\text{NP3}_{13}  &=\epsilon^3 I^\text{NP3}_{0,0,1,0,1,1,2,0,0},\quad&
\M^\text{NP3}_{14}  &=\epsilon^3 I^\text{NP3}_{0,1,0,1,2,0,1,0,0},\quad\nonumber\\
\M^\text{NP3}_{15}  &=\epsilon^3 I^\text{NP3}_{1,0,0,0,1,1,2,0,0},\quad&
\M^\text{NP3}_{16}  &=\epsilon^2 I^\text{NP3}_{1,0,0,0,1,1,3,0,0},\quad\nonumber\\
\M^\text{NP3}_{17}  &=\epsilon^3 I^\text{NP3}_{0,1,1,1,0,2,0,0,0},\quad&
\M^\text{NP3}_{18}  &=\epsilon^2 I^\text{NP3}_{0,1,1,1,0,3,0,0,0},\quad\nonumber\\
\M^\text{NP3}_{19}  &=\epsilon^3 I^\text{NP3}_{0,1,1,0,0,2,1,0,0},\quad&
\M^\text{NP3}_{20}  &=\epsilon^2 I^\text{NP3}_{0,1,1,0,0,3,1,0,0},\quad\nonumber\\
\M^\text{NP3}_{21}  &=\epsilon^3 I^\text{NP3}_{1,1,0,0,0,2,1,0,0},\quad&
\M^\text{NP3}_{22}  &=\epsilon^2 I^\text{NP3}_{1,1,0,0,0,3,1,0,0},\quad\nonumber\\
\M^\text{NP3}_{23}  &=\epsilon^4 I^\text{NP3}_{0,1,1,1,1,0,1,0,0},\quad&
\M^\text{NP3}_{24}  &=\epsilon^4 I^\text{NP3}_{0,0,1,1,1,1,1,0,0},\quad\nonumber\\
\M^\text{NP3}_{25}  &=\epsilon^2 I^\text{NP3}_{0,0,1,1,1,1,2,0,0},\quad&
\M^\text{NP3}_{26}  &=\epsilon^3 I^\text{NP3}_{1,0,1,1,1,0,2,0,0},\quad\nonumber\\
\M^\text{NP3}_{27}  &=\epsilon^2 I^\text{NP3}_{1,0,1,1,1,0,3,0,0},\quad&
\M^\text{NP3}_{28}  &=\epsilon^3 I^\text{NP3}_{1,1,1,1,0,2,0,0,0},\quad\nonumber\\
\M^\text{NP3}_{29}  &=\epsilon^2 I^\text{NP3}_{1,1,1,1,0,3,0,0,0},\quad&
\M^\text{NP3}_{30}  &=\epsilon^4 I^\text{NP3}_{1,1,0,1,1,0,1,0,0},\quad\nonumber\\
\M^\text{NP3}_{31}  &=\epsilon^3 I^\text{NP3}_{1,1,0,1,1,0,2,0,0},\quad&
\M^\text{NP3}_{32}  &=\epsilon^4 I^\text{NP3}_{1,1,0,1,1,1,0,0,0},\quad\nonumber\\
\M^\text{NP3}_{33}  &=\epsilon^3 I^\text{NP3}_{1,1,0,1,1,2,0,0,0},\quad&
\M^\text{NP3}_{34}  &=\epsilon^4 I^\text{NP3}_{0,1,0,1,1,1,1,0,0},\quad\nonumber\\
\M^\text{NP3}_{35}  &=\epsilon^3 I^\text{NP3}_{0,1,0,1,1,1,2,0,0},\quad&
\M^\text{NP3}_{36}  &=\epsilon^4 I^\text{NP3}_{0,1,1,1,0,1,1,0,0},\quad\nonumber\\
\M^\text{NP3}_{37}  &=\epsilon^3 I^\text{NP3}_{0,2,1,1,0,1,1,0,0},\quad&
\M^\text{NP3}_{38}  &=\epsilon^4 I^\text{NP3}_{1,0,1,0,1,1,1,0,0},\quad\nonumber\\
\M^\text{NP3}_{39}  &=\epsilon^3 I^\text{NP3}_{1,0,1,0,1,1,2,0,0},\quad&
\M^\text{NP3}_{40}  &=\epsilon^3 I^\text{NP3}_{1,0,1,0,1,2,1,0,0},\quad\nonumber\\
\M^\text{NP3}_{41}  &=\epsilon^4 I^\text{NP3}_{1,1,1,0,0,1,1,0,0},\quad&
\M^\text{NP3}_{42}  &=\epsilon^3 I^\text{NP3}_{1,1,1,0,0,2,1,0,0},\quad\nonumber\\
\M^\text{NP3}_{43}  &=\epsilon^3 I^\text{NP3}_{1,1,1,0,0,1,2,0,0},\quad&
\M^\text{NP3}_{44}  &=\epsilon^4 I^\text{NP3}_{1,1,0,0,1,1,1,0,0},\quad\nonumber\\
\M^\text{NP3}_{45}  &=\epsilon^3 I^\text{NP3}_{1,1,0,0,1,1,2,0,0},\quad&
\M^\text{NP3}_{46}  &=\epsilon^3 I^\text{NP3}_{1,1,0,0,1,2,1,0,0},\quad\nonumber\\
\M^\text{NP3}_{47}  &=\epsilon^3 I^\text{NP3}_{2,1,0,0,1,1,1,0,0},\quad&
\M^\text{NP3}_{48}  &=\epsilon^4 I^\text{NP3}_{0,1,1,1,1,1,1,0,0},\quad\nonumber\\
\M^\text{NP3}_{49}  &=\epsilon^4 I^\text{NP3}_{1,0,1,1,1,1,1,0,0},\quad&
\M^\text{NP3}_{50}  &=\epsilon^4 I^\text{NP3}_{1,1,1,0,1,1,1,0,0},\quad\nonumber\\
\M^\text{NP3}_{51}  &=\epsilon^4 I^\text{NP3}_{1,1,1,0,1,1,1,-1,0},\quad&
\M^\text{NP3}_{52}  &=\epsilon^4 I^\text{NP3}_{1,1,1,1,0,1,1,0,0},\quad\nonumber\\
\M^\text{NP3}_{53}  &=\epsilon^4 I^\text{NP3}_{1,1,0,1,1,1,1,0,0},\quad&
\M^\text{NP3}_{54}  &=\epsilon^4 I^\text{NP3}_{1,1,-1,1,1,1,1,0,0},\quad\nonumber\\
\M^\text{NP3}_{55}  &=\epsilon^4 I^\text{NP3}_{1,1,0,1,1,1,1,-1,0},\quad&
\M^\text{NP3}_{56}  &=\epsilon^4 I^\text{NP3}_{1,1,0,1,1,1,1,0,-1},\quad\nonumber\\
\M^\text{NP3}_{57}  &=\epsilon^4 I^\text{NP3}_{1,1,1,1,1,1,1,0,0},\quad&
\M^\text{NP3}_{58}  &=\epsilon^4 I^\text{NP3}_{1,1,1,1,1,1,1,-1,0},\quad\nonumber\\
\M^\text{NP3}_{59}  &=\epsilon^4 I^\text{NP3}_{1,1,1,1,1,1,1,0,-1},\quad&
\M^\text{NP3}_{60}  &=\epsilon^4 I^\text{NP3}_{1,1,1,1,1,1,1,-1,-1}.
\end{align}
The corresponding topology diagrams are displayed in figure \ref{fig:NP3_MIs} in the appendix.

The canonical basis $F^\text{NP3}_{i},~i=1,\dots,60,$ can be constructed as linear combinations of  $\M^\text{NP3}_{i}$,
\begin{align}
	F^\text{NP3}_1 &= \M^\text{NP3}_1,\quad F^\text{NP3}_2 =\M^\text{NP3}_2 (-s),\quad F^\text{NP3}_3 = \M^\text{NP3}_3 m_t^2,\quad F^\text{NP3}_4 = \M^\text{NP3}_4 (-s),\quad
	\nonumber\\
	F^\text{NP3}_5 &= \M^\text{NP3}_5 (-s),\quad
	F^\text{NP3}_6 =\frac{\M^\text{NP3}_5 r^\text{NP3}_1}{2}+\M^\text{NP3}_6 r^\text{NP3}_1,\quad F^\text{NP3}_7 =\M^\text{NP3}_7 m_t^2,\quad 
	\nonumber\\
	F^\text{NP3}_8 &= \M^\text{NP3}_8 (m_t^2-s-u),\quad 
	F^\text{NP3}_9 = -2 \M^\text{NP3}_8 m_t^2-\M^\text{NP3}_9 (s+u),\quad
	F^\text{NP3}_{10}= u\M^\text{NP3}_{10},\quad 
	\nonumber\\
	F^\text{NP3}_{11} &= \M^\text{NP3}_{11} (u-m_t^2)-2 \M^\text{NP3}_{10}m_t^2,\quad
	F^\text{NP3}_{12}= \M^\text{NP3}_{12} (m_t^2-s),\quad F^\text{NP3}_{13} = \M^\text{NP3}_{13}(m_t^2-s),\quad
	\nonumber\\	
	F^\text{NP3}_{14} &= (m_t^2 - s) \M^\text{NP3}_{14},\quad F^\text{NP3}_{15} = -\M^\text{NP3}_{15} (s+u),\quad 	
	F^\text{NP3}_{16} = -\M^\text{NP3}_{16} (s+u)m_t^2, \quad 
	\nonumber\\	
	F^\text{NP3}_{17} &= \M^\text{NP3}_{17} (m_t^2-s),\quad
	F^\text{NP3}_{18} = \M^\text{NP3}_{18} (m_t^2-s)m_t^2,\quad F^\text{NP3}_{19} = \M^\text{NP3}_{19}\left(m_t^2-s\right),\quad
	\nonumber\\ 
	F^\text{NP3}_{20} &= \M^\text{NP3}_{20} \left(m_t^2-s\right)m_t^2,\quad 
	F^\text{NP3}_{21} = u \M^\text{NP3}_{21},\quad
	F^\text{NP3}_{22} = \M^\text{NP3}_{22}u\,m_t^2,\quad, 
	\nonumber\\
	F^\text{NP3}_{23} &=-\M^\text{NP3}_{23} \left(m_t^2-s\right),\quad
	F^\text{NP3}_{24} =-\M^\text{NP3}_{24} \left(m_t^2-s\right),\quad
	\nonumber\\
	F^\text{NP3}_{25} &= 
	\frac{\M^\text{NP3}_1 m_t^2}{2 \left(m_t^2-s\right)}
	+\frac{3 \M^\text{NP3}_5 s m_t^2}{2 \left(s-m_t^2\right)}	
	+\frac{\M^\text{NP3}_7 m_t^4}{2 \left(s-m_t^2\right)}
	+\M^\text{NP3}_{13} m_t^2
	-2 \M^\text{NP3}_{17} m_t^2	
	-2 \M^\text{NP3}_{18} m_t^4
	\nonumber\\&\quad
	+4 \M^\text{NP3}_{24} m_t^2
	+\M^\text{NP3}_{25} sm_t^2,\quad
	\nonumber\\
	F^\text{NP3}_{26} &= \M^\text{NP3}_{26} s \left(-m_t^2+s+u\right),\quad F^\text{NP3}_{27} = \M^\text{NP3}_{26}sm_t^2+\M^\text{NP3}_{27} s (s+u)m_t^2,\quad
	\nonumber\\ 
	F^\text{NP3}_{28} &= \M^\text{NP3}_{28} (-s) u,\quad
	F^\text{NP3}_{29} = \M^\text{NP3}_{28}s m_t^2+\M^\text{NP3}_{29} s \left(m_t^2-u\right)m_t^2,\quad 
	F^\text{NP3}_{30} = \M^\text{NP3}_{30} u,\quad 
	\nonumber\\
	F^\text{NP3}_{31} &= \M^\text{NP3}_{31} um_t^2,\quad F^\text{NP3}_{32} = \M^\text{NP3}_{32} (s+u),\quad 
	F^\text{NP3}_{33} = \M^\text{NP3}_{33} (-s)m_t^2,\quad 
	\nonumber\\
	F^\text{NP3}_{34} &= \M^\text{NP3}_{34}\left(m_t^2-s\right) ,\quad F^\text{NP3}_{35} = \M^\text{NP3}_{35}\left(m_t^2-s\right)m_t^2 ,\quad F^\text{NP3}_{36}= \M^\text{NP3}_{36}\left(m_t^2-s\right),\quad
	\nonumber\\
	F^\text{NP3}_{37}&= \M^\text{NP3}_{37}\left(m_t^2-s\right)m_t^2,\quad F^\text{NP3}_{38} = \M^\text{NP3}_{38}\left(m_t^2-u\right),\quad 
	F^\text{NP3}_{39} = \M^\text{NP3}_{39}r^\text{NP3}_2,\quad
	\nonumber\\
	F^\text{NP3}_{40}&=\frac{1}{2} \M^\text{NP3}_{39} \left(s \left(u-2 m_t^2\right)-2 u m_t^2+2 m_t^4+s^2\right)+\M^\text{NP3}_{40} m_t^2 \left(m_t^2-s-u\right),\quad
	\nonumber\\
	F^\text{NP3}_{41}&=\M^\text{NP3}_{41}\left(m_t^2-s-u\right),\quad F^\text{NP3}_{42}=\M^\text{NP3}_{42} r^\text{NP3}_3,\quad 
	\nonumber\\
	F^\text{NP3}_{43} &= \frac{\M^\text{NP3}_{42} \left(s-2 m_t^2\right) \left(u-m_t^2\right)}{2}+\M^\text{NP3}_{43} \left(m_t^2-s-u\right)m_t^2,\quad
	F^\text{NP3}_{44}= \M^\text{NP3}_{44} \left(m_t^2-s\right),\quad
	\nonumber\\ 
	F^\text{NP3}_{45} &= \M^\text{NP3}_{45} m_t^2 \left(m_t^2-s\right),\quad
	F^\text{NP3}_{46}= \M^\text{NP3}_{46} \left(m_t^2-s-u\right)m_t^2,\quad
	\nonumber\\
	F^\text{NP3}_{47} &= \M^\text{NP3}_{47} (s+u) \left(m_t^2-u\right),\quad
	F^\text{NP3}_{48} = -\M^\text{NP3}_{48} \left(m_t^2+s\right) \left(m_t^2-s\right),\quad 
	\nonumber\\
	F^\text{NP3}_{49} &= \M^\text{NP3}_{49} s (s+u),\quad
	F^\text{NP3}_{50} = \M^\text{NP3}_{50} \left(m_t^2-u\right)m_t^2 ,\quad 
	\nonumber\\
	F^\text{NP3}_{51} &=
	\frac{\M^\text{NP3}_1 \left(-m_t^2+s+u\right)}{4 \left(s-m_t^2\right)}
	-\frac{3 \M^\text{NP3}_5 s \left(-m_t^2+s+u\right)}{4 \left(s-m_t^2\right)}    
	-\frac{\M^\text{NP3}_7 m_t^2 \left(-m_t^2+s+u\right)}{4 \left(s-m_t^2\right)}
	\nonumber\\&\quad
	+\M^\text{NP3}_{13} \left(m_t^2-s-u\right)
	+\M^\text{NP3}_{19} \left(m_t^2-s-u\right)
	+\M^\text{NP3}_{20} m_t^2 \left(m_t^2-s-u\right)	
	\nonumber\\&\quad
	+\M^\text{NP3}_{50} \left(u-m_t^2\right) \left(-m_t^2+s+u\right)
	+\M^\text{NP3}_{51} \left(m_t^2-s-u\right),\quad
	\nonumber\\
	F^\text{NP3}_{52} & = \M^\text{NP3}_{52} s \,u,\quad F^\text{NP3}_{53} = \M^\text{NP3}_{53} r^\text{NP3}_4,\quad 
	\nonumber\\
	F^\text{NP3}_{54} &= 
	\M^\text{NP3}_{34} s	
	-\M^\text{NP3}_{44} (s+u)
	+ \M^\text{NP3}_{54} (s+u)
	-\M^\text{NP3}_{55} s,\quad
	\nonumber\\
	F^\text{NP3}_{55} & = \M^\text{NP3}_{34} \left(-m_t^2+s+u\right)+\M^\text{NP3}_{53} u (s+u)-\M^\text{NP3}_{55} u,\quad
	\nonumber\\
	F^\text{NP3}_{56}& =
	-\frac{\M^\text{NP3}_7 m_t^2 \left(s-m_t^2\right)}{4 u}
	-\frac{\M^\text{NP3}_{10} \left(s-m_t^2\right) \left(m_t^2+u\right)}{2 u}	
	+\frac{\M^\text{NP3}_{11} \left(s-m_t^2\right) \left(u-m_t^2\right)}{4 u}
	\nonumber\\&\quad
	+\M^\text{NP3}_{21} \left(m_t^2-s\right)	
	+\M^\text{NP3}_{22}\left(m_t^2-s\right)m_t^2
	+\M^\text{NP3}_{30} \left(m_t^2-s\right)+\M^\text{NP3}_{32} \left(s-m_t^2\right)\nonumber\\&\quad
	+\M^\text{NP3}_{34} \left(s-m_t^2\right)
	+\M^\text{NP3}_{55} \left(m_t^2-s\right)+\M^\text{NP3}_{56} \left(m_t^2-s\right),\quad
	\nonumber\\
	F^\text{NP3}_{57}&= 
	\M^\text{NP3}_{48} s (s+u)
	+\M^\text{NP3}_{50} m_t^2 \left(u-m_t^2\right)
	+\M^\text{NP3}_{53} (s+u) \left(m_t^2+s\right)
	\nonumber\\&\quad
	+\M^\text{NP3}_{57} s (s+u) \left(u-m_t^2\right)
	-\M^\text{NP3}_{58} s (s+u),\quad
	\nonumber\\
	F^\text{NP3}_{58}&=
	\M^\text{NP3}_{48} m_t^2 \left(-m_t^2-s\right)
	+\M^\text{NP3}_{50} \left(s m_t^2+u m_t^2-m_t^4-s u\right)
	-\M^\text{NP3}_{53} (s+u) \left(m_t^2+s\right)
	\nonumber\\&\quad
	+\M^\text{NP3}_{58} s \left(m_t^2+s\right),\quad
	\nonumber\\
	F^\text{NP3}_{59}&=
	\M^\text{NP3}_{48} m_t^2 \left(m_t^2-s\right)
	+\M^\text{NP3}_{52} s \left(-m_t^2+s+u\right)
	-\M^\text{NP3}_{53} (s+u) \left(s-m_t^2\right)
	\nonumber\\&\quad
	-\M^\text{NP3}_{59} s \left(s-m_t^2\right),\quad
	\nonumber\\
	F^\text{NP3}_{60}&= 
	-\frac{\M^\text{NP3}_1 \left(-2 m_t^2 (s-2 u)+m_t^4+s (s+u)\right)}{8 \left(s-m_t^2\right)^2}	
	+\frac{\M^\text{NP3}_3 m_t^4}{4 \left(m_t^2-s\right)}
	\nonumber\\&\quad
	+\frac{\M^\text{NP3}_5 s \left(m_t^2 (12 u-7 s)+4 m_t^4+3 s (s+u)\right)}{8 \left(s-m_t^2\right)^2}	
	\nonumber\\&\quad
	+\frac{\M^\text{NP3}_7 \left(m_t^4 \left(-6 s^2-5 s u+4 u^2\right)+s m_t^2 \left(2 s^2+3 s u+u^2\right)+2 m_t^6 (3 s+u)-2 m_t^8\right)}{8 u \left(s-m_t^2\right)^2}
	\nonumber\\&\quad	
	+\frac{\M^\text{NP3}_8 m_t^2 \left(m_t^2-2 (s+u)\right)}{4 \left(m_t^2-s\right)}
	+\frac{\M^\text{NP3}_9 m_t^2 (s+u)}{8 \left(s-m_t^2\right)}
	\nonumber\\&\quad
	+\frac{\M^\text{NP3}_{10} \left(m_t^2 \left(2 s^2-2 s u-u^2\right)-4 s m_t^4+2 m_t^6+2 s u (s+u)\right)}{4 u \left(s-m_t^2\right)}
	\nonumber\\&\quad
	-\frac{\M^\text{NP3}_{11} \left(u-m_t^2\right) \left(-m_t^2+s+u\right)}{4 u}
	\nonumber\\&\quad
	+\frac{\M^\text{NP3}_{12} \left(s m_t^2 (u-3 s)+4 s m_t^4-2 m_t^6+s^2 (s+u)\right)}{2 \left(s-m_t^2\right)^2}
	\nonumber\\&\quad
	+\frac{\M^\text{NP3}_{13} \left(m_t^2 (8 u-3 s)+3 s (s+u)\right)}{4 \left(s-m_t^2\right)}
	\nonumber\\&\quad
	+\frac{\M^\text{NP3}_{17} \left(s m_t^2 (8 s-u)-9 s m_t^4+4 m_t^6-3 s^2 (s+u)\right)}{2 \left(s-m_t^2\right)^2}
	\nonumber\\&\quad
	+\frac{\M^\text{NP3}_{18} \left(-3 s^2 m_t^2 (s+u)+s m_t^4 (8 s-u)-9 s m_t^6+4 m_t^8\right)}{2 \left(s-m_t^2\right)^2}
	\nonumber\\&\quad
	+\frac{\M^\text{NP3}_{19} \left(m_t^2 (s+8 u)-2 m_t^4+s (s+u)\right)}{4 \left(s-m_t^2\right)}+\frac{\M^\text{NP3}_{20} \left(s m_t^2 (s+u)-m_t^4 (s-4 u)\right)}{2 \left(s-m_t^2\right)}
	\nonumber\\&\quad
	+\frac{\M^\text{NP3}_{21} \left(-m_t^2 (4 s+u)+2 m_t^4+2 s (s+u)\right)}{2 \left(s-m_t^2\right)}
	\nonumber\\&\quad
	+\frac{\M^\text{NP3}_{22} \left(-m_t^4 (4 s+u)+2 s m_t^2 (s+u)+2 m_t^6\right)}{2 \left(s-m_t^2\right)}
	\nonumber\\&\quad
	+\frac{\M^\text{NP3}_{23} s \left(-m_t^2+s+u\right)}{2 \left(s-m_t^2\right)}+\frac{\M^\text{NP3}_{24} \left(s m_t^2 (2 s-u)-2 s m_t^4+m_t^6-s^2 (s+u)\right)}{\left(s-m_t^2\right)^2}
	\nonumber\\&\quad
	+\M^\text{NP3}_{30}\left(-m_t^2+s+u\right)
	+\M^\text{NP3}_{32} m_t^2
	+\M^\text{NP3}_{34} \left(m_t^2-s-u\right)
	\nonumber\\&\quad	
	-\frac{\M^\text{NP3}_{35} s m_t^2 \left(-m_t^2+s+u\right)}{2 \left(s-m_t^2\right)}	
    -\frac{\M^\text{NP3}_{36} s \left(-m_t^2+s+u\right)}{s-m_t^2}	
	\nonumber\\&\quad
	-\frac{\M^\text{NP3}_{37} s m_t^2 \left(-m_t^2+s+u\right)}{4 \left(s-m_t^2\right)}	
	+\frac{3 \M^\text{NP3}_{38} m_t^2 \left(m_t^2-u\right)}{2 \left(s-m_t^2\right)}
	\nonumber\\&\quad
	+\frac{\M^\text{NP3}_{39} \left(-2 m_t^4 (s+u)+s m_t^2 (s+u)+2 m_t^6\right)}{s-m_t^2}
	-\frac{2 \M^\text{NP3}_{40} m_t^4 \left(m_t^2-s-u\right)}{m_t^2-s}
	\nonumber\\&\quad
	-\frac{\M^\text{NP3}_{41} m_t^2 \left(m_t^2-s-u\right)}{m_t^2-s}	
	+\frac{\M^\text{NP3}_{42} m_t^2 \left(2 m_t^2-s\right) \left(m_t^2-u\right)}{4 \left(s-m_t^2\right)}
	+\frac{\M^\text{NP3}_{43} m_t^4 \left(m_t^2-s-u\right)}{2 \left(s-m_t^2\right)}
	\nonumber\\&\quad
	-\frac{1}{2} \M^\text{NP3}_{44} m_t^2
	+\frac{1}{2} \M^\text{NP3}_{45} m_t^4	
	+\frac{\M^\text{NP3}_{46} m_t^4 \left(m_t^2-s-u\right)}{m_t^2-s}
	-\frac{\M^\text{NP3}_{47} m_t^2 (s+u) \left(u-m_t^2\right)}{2 \left(s-m_t^2\right)}
	\nonumber\\&\quad
	+\frac{\M^\text{NP3}_{48} s \left(m_t^4 (5 s+u)+s m_t^2 (2 u-3 s)-3 m_t^6+s^2 (s+u)\right)}{2 \left(s-m_t^2\right)^2}
	\nonumber\\&\quad
	-\frac{\M^\text{NP3}_{50} m_t^2 \left(m_t^2-u\right) \left(3 m_t^2-4 (s+u)\right)}{2 \left(s-m_t^2\right)}
	+\frac{2 \M^\text{NP3}_{51} m_t^2 \left(m_t^2-s-u\right)}{m_t^2-s}
	\nonumber\\&\quad
	+\M^\text{NP3}_{52}\left(-m_t^2+s+u\right)s-\M^\text{NP3}_{53} (s+u) \left(-m_t^2+s+u\right)
	+\M^\text{NP3}_{55} \left(-m_t^2+s+u\right)
	\nonumber\\&\quad
	-\M^\text{NP3}_{56} m_t^2	
	-\M^\text{NP3}_{59} s \left(-m_t^2+s+u\right)+\M^\text{NP3}_{60} s,
\end{align}
where the four square roots are given by
\begin{align}
	r^\text{NP3}_1 &=\sqrt{s(s-4m_t^2)},\nonumber\\
	r^\text{NP3}_2 &=\sqrt{s(s+u)\left(s(s+u)+4m_t^4-4m_t^2(s+u)\right)},\nonumber\\
	r^\text{NP3}_3 &=\sqrt{s(u−m_t^2)\left(s(u−m_t^2)−4m_t^2u\right)},\nonumber\\
	r^\text{NP3}_4 &=\sqrt{m_t^2(s+u)\left(m_t^2(s+u)-4su\right)}.
\end{align}
\subsubsection{Rationalization of square roots}
Since we have two free 
dimensionless variables $s/m_t^2$ and $u/m_t^2$, it is impossible to rationalize four square roots simultaneously \cite{Besier:2020hjf}. Moreover, we cannot even rationalize $r^\text{NP3}_2$ and $r^\text{NP3}_3$ simultaneously. According to the square root dependence, the canonical basis $F^\text{NP3}_{i}$ can be categorized into several groups. The analytical results of 
\begin{align}
\{ &F^\text{NP3}_{1}, F^\text{NP3}_{2}, F^\text{NP3}_{3}, F^\text{NP3}_{4}, F^\text{NP3}_{7}, F^\text{NP3}_{8}, F^\text{NP3}_{9}, F^\text{NP3}_{10}, F^\text{NP3}_{11},F^\text{NP3}_{12}, F^\text{NP3}_{14}, F^\text{NP3}_{15}, F^\text{NP3}_{16}, 
\nonumber\\&
F^\text{NP3}_{17}, F^\text{NP3}_{18}, F^\text{NP3}_{21},F^\text{NP3}_{22},F^\text{NP3}_{23}, F^\text{NP3}_{26}, F^\text{NP3}_{27}, F^\text{NP3}_{28}, F^\text{NP3}_{29}, F^\text{NP3}_{30}, F^\text{NP3}_{31}, F^\text{NP3}_{32},F^\text{NP3}_{33}, 
\nonumber\\&
F^\text{NP3}_{34}, F^\text{NP3}_{35}, F^\text{NP3}_{44}, F^\text{NP3}_{45}, F^\text{NP3}_{46}, F^\text{NP3}_{47}\}
\end{align}
do not depend on any square roots, and thus we simply define dimensionless variables by
\begin{align}
    s = m_t^2(1-y-z),\quad t = m_t^2y,\quad u = m_t^2z.
\end{align}
The analytical results of 
\begin{align}\label{eq:c2}
	\{F^\text{NP3}_{5}, F^\text{NP3}_{6}, F^\text{NP3}_{13}, F^\text{NP3}_{19}, F^\text{NP3}_{20}, F^\text{NP3}_{24}, F^\text{NP3}_{25}, F^\text{NP3}_{36}, F^\text{NP3}_{37}, F^\text{NP3}_{48}\}
\end{align}
 depend only on $r^\text{NP3}_1$, and we use the standard transformation
\begin{align}\label{eq:r2}
	s = -\frac{(y_2-1)^2}{y_2}m_t^2
\end{align}
to rationalize this square root. Then we have
\begin{align}
	r^\text{NP3}_1 = \frac{(y_2-1) (y_2+1)}{y_2}m_t^2.
\end{align}
The analytical results of 
\begin{align}
	\{F^\text{NP3}_{53},F^\text{NP3}_{54},F^\text{NP3}_{55},F^\text{NP3}_{56}\}
\end{align}
depend only on $r^\text{NP3}_4$, which can be rationalized by defining
\begin{align}
	s = \frac{\left(y_3-1\right) \left(y_3+1\right)}{2 y_3 \left(y_3-z_4\right)}m_t^2,\quad
	u = \frac{\left(y_3-1\right) \left(y_3+1\right)}{2 y_3 \left(y_3+z_4\right)}m_t^2\,.
\end{align}
Then we have
\begin{align}
	r^\text{NP3}_4  =\frac{\left(y_3-1\right) \left(y_3+1\right)}{y_3 \left(y_3-z_4\right) \left(y_3+z_4\right)}m_t^4.
\end{align}
The analytical results of
\begin{align}
\{F^\text{NP3}_{38},F^\text{NP3}_{39},F^\text{NP3}_{40},F^\text{NP3}_{49}\}
\end{align}
depend on both $r^\text{NP3}_1$ and $r^\text{NP3}_2$.
Applying the transformation 
\begin{align}
	s =  -\frac{(y_2-1)^2}{y_2}m_t^2, \quad u =  \frac{\left(y_2^2-z_2\right) \left(y_2^2 z_2-1\right)}{y_2 (y_2+1)^2 z_2}m_t^2,
\end{align}
they are converted to the rational form,
\begin{align}
	r^\text{NP3}_1 = \frac{(y_2-1) (y_2+1)}{y_2}m_t^2,\quad
	r^\text{NP3}_2 = \frac{(y_2-1) (z_2+1)(z_2-1)}{(y_2+1) z_2}m_t^4.
\end{align}
The analytical results of
\begin{align}
	\{F^\text{NP3}_{41},F^\text{NP3}_{42},F^\text{NP3}_{43},F^\text{NP3}_{52}\}
\end{align}
depend on both $r^\text{NP3}_1$ and $r^\text{NP3}_3$.
We define 
\begin{align}
	s =  -\frac{(y_2-1)^2}{y_2}m_t^2, \quad u =  \frac{(y_2+z_3) (y_2 z_3+1)}{(y_2+1)^2 z_3}m_t^2,
\end{align}
to rationalize the two square roots simultaneously,
\begin{align}\label{eq:r2r4}
	r^\text{NP3}_1 = \frac{(y_2-1) (y_2+1)}{y_2}m_t^2,\quad
	r^\text{NP3}_3 = \frac{(y_2-1) (z_3+1)(z_3-1)}{(y_2+1) z_3}m_t^4.
\end{align}

As for the remaining canonical basis, 
$F^\text{NP3}_{50}$ and $F^\text{NP3}_{51}$ depend on $r^\text{NP3}_1$, $r^\text{NP3}_2$ and $r^\text{NP3}_3$, while
$F^\text{NP3}_{57}$, $F^\text{NP3}_{58}$, $F^\text{NP3}_{59}$ and $F^\text{NP3}_{60}$ depend on all the four square roots, i.e., $r^\text{NP3}_1$, $r^\text{NP3}_2$, $r^\text{NP3}_3$ and $r^\text{NP3}_4$.
We list this dependence at different weights in table \ref{tab:NP3roots}.

We find that $r^\text{NP3}_{2}$ starts from the sector of $\{F^\text{NP3}_{38},F^\text{NP3}_{39},F^\text{NP3}_{40}\}$.
Specifically, it appears firstly at weight two in $F^\text{NP3}_{39}$. 
The square root $r^\text{NP3}_{3}$ exists in the sector of $\{F^\text{NP3}_{41},F^\text{NP3}_{42},F^\text{NP3}_{43}\}$ and  appears initially at weight two in $F^\text{NP3}_{42}$. 
The last square root $r^\text{NP3}_{4}$ belongs to the sector of $\{F^\text{NP3}_{53}, F^\text{NP3}_{54}, F^\text{NP3}_{55}, F^\text{NP3}_{56}\}$ and contributes firstly to the weight-three result of $F^\text{NP3}_{53}$. 
Since the square roots $r^\text{NP3}_2$ and $r^\text{NP3}_3$ can not be rationalized simultaneously, the results of  $F^\text{NP3}_{50}$, $F^\text{NP3}_{51}$, $F^\text{NP3}_{57}$, $F^\text{NP3}_{58}$, $F^\text{NP3}_{59}$ and $F^\text{NP3}_{60}$ cannot be written in multiple polylogarithms from weight three. 
However, we find that this problem can be avoided at weight three by choosing proper linear combinations of $F^\text{NP3}_i$ as the new canonical basis.

\begin{table*}[ht]
	\centering
	\begin{tabular}{|c|c|c|c|}
		\hline
    Sectors & Weight 2 & Weight 3 & Weight 4 \\ 
    \hline
    $F^\text{NP3}_{38},F^\text{NP3}_{40}$  & $r^\text{NP3}_{1}$ & $r^\text{NP3}_{1},r^\text{NP3}_{2}$ & $r^\text{NP3}_{1},r^\text{NP3}_{2}$ \\
    \cline{2-4}
    $F^\text{NP3}_{39}$  & $r^\text{NP3}_{1}, r^\text{NP3}_{2}$ & $r^\text{NP3}_{1},r^\text{NP3}_{2}$ & $r^\text{NP3}_{1},r^\text{NP3}_{2}$ \\
    \hline 
    $F^\text{NP3}_{41},F^\text{NP3}_{43}$  & $r^\text{NP3}_{1}$ & $r^\text{NP3}_{1}, r^\text{NP3}_{3}$ & $r^\text{NP3}_{1},r^\text{NP3}_{3}$ \\
    \cline{2-4}       
    $F^\text{NP3}_{42}$  & $r^\text{NP3}_{1}, r^\text{NP3}_{3}$ & $r^\text{NP3}_{1},r^\text{NP3}_{3}$ & $r^\text{NP3}_{1},r^\text{NP3}_{3}$ \\
    \hline        
    $F^\text{NP3}_{54},F^\text{NP3}_{55},F^\text{NP3}_{56}$  & --- & --- & $r^\text{NP3}_{4}$ \\
    \cline{2-4}        
    $F^\text{NP3}_{53}$  & --- & $r^\text{NP3}_{4}$ & $r^\text{NP3}_{4}$ \\
    \hline            
    $F^\text{NP3}_{50},F^\text{NP3}_{51}$ & ---  & $r^\text{NP3}_{1},r^\text{NP3}_{2},r^\text{NP3}_{3}$ & $r^\text{NP3}_{1},r^\text{NP3}_{2},r^\text{NP3}_{3}$ \\ 
    \hline        
    $g^\text{NP3}_{50},g^\text{NP3}_{51}$ & ---  & $r^\text{NP3}_{1},r^\text{NP3}_{3}$ & $r^\text{NP3}_{1},r^\text{NP3}_{2},r^\text{NP3}_{3}$ \\ \hline    
    $F^\text{NP3}_{57},F^\text{NP3}_{58},F^\text{NP3}_{59},F^\text{NP3}_{60}$  & --- & $r^\text{NP3}_{1},r^\text{NP3}_{2},r^\text{NP3}_{3}$ & $r^\text{NP3}_{1},r^\text{NP3}_{2},r^\text{NP3}_{3},r^\text{NP3}_{4}$ \\
    \hline           
    $g^\text{NP3}_{57},g^\text{NP3}_{58},g^\text{NP3}_{59},g^\text{NP3}_{60}$  & --- & $r^\text{NP3}_{1},r^\text{NP3}_{3}$ & $r^\text{NP3}_{1},r^\text{NP3}_{2},r^\text{NP3}_{3}$ \\
    \hline               
\end{tabular}
\caption{Dependence of the basis integrals in some sectors on the square roots at different weights.}
\label{tab:NP3roots}
\end{table*}

\subsubsection{Linear combinations of canonical basis} 
Here we choose $F^\text{NP3}_{51}$ as an example to demonstrate our method. The differential equation  of $F^\text{NP3}_{51}$ with respect to $s$ is  
\begin{align}
	\frac{\partial F^\text{NP3}_{51}}{\partial s} = \epsilon\bigg(&-\frac{F^\text{NP3}_{39} \left(u-m_t^2\right) \left(2 m_t^2+s+u\right)}{2 r^\text{NP3}_2 \left(-m_t^2+s+u\right)}+\frac{F^\text{NP3}_{42} \left(u-m_t^2\right)}{2 r^\text{NP3}_3}
	\nonumber\\&
	-\frac{F^\text{NP3}_{40} \left(s \left(u-m_t^2\right)+3 m_t^2 \left(u-m_t^2\right)+4 s^2\right)}{s \left(s-m_t^2\right) \left(-m_t^2+s+u\right)}+\frac{3 F^\text{NP3}_{38} m_t^2}{s \left(m_t^2-s\right)}+\cdots\bigg),
\end{align}
where $F^\text{NP3}_{39}$ starts at weight two. 
The omitted terms do not contain any square roots.
By observing the differential equation of $F^\text{NP3}_{38}$,
\begin{align}
	\frac{\partial F^\text{NP3}_{38}}{\partial s} =& \epsilon\bigg(\frac{F^\text{NP3}_{39} \left(u-m_t^2\right) \left(2 m_t^2+s+u\right)}{2 r^\text{NP3}_2 \left(-m_t^2+s+u\right)}-\frac{2F^\text{NP3}_{38} \left(-m_t^2+2 s+u\right)}{s \left(-m_t^2+s+u\right)}\nonumber\\
	&\quad-\frac{F^\text{NP3}_{40} \left(u-m_t^2\right)}{s \left(-m_t^2+s+u\right)}+\cdots\bigg),
\end{align}
we find that $F^\text{NP3}_{39}/r_2^{\rm NP3}$ disappears in the differential equation of the combination of  $F^\text{NP3}_{51}$ and $F^\text{NP3}_{38}$,
\begin{align}
	\frac{\partial \left(F^\text{NP3}_{51}+F^\text{NP3}_{38}\right)}{\partial s} = \epsilon\bigg(&\frac{F^\text{NP3}_{42} \left(u-m_t^2\right)}{2 r^\text{NP3}_3}
-\frac{2 F^\text{NP3}_{40} \left(s \left(u-m_t^2\right)+u m_t^2-m_t^4+2 s^2\right)}{s \left(s-m_t^2\right) \left(-m_t^2+s+u\right)}
\nonumber\\&
-\frac{F^\text{NP3}_{38} \left(s \left(2 u-3 m_t^2\right)+u m_t^2-m_t^4+4 s^2\right)}{s \left(s-m_t^2\right) \left(-m_t^2+s+u\right)}+\cdots\bigg).
\label{eq:g51diff}
\end{align}
This implies that it is convenient to define
\begin{align}
	g^\text{NP3}_{51} \equiv  F^\text{NP3}_{51}+ F^\text{NP3}_{38} .
\end{align}
Then we rationalize $r^\text{NP3}_1$ and $r^\text{NP3}_3$ according to eq.(\ref{eq:r2r4}).
Consequently, the weight-three result of $g^\text{NP3}_{51}$ can be written in terms of multiple polylogarithms. 
Its dependence on the square roots at different weights is also shown in table \ref{tab:NP3roots}.

The weight-four result of $g^\text{NP3}_{51}$, however, can not be expressed in  multiple polylogarithms from the solution of the differential equations, since $F^\text{NP3}_{38}$ and $F^\text{NP3}_{40}$ in eq.(\ref{eq:g51diff}) contain $r^\text{NP3}_2$ at weight three.
Notice that this does not mean that they can not be evaluated to multiple polylogarithms with other methods, such as the direct integration in Feynman parameter space as shown in \cite{Heller:2019gkq,Kreer:2021sdt}.
In this work, we use numerical integration to obtain the values of $g^\text{NP3}_{51}$ at weight four, which is a single integral over the weight-three multiple polylogarithms. 

Similarly to $g^\text{NP3}_{51}$, we define 
\begin{align}
	g^\text{NP3}_{50} &\equiv  F^\text{NP3}_{50}+ F^\text{NP3}_{38} ,\nonumber\\
	g^\text{NP3}_{57} &\equiv F^\text{NP3}_{57} + 2 F^\text{NP3}_{55} − F^\text{NP3}_{56},\nonumber\\
	g^\text{NP3}_{58} &\equiv F^\text{NP3}_{58} + 2 F^\text{NP3}_{50},\nonumber\\
	g^\text{NP3}_{59} &\equiv F^\text{NP3}_{59} - F^\text{NP3}_{38} +2F^\text{NP3}_{49} -F^\text{NP3}_{56},\nonumber\\
	g^\text{NP3}_{60} &\equiv F^\text{NP3}_{60} + 2F^\text{NP3}_{49}
\end{align}
as the new canonical basis. Then their differential equations  contain only $r^\text{NP3}_1$ and $r^\text{NP3}_3$ and the corresponding results can be written as multiple polylogarithms  at weight three. 
The weight-four results are obtained as single integrals.

\subsubsection{Boundary conditions and numerical check}
The analytical results of $F^\text{NP3}_1$, $F^\text{NP3}_3$ and $F^\text{NP3}_7$ can be directly obtained,
\begin{align}
	F^\text{NP3}_1 &=  1,\nonumber\\
	F^\text{NP3}_3 &= -\frac{1}{4}-\frac{\pi ^2}{6}\epsilon^2-2 \zeta (3)\epsilon^3-\frac{8\pi ^4}{45}\epsilon^4+ {\cal O}(\epsilon^5),\nonumber\\
	F^\text{NP3}_7 &= 1+\frac{\pi ^2}{3}\epsilon^2-2 \zeta (3)\epsilon^3+\frac{\pi ^4}{10}\epsilon^4+ {\cal O}(\epsilon^5).
\end{align}
And the boundary conditions of $F^\text{NP3}_2$, $F^\text{NP3}_4$, $F^\text{NP3}_9$ and $F^\text{NP3}_{11}$ are easy to compute,
\begin{align}
	F^\text{NP3}_2|_{s=m_t^2} &= -1-2 i \pi\epsilon+\frac{7 \pi ^2}{3}\epsilon^2 + \left(10 \zeta (3)+2 i \pi ^3\right)\epsilon^3+\left(-\frac{109 \pi ^4}{90}+20 i \pi  \zeta (3)\right)\epsilon^4+ {\cal O}(\epsilon^5),\nonumber\\
	F^\text{NP3}_4|_{s=m_t^2} &= 1 +i \pi  \epsilon -\frac{2 \pi ^2 \epsilon ^2}{3}-\left(2 \zeta (3)+\frac{i \pi ^3}{3}\right) \epsilon ^3 + \left(\frac{\pi ^4}{10}-2 i \pi  \zeta (3)\right) \epsilon ^4+ {\cal O}(\epsilon^5),\nonumber\\
	F^\text{NP3}_9|_{t=0} &= F^\text{NP3}_{11}|_{u=0} = 1+\frac{\pi ^2}{3}\epsilon^2-2\zeta (3)\epsilon^3+\frac{\pi^4}{10}\epsilon^4+ {\cal O}(\epsilon^5).
\end{align}
The boundary conditions of $F^\text{NP3}_{15}$, $F^\text{NP3}_{16}$, $F^\text{NP3}_{32}$, $F^\text{NP3}_{39}$ and $F^\text{NP3}_{49}$ can be found at $t = m_t^2$,
\begin{align}
	F^\text{NP3}_{15}|_{t=m_t^2} = F^\text{NP3}_{16}|_{t=m_t^2} = F^\text{NP3}_{32}|_{t=m_t^2} = F^\text{NP3}_{39}|_{t=m_t^2} =F^\text{NP3}_{49}|_{t=m_t^2} = 0.
\end{align}
The boundary condition of $F^\text{NP3}_{42}$ is fixed at $u = m_t^2$,
\begin{align}
	F^\text{NP3}_{42}|_{u=m_t^2} = 0.
\end{align}
The boundary conditions of $F^\text{NP3}_{5}$, $F^\text{NP3}_{6}$, $F^\text{NP3}_{25}$ are set at $s = 0$,
\begin{align}
	F^\text{NP3}_{5}|_{s=0} &= F^\text{NP3}_{6}|_{s=0} = 0,\nonumber\\
	F^\text{NP3}_{25}|_{s=0} &= \frac{1}{2} (\M^\text{NP3}_1	-\M^\text{NP3}_7
	+2 \M^\text{NP3}_{13}-4 \M^\text{NP3}_{17}-4 \M^\text{NP3}_{18}+8 \M^\text{NP3}_{24}
	)|_{s=0}.
\end{align}
Lastly, the other boundary conditions  can be found in the regular point
$(s = m_t^2, u = 0, t = 0)$.
From eq.(\ref{eq:r2}),  $s = m_t^2$ corresponds to $y_2 = (1\pm\sqrt{3}i)/2$, and therefore $(\pm1\pm\sqrt{3}i)/2$ appear in the analytical results of the integrals in eq.(\ref{eq:c2}). The expressions can be significantly reduced after applying the relations among such multiple polylogarithms
with arguments of $\{0,\pm1,(\pm1\pm\sqrt{3}i)/2\}$  \cite{Henn:2015sem}. 

Now, solving the differential equation, we obtain the analytical results of basis integrals in terms of multiple polylogarithms or single integrals over them. 
The numerical integration of $g^\text{NP3}_{50}$, $g^\text{NP3}_{51}$, $g^\text{NP3}_{57}$, $g^\text{NP3}_{58}$, $g^\text{NP3}_{59}$ and $g^\text{NP3}_{60}$ at weight four can be performed by using the Newton-Cotes formulas \cite{Hildebrand}.
As an example, we show the numerical  result of $g^\text{NP3}_{57}$ at weight four at the phase space point $(m_t^2 = 1, s = 10, u = -9/4, t = -27/4)$
and the comparison with the result using {\tt AMFlow}, which is considered as the exact evaluation,  
\begin{align}
	g^{\text{NP3},\text{analytic}}_{57} &=  \cdots+(10.61162043 + 49.10659750 i)\epsilon^4,\nonumber\\
	g^{\text{NP3},\text{AMFlow}}_{57} &= \cdots+(10.61162046 + 49.10659740 i)\epsilon^4\,.
\end{align} 
Here we have used the 5-point Newton-Cotes formula. One can see that the difference is very small.
The results of the other basis integrals are provided in the ancillary file.

\subsection{NP4 integral family}
\subsubsection{Canonical basis}
In the limit of $m_W \to 0$, the integrals in the NP4 family can be reduced to 48 MIs after considering the symmetries between  integrals. For comparison, there are 56 MIs if $m_W \neq0$. We first select the MIs whose differential equations have coefficients linear in $\epsilon$,
\begin{align}
\M^\text{NP4}_1&=\epsilon^2I^\text{NP4}_{0,0,0,2,0,0,2,0,0},\quad&
\M^\text{NP4}_2&=\epsilon^2I^\text{NP4}_{0,0,1,2,0,0,2,0,0}, \quad\nonumber\\
\M^\text{NP4}_3&=\epsilon^2I^\text{NP4}_{1,0,0,0,2,2,0,0,0}, \quad&
\M^\text{NP4}_4&=\epsilon^2I^\text{NP4}_{0,0,0,2,2,1,0,0,0}, \quad\nonumber\\
\M^\text{NP4}_5&=\epsilon^2I^\text{NP4}_{0,1,0,2,2,0,0,0,0},\quad&
\M^\text{NP4}_6&=\epsilon^2I^\text{NP4}_{0,2,0,1,2,0,0,0,0}, \quad\nonumber\\
\M^\text{NP4}_7&=(1-\epsilon)\epsilon I^\text{NP4}_{1,0,0,0,0,2,2,0,0},\quad&
\M^\text{NP4}_8&=\epsilon^2I^\text{NP4}_{1,2,0,0,0,0,2,0,0}, \quad\nonumber\\
\M^\text{NP4}_9&=\epsilon^2I^\text{NP4}_{2,2,0,0,0,0,1,0,0}, \quad&
\M^\text{NP4}_{10}&=\epsilon^3I^\text{NP4}_{0,0,1,1,2,1,0,0,0}, \quad\nonumber\\
\M^\text{NP4}_{11}&=\epsilon^3I^\text{NP4}_{1,0,0,0,1,2,1,0,0},\quad&
\M^\text{NP4}_{12}&=\epsilon^3I^\text{NP4}_{0,1,0,2,1,0,1,0,0}, \quad\nonumber\\
\M^\text{NP4}_{13}&=\epsilon^2I^\text{NP4}_{0,1,0,3,1,0,1,0,0},\quad&
\M^\text{NP4}_{14}&=\epsilon^3I^\text{NP4}_{0,1,1,1,0,0,2,0,0}, \quad\nonumber\\
\M^\text{NP4}_{15}&=\epsilon^2I^\text{NP4}_{0,1,1,1,0,0,3,0,0}, \quad&
\M^\text{NP4}_{16}&=\epsilon^2I^\text{NP4}_{0,1,1,2,0,0,2,0,0}, \quad\nonumber\\
\M^\text{NP4}_{17}&=\epsilon^3I^\text{NP4}_{1,0,0,1,1,2,0,0,0},\quad&
\M^\text{NP4}_{18}&=\epsilon^3I^\text{NP4}_{1,1,0,1,0,0,2,0,0}, \quad\nonumber\\
\M^\text{NP4}_{19}&=\epsilon^2I^\text{NP4}_{1,1,0,1,0,0,3,0,0},\quad&
\M^\text{NP4}_{20}&=\epsilon^3I^\text{NP4}_{1,0,1,1,2,1,0,0,0}, \quad\nonumber\\
\M^\text{NP4}_{21}&=\epsilon^4I^\text{NP4}_{1,1,0,1,1,1,0,0,0}, \quad&
\M^\text{NP4}_{22}&=\epsilon^3I^\text{NP4}_{1,1,0,2,1,1,0,0,0}, \quad\nonumber\\
\M^\text{NP4}_{23}&=\epsilon^4I^\text{NP4}_{0,0,1,1,1,1,1,0,0},\quad&
\M^\text{NP4}_{24}&=\epsilon^3I^\text{NP4}_{0,0,1,2,1,1,1,0,0}, \quad\nonumber\\
\M^\text{NP4}_{25}&=\epsilon^4I^\text{NP4}_{1,0,0,1,1,1,1,0,0},\quad&
\M^\text{NP4}_{26}&=\epsilon^3I^\text{NP4}_{1,0,0,2,1,1,1,0,0}, \quad\nonumber\\
\M^\text{NP4}_{27}&=\epsilon^4I^\text{NP4}_{1,0,1,0,1,1,1,0,0}, \quad&
\M^\text{NP4}_{28}&=\epsilon^3I^\text{NP4}_{1,0,1,0,1,1,2,0,0}, \quad\nonumber\\
\M^\text{NP4}_{29}&=\epsilon^4I^\text{NP4}_{1,1,0,0,1,1,1,0,0},\quad&
\M^\text{NP4}_{30}&=\epsilon^3I^\text{NP4}_{1,1,0,0,1,1,2,0,0}, \quad\nonumber\\
\M^\text{NP4}_{31}&=\epsilon^4I^\text{NP4}_{1,1,1,0,0,1,1,0,0},\quad&
\M^\text{NP4}_{32}&=\epsilon^3I^\text{NP4}_{1,1,1,0,0,1,2,0,0}, \quad\nonumber\\
\M^\text{NP4}_{33}&=(1-2\epsilon)\epsilon^3I^\text{NP4}_{1,1,1,1,0,0,1,0,0}, \quad&
\M^\text{NP4}_{34}&=\epsilon^3I^\text{NP4}_{1,1,1,1,0,0,2,0,0}, \quad\nonumber\\
\M^\text{NP4}_{35}&=\epsilon^4I^\text{NP4}_{1,1,0,1,1,0,1,0,0},\quad&
\M^\text{NP4}_{36}&=\epsilon^3I^\text{NP4}_{1,1,0,1,1,0,2,0,0}, \quad\nonumber\\
\M^\text{NP4}_{37}&=\epsilon^3I^\text{NP4}_{2,1,0,1,1,0,1,0,0},\quad&
\M^\text{NP4}_{38}&=\epsilon^3I^\text{NP4}_{1,1,0,2,1,0,1,0,0}, \quad\nonumber\\
\M^\text{NP4}_{39}&=\epsilon^4I^\text{NP4}_{0,1,1,1,1,1,1,0,0}, \quad&
\M^\text{NP4}_{40}&=\epsilon^4I^\text{NP4}_{1,0,1,1,1,1,1,0,0}, \quad\nonumber\\
\M^\text{NP4}_{41}&=\epsilon^4I^\text{NP4}_{1,0,1,1,1,1,1,0,-1},\quad&
\M^\text{NP4}_{42}&=\epsilon^4I^\text{NP4}_{1,1,0,1,1,1,1,0,0}, \quad\nonumber\\
\M^\text{NP4}_{43}&=\epsilon^4I^\text{NP4}_{1,1,-1,1,1,1,1,0,0},\quad&
\M^\text{NP4}_{44}&=\epsilon^4I^\text{NP4}_{1,1,0,1,1,1,1,-1,0}, \quad\nonumber\\
\M^\text{NP4}_{45}&=\epsilon^4I^\text{NP4}_{1,1,0,1,1,1,1,0,-1}, \quad&
\M^\text{NP4}_{46}&=\epsilon^4I^\text{NP4}_{1,1,1,1,1,1,1,0,0}, \quad\nonumber\\
\M^\text{NP4}_{47}&=\epsilon^4I^\text{NP4}_{1,1,1,1,1,1,1,-1,0},\quad&
\M^\text{NP4}_{48}&=\epsilon^4I^\text{NP4}_{1,1,1,1,1,1,1,0,-1}.
\end{align}
The corresponding topology diagrams are displayed in figure \ref{fig:NP4_MIs} in the appendix.

The canonical basis $F^\text{NP4}_{i},~i=1,\dots,48,$ can be constructed as  linear combinations of  $\M^\text{NP4}_{i}$,
\begin{align}
	F^\text{NP4}_1&=\M^\text{NP4}_1,\quad F^\text{NP4}_2=\M^\text{NP4}_2 \left(m_t^2-s-u\right),\quad F^\text{NP4}_3=\M^\text{NP4}_3 (-s),\quad 
	F^\text{NP4}_4=\M^\text{NP4}_4 m_t^2,\quad 
	\nonumber\\
	F^\text{NP4}_5&=\M^\text{NP4}_5 \left(m_t^2-s-u\right),\quad 
	F^\text{NP4}_6=-2 \M^\text{NP4}_5 m_t^2-\M^\text{NP4}_6 (s+u),\quad
	F^\text{NP4}_7=\M^\text{NP4}_7 m_t^2,\quad 
	\nonumber\\
	F^\text{NP4}_8&=\M^\text{NP4}_8 u,\quad F^\text{NP4}_9=-2 \M^\text{NP4}_8m_t^2+\M^\text{NP4}_9 \left(u-m_t^2\right),\quad
	F^\text{NP4}_{10}=\M^\text{NP4}_{10} (s+u),\quad
	\nonumber\\
	F^\text{NP4}_{11}&=\M^\text{NP4}_{11} \left(m_t^2-s\right),\quad
	F^\text{NP4}_{12}=\M^\text{NP4}_{12} (s+u),\quad F^\text{NP4}_{13}=\M^\text{NP4}_{13} m_t^2 (s+u),\quad
	\nonumber\\
	F^\text{NP4}_{14}&=\M^\text{NP4}_{14} (s+u),\quad F^\text{NP4}_{15}=\M^\text{NP4}_{15} m_t^2 (s+u),\quad
	\nonumber\\
	F^\text{NP4}_{16}&=3 \M^\text{NP4}_{14} m_t^2+2 \M^\text{NP4}_{15} m_t^4-\M^\text{NP4}_{16} m_t^2 \left(-m_t^2-s-u\right),\quad 
	F^\text{NP4}_{17}=\M^\text{NP4}_{17} \left(m_t^2-s\right),\quad
	\nonumber\\
	F^\text{NP4}_{18}&=\M^\text{NP4}_{18} u,\quad
	F^\text{NP4}_{19}=\M^\text{NP4}_{19} u m_t^2,\quad
	F^\text{NP4}_{20}=\M^\text{NP4}_{20} s (s+u),\quad F^\text{NP4}_{21}=\M^\text{NP4}_{21} u,\quad
	\nonumber\\ 
	F^\text{NP4}_{22}&=\M^\text{NP4}_{22} u m_t^2,\quad F^\text{NP4}_{23}=\M^\text{NP4}_{23} (s+u),\quad
	F^\text{NP4}_{24}=\M^\text{NP4}_{24} m_t^2 (s+u),\quad
	\nonumber\\ 
	F^\text{NP4}_{25}&=\M^\text{NP4}_{25} \left(m_t^2-s\right),\quad
	F^\text{NP4}_{26}=\M^\text{NP4}_{26} m_t^2 \left(m_t^2-s\right),\quad 
	F^\text{NP4}_{27}=\M^\text{NP4}_{27} \left(m_t^2-u\right),\quad
	\nonumber\\
	F^\text{NP4}_{28}&=\M^\text{NP4}_{28} s m_t^2,\quad 
	F^\text{NP4}_{29}=\M^\text{NP4}_{29} (s+u),\quad 
	F^\text{NP4}_{30}=\M^\text{NP4}_{30} (-s) m_t^2,\quad F^\text{NP4}_{31}=\M^\text{NP4}_{31} (-s),\quad
	\nonumber\\
	F^\text{NP4}_{32}&=\M^\text{NP4}_{32} (s+u) \left(m_t^2-u\right),\quad
	F^\text{NP4}_{33}=\left(\M^\text{NP4}_{33}-\M^\text{NP4}_{34} m_t^2\right) \left(m_t^2-s-u\right),\quad 
	\nonumber\\
	F^\text{NP4}_{34}&=\M^\text{NP4}_{34} \left(\left(-m_t^2-s-u\right) \left(m_t^2-s-u\right)-s (s+u)\right),\quad 
	\nonumber\\ 
	F^\text{NP4}_{35}&=\M^\text{NP4}_{35} \left(m_t^2-s\right),\quad 
	F^\text{NP4}_{36}=\M^\text{NP4}_{36} m_t^2 \left(m_t^2-s-u\right),\quad
	\nonumber\\ 
	F^\text{NP4}_{37}&=\M^\text{NP4}_{37} m_t^2 \left(m_t^2-u\right)+\M^\text{NP4}_{38} m_t^2 \left(2 m_t^2-s-u\right),\quad
	F^\text{NP4}_{38}=\M^\text{NP4}_{38} m_t^2 \left(m_t^2-s\right),\quad
	\nonumber\\
	F^\text{NP4}_{39}&=\M^\text{NP4}_{39} (s+u)^2,\quad
	F^\text{NP4}_{40}=\M^\text{NP4}_{40} r^\text{NP4}_1,\quad
	\nonumber\\
	F^\text{NP4}_{41}&=-\M^\text{NP4}_{23} m_t^2+\M^\text{NP4}_{25} \left(m_t^2-s-u\right)+\M^\text{NP4}_{27} s+\M^\text{NP4}_{40} s m_t^2+\M^\text{NP4}_{41} \left(m_t^2-s-u\right),\quad 
	\nonumber\\
	F^\text{NP4}_{42}&=\M^\text{NP4}_{42} r^\text{NP4}_2,
	\nonumber\\
	F^\text{NP4}_{43}&=-\frac{\M^\text{NP4}_4 m_t^2 \left(m_t^2-s-u\right)}{-s-u}-\frac{\M^\text{NP4}_5 \left(m_t^2-s-u\right) \left(2 m_t^2-s-u\right)}{2 (s+u)}
	\nonumber\\&\quad
	-\frac{1}{4} \M^\text{NP4}_6 \left(m_t^2-s-u\right)+\M^\text{NP4}_{12} \left(m_t^2-s-u\right)+\M^\text{NP4}_{13} m_t^2 \left(m_t^2-s-u\right)-\M^\text{NP4}_{29} s
	\nonumber\\&\quad
	-\M^\text{NP4}_{42} u m_t^2-\M^\text{NP4}_{43} u+\M^\text{NP4}_{45} \left(m_t^2-s-u\right),\quad
	\nonumber\\
	F^\text{NP4}_{44}&=\frac{\M^\text{NP4}_4 sm_t^2}{s+u}-\frac{\M^\text{NP4}_5 s \left(2 m_t^2-s-u\right)}{2 (s+u)}-\frac{\M^\text{NP4}_6 s}{4}+\frac{\M^\text{NP4}_7 s m_t^2}{4 u}
	-\frac{\M^\text{NP4}_8 s \left(-m_t^2-u\right)}{2 u}
	\nonumber\\&\quad
	+\frac{\M^\text{NP4}_9 s \left(m_t^2-u\right)}{4 u}
	+\M^\text{NP4}_{12} s
	+\M^\text{NP4}_{13} s m_t^2+\M^\text{NP4}_{18} s+\M^\text{NP4}_{19} s m_t^2+\M^\text{NP4}_{21} s-\M^\text{NP4}_{29} s
	\nonumber\\&\quad
	-\M^\text{NP4}_{35} (s+u)
	-\M^\text{NP4}_{42} s u+\M^\text{NP4}_{43} s+\M^\text{NP4}_{44} (s+u)
	+\M^\text{NP4}_{45} s,\quad
	\nonumber\\
	F^\text{NP4}_{45}&=-\frac{\M^\text{NP4}_4 \left(s-m_t^2\right)m_t^2}{s+u}+\frac{\M^\text{NP4}_5 \left(s-m_t^2\right) \left(2 m_t^2-s-u\right)}{2 (s+u)}+\frac{1}{4} \M^\text{NP4}_6 \left(s-m_t^2\right)
	\nonumber\\&\quad
	+\M^\text{NP4}_{12} \left(m_t^2-s\right)
	+\M^\text{NP4}_{13} m_t^2 \left(m_t^2-s\right)
	+\M^\text{NP4}_{42} u \left(s-m_t^2\right)+\M^\text{NP4}_{45} \left(m_t^2-s\right),
	\nonumber\\
	F^\text{NP4}_{46}&=-\frac{3 \M^\text{NP4}_3 s \left(-m_t^2+4 \epsilon  (s+u)+u\right)}{8 (4 \epsilon +1) \left(u-m_t^2\right)}-\frac{\M^\text{NP4}_5 \left(2 \left(m_t^2-s-u\right)+m_t^2\right)}{16 \epsilon +4}-\frac{\M^\text{NP4}_6 (s+u)}{8 (4\epsilon +1)}
	\nonumber\\&\quad
	-\frac{\M^\text{NP4}_7 \left(u m_t^2-m_t^4+4 \epsilon  (s+u)m_t^2\right)}{8 (4 \epsilon +1) \left(u-m_t^2\right)}-\frac{\M^\text{NP4}_8 \epsilon  (s+u) \left(m_t^2+2 u\right)}{(4 \epsilon +1) \left(u-m_t^2\right)}
	\nonumber\\&\quad
	+\frac{\M^\text{NP4}_9 \left(\epsilon  m_t^2-\epsilon  \left(m_t^2-s-u\right)\right)}{2(4\epsilon +1)}
	+\frac{3\M^\text{NP4}_{27} \left(m_t^2- u\right)}{2 (4\epsilon +1)}
	+\frac{\M^\text{NP4}_{28} s m_t^2}{4 \epsilon +1}
	\nonumber\\&\quad
	+\frac{6 \M^\text{NP4}_{29} \epsilon  (s+u)^2}{(4 \epsilon +1) \left(u-m_t^2\right)}
	+\frac{4 \M^\text{NP4}_{30} s m_t^2\epsilon  (s+u)}{(4 \epsilon +1) \left(u-m_t^2\right)}
	\nonumber\\&\quad
	-\frac{\M^\text{NP4}_{32} (s+u) \left(4 \epsilon  \left(m_t^2-s-u\right)+m_t^2-u\right)}{4(4\epsilon +1)}
	\nonumber\\&\quad
	+\M^\text{NP4}_{39} \left(3 m_t^2 \left(m_t^2-s-u\right)-2 \left(m_t^2-s-u\right)^2-m_t^4\right)
	\nonumber\\&\quad
	+\M^\text{NP4}_{42} \left(m_t^2 \left(m_t^2-s-u\right)-u \left(m_t^2-s-u\right)+u m_t^2-m_t^4\right)
	\nonumber\\&\quad
	-\M^\text{NP4}_{46} s m_t^2 (s+u)
	-\M^\text{NP4}_{48} (s+u) \left(m_t^2-s-u\right),
	\quad\nonumber\\
	F^\text{NP4}_{47}&=
	\frac{\M^\text{NP4}_1 \left(m_t^2-s-u\right)}{2 \left(u-m_t^2\right)}
	-\frac{\M^\text{NP4}_2 \left(m_t^2-s-u\right)^2}{u-m_t^2}
	+\frac{3 \M^\text{NP4}_4 \left(m_t^2-s-u\right)m_t^2}{2 \left(u-m_t^2\right)}
	\nonumber\\&\quad
	+\frac{\M^\text{NP4}_8 \left(m_t^2-s-u\right)m_t^2}{2 \left(u-m_t^2\right)}
	-\frac{1}{4} \M^\text{NP4}_9 \left(m_t^2-s-u\right)
	\nonumber\\&\quad
	-\frac{3 \M^\text{NP4}_{14} \left(-2 m_t^2-s-u\right) \left(m_t^2-s-u\right)}{2 \left(u-m_t^2\right)}
	\nonumber\\&\quad
	-\frac{\M^\text{NP4}_{15} \left(-2 m_t^2-s-u\right) \left(m_t^2-s-u\right)m_t^2}{u-m_t^2}
	\nonumber\\&\quad
	-\frac{\M^\text{NP4}_{16} \left(-m_t^2-s-u\right) \left(m_t^2-s-u\right)m_t^2}{u-m_t^2}
	\nonumber\\&\quad
	+\frac{\M^\text{NP4}_{18} u \left(m_t^2-s-u\right)}{2 m_t^2-2 u}
	-\frac{\M^\text{NP4}_{19} u m_t^2\left(m_t^2-s-u\right)}{u-m_t^2}
	\nonumber\\&\quad
	+\frac{1}{2} \M^\text{NP4}_{32} (s+u) \left(m_t^2-s-u\right)
	+\frac{3 \M^\text{NP4}_{33} \left(m_t^2-s-u\right)^2}{2 m_t^2-2 u}
	\nonumber\\&\quad
	+\frac{\M^\text{NP4}_{34} \left(m_t^2-s-u\right) \left(\left(3 m_t^2+u\right) \left(m_t^2-s-u\right)-u m_t^2+m_t^4\right)}{2 \left(u-m_t^2\right)}
	\nonumber\\&\quad
	-\M^\text{NP4}_{40} m_t^2 (s+u)
	+\M^\text{NP4}_{42} s (s+u)
	+\M^\text{NP4}_{47} (s+u)^2,\quad
	\nonumber\\
	F^\text{NP4}_{48}&=\M^\text{NP4}_{39} (s+u) \left(u-m_t^2\right)-\M^\text{NP4}_{42} (s+u) \left(u-m_t^2\right)-\M^\text{NP4}_{46} s (s+u) \left(u-m_t^2\right)
	\nonumber\\&\quad
	-\M^\text{NP4}_{48} (s+u) \left(u-m_t^2\right),
\end{align}
where
\begin{align}
	r^\text{NP4}_1 &= \sqrt{s (s+u) \left(-4 m_t^2 (s+u)+4 m_t^4+s (s+u)\right)},\nonumber\\
	r^\text{NP4}_2 &= \sqrt{(s+u) \left(m_t^2 (s+u)-4 s \,u\right)}.
\end{align}

\subsubsection{Rationalization of square roots}

With the help of {\tt RationalizeRoots} \cite{Besier:2018jen,Besier:2019kco}, 
we find the variable transformation
\begin{align}
s = \frac{2(-y_4+z_5-1)}{(z_5-1)(z_5+1)}m_t^2,\quad u = \frac{2 y_4(2 z_5-y_4)}{(z_5-1)(z_5+1) (-y_4+z_5+1)}m_t^2,
\label{eq:np4r1}
\end{align} 
which can rationalize $r^\text{NP4}_1$, 
\begin{align}
r^\text{NP4}_1 = \frac{4 z_5(y_4-z_5+1)}{(z_5-1)(z_5+1)(y_4-z_5-1)}m_t^4.
\end{align}
Similarly, 
after defining the new variables
\begin{align}
s = \frac{y_5+z_6}{z_6(y_5+z_6+1)}m_t^2,\quad u =\frac{y_5+z_6}{(y_5+1)(y_5+z_6+1)}m_t^2,
\end{align} 
The square root $r^\text{NP4}_2$ becomes
\begin{align}
r^\text{NP4}_2 = \frac{(1-y_5-z_6)(y_5+z_6)}{(y_5+1) z_6(y_5+z_6+1)}m_t^4.
\end{align}
Though the above two square roots can be rationalized  simultaneously according to the criteria in \cite{Besier:2020hjf}, the  solution  of the resulting differential equations has a complicated form.
One needs to determine the roots of a polynomial of degree four in order to take partial fraction in the coefficients of the differential equations in order to transform them into the canonical form.

\subsubsection{Linear combinations of canonical basis}
The differential equations of the basis integrals in the top sector  $\{F^\text{NP4}_{46},F^\text{NP4}_{47},F^\text{NP4}_{48}\}$ contain both $r^\text{NP4}_1$ and 
$r^\text{NP4}_2$, while the other basis integrals involve at most only one square root. 
The analytical results of these integrals  written in terms of multiple polylogarithms would be too lengthy and cumbersome. 
We find that this problem of rationalizing two square roots  simultaneously can be bypassed by choosing proper linear combinations of the canonical basis. Explicitly, we define the new basis
\begin{align}
g^\text{NP4}_{46} &\equiv F^\text{NP4}_{46} + 3F^\text{NP4}_{43} -4F^\text{NP4}_{45},\nonumber\\
g^\text{NP4}_{47} &\equiv F^\text{NP4}_{47} + F^\text{NP4}_{48}- F^\text{NP4}_{41} -3F^\text{NP4}_{43}+4F^\text{NP4}_{45},\nonumber\\
g^\text{NP4}_{48} &\equiv F^\text{NP4}_{48}  + F^\text{NP4}_{43} - F^\text{NP4}_{45} + F^\text{NP4}_{46}.	
\end{align}
Then the differential equations of $g^\text{NP4}_{46}$ and $g^\text{NP4}_{47}$ do not contain any square roots any more, while the differential equations of $g^\text{NP4}_{48}$ contain  only $r^\text{NP4}_1$. 
Simultaneously, 
the square root $r^\text{NP4}_2$ exists only in the differential equations of $F^\text{NP4}_{42},F^\text{NP4}_{43}$ and $F^\text{NP4}_{45}$,
and it is needed from the weight-three result of $F^\text{NP4}_{42}$.
As a consequence, $F^\text{NP4}_{43}$ and $F^\text{NP4}_{45}$ depends on $r^\text{NP4}_2$ from weight four
\footnote{The basis integral $F^\text{NP4}_{44}$ in the same sector relies on $r^\text{NP4}_2$ from weight five.}.
The analytic results of $g^\text{NP4}_{46}$, $g^\text{NP4}_{47}$ and $g^\text{NP4}_{48}$ depend on $r^\text{NP4}_2$ only implicitly via $F^\text{NP4}_{43}$, $F^\text{NP4}_{44}$ and $F^\text{NP4}_{45}$, and thus are free of $r^\text{NP4}_2$  at weight four.
The parameterization in eq.(\ref{eq:np4r1}) is sufficient to express $g^\text{NP4}_{46}$, $g^\text{NP4}_{47}$ and $g^\text{NP4}_{48}$ in terms of multiple polylogarithms up to weight four.

\subsubsection{Boundary conditions and numerical check}
The analytical results of single scale integrals  $F^\text{NP4}_1$, $F^\text{NP4}_3$ and $F^\text{NP4}_7$ can be  obtained by direct calculation,
\begin{align}
	F^\text{NP4}_1 &=  1,\nonumber\\
	F^\text{NP4}_4 &= -\frac{1}{4}-\frac{\pi ^2}{6}\epsilon^2-2 \zeta (3)\epsilon^3-\frac{8\pi ^4}{45}\epsilon^4+ {\cal O}(\epsilon^5),\nonumber\\
	F^\text{NP4}_7 &= 1+\frac{\pi ^2}{3}\epsilon^2-2 \zeta (3)\epsilon^3+\frac{\pi ^4}{10}\epsilon^4+ {\cal O}(\epsilon^5).
\end{align}
And the boundary conditions of $F^\text{NP4}_3$, $F^\text{NP4}_6$ and $F^\text{NP4}_{9}$ are computed at special kinematic points,
\begin{align}
	F^\text{NP4}_3|_{s=m_t^2} &= -1-2 i \pi\epsilon+\frac{7 \pi ^2}{3}\epsilon^2 + \left(10 \zeta (3)+2 i \pi ^3\right)\epsilon^3+\left(-\frac{109 \pi ^4}{90}+20 i \pi  \zeta (3)\right)\epsilon^4+ {\cal O}(\epsilon^5),\nonumber\\
	F^\text{NP4}_6|_{t=0} &= F^\text{NP4}_{9}|_{u=0} = 1+\frac{\pi ^2}{3}\epsilon^2-2\zeta (3)\epsilon^3+\frac{\pi^4}{10}\epsilon^4+ {\cal O}(\epsilon^5).
\end{align}
The boundary conditions of $F^\text{NP4}_{14}$, $F^\text{NP4}_{15}$, $F^\text{NP4}_{23}$ are fixed at $t = m_t^2$,
\begin{align}
	F^\text{NP4}_{14}|_{t=m_t^2} = F^\text{NP4}_{15}|_{t=m_t^2} = F^\text{NP4}_{23}|_{t=m_t^2} = F^\text{NP4}_{29}|_{t=m_t^2} = 0.
\end{align}
The boundary conditions of $F^\text{NP4}_{27}$ and $F^\text{NP4}_{31}$ are determined at $u = m_t^2$ and $s = 0$, respectively,
\begin{align}
	F^\text{NP4}_{27}|_{u=m_t^2} =0, \quad F^\text{NP4}_{31}|_{s=0} = 0.
\end{align}
The remaining boundary conditions can be specified using the regularity property at the kinematic point 
$(s = m_t^2, u = 0, t = 0)$.

The solution of the differential equations can be expressed fully in terms of multiple polylogarithms.
All the analytical results have been checked with the numerical packages {\tt FIESTA} \cite{Smirnov:2021rhf} and {\tt AMFlow} \cite{Liu:2022chg}. Here we present the result of $g^\text{NP4}_{46}$ at a physical point $(m_t^2 = 1, s = 10, u = -9/4, t = -27/4)$,
\begin{align}
	g^\text{NP4,\text{analytic}}_{46} &= -0.312500 + (2.32868 - 0.785398 i)\epsilon + (-5.55281 + 7.88406 i) \epsilon^2 
	\nonumber\\
	&\quad- (5.90811 + 47.4396 i) \epsilon^3 + (201.574 - 88.2022 i) \epsilon^4+ {\cal O}(\epsilon^5).
\end{align}
The results of the other basis integrals are provided in the ancillary file.

\subsection{The behavior of $m_W^2$ expansion}
In this section, we have focused on the MIs with $m_W^2=0$.
The full MIs  $I(m_W^2)$ can be considered as a Taylor series of $m_W^2$,
i.e.,
\begin{eqnarray}
	I\left(m_W^2\right) & = &  I(m_W^2=0)+ \sum_{n=1}^{\infty}\frac{(m_W^2)^n}{n!}\left.\frac{\partial^n I}{\partial (m_W^2)^n}\right|_{m_W^2=0} \nonumber \\
	&=& I(m_W^2=0)+\sum_{n=1}^{\infty} I^{(n)}\times  m_W^{2n},
	\label{eq:Iexpansion}	
\end{eqnarray}
where we have suppressed the other variables in the arguments of $I(m_W^2)$.
The coefficient at each order $I^{(n)}$ is a linear combination of the MIs with $m_W^2=0$ because we can transform the differential operator to
\begin{eqnarray}
	\frac{\partial}{\partial(m_W^2)} = \left(c_1 k_1 + c_2 k_2 + c_3 k_3\right)\cdot \frac{\partial}{\partial k_3},
	\label{eq:partialk}
\end{eqnarray}
where
\begin{eqnarray}
	c_1 &=& \frac{u \left(-m_W^2+2 s+u\right)+m_t^2 \left(m_W^2-u\right)}{2 s \left(u \left(-m_W^2+s+u\right)+m_t^2 \left(m_W^2-u\right)\right)},\nonumber\\
	c_2 &=& \frac{\left(u-m_t^2\right) \left(-m_t^2+s+u\right)}{2 s \left(u \left(-m_W^2+s+u\right)+m_t^2 \left(m_W^2-u\right)\right)},\nonumber\\
	c_3 &=& \frac{m_t^2-u}{2 \left(u \left(-m_W^2+s+u\right)+m_t^2 \left(m_W^2-u\right)\right)}.
\end{eqnarray}
After applying the differential operator, the integrals can again be reduced back to a linear combination of the MIs.
Notice that the coefficients $c_i,i=1,2,3,$ must be expanded before taking the limit $m_W^2\to 0$.

\begin{figure}[ht]
	\centering
	\begin{minipage}{0.45\linewidth}
		\centering
		\includegraphics[width=1.0\linewidth]{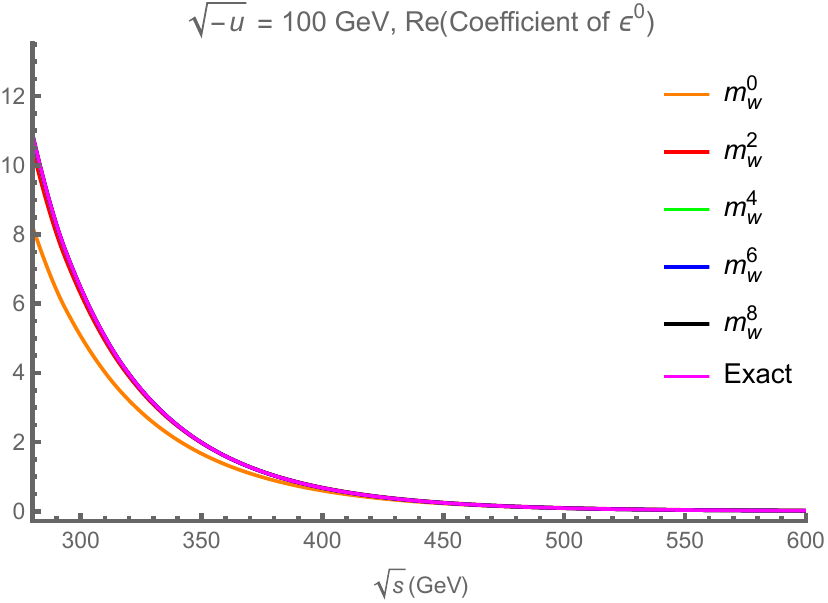}
	\end{minipage}
	\begin{minipage}{0.45\linewidth}
		\centering
		\includegraphics[width=1.0\linewidth]{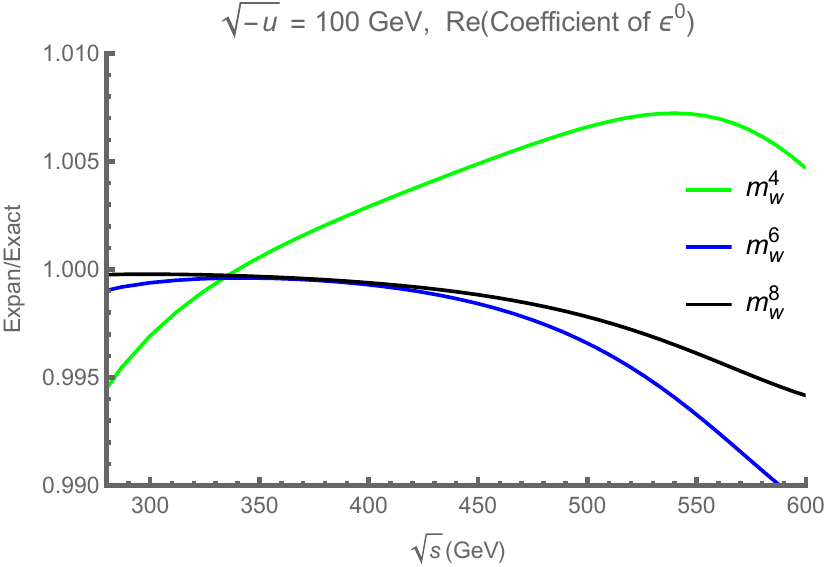}
	\end{minipage}
	\begin{minipage}{0.45\linewidth}
	\centering
	\includegraphics[width=1.0\linewidth]{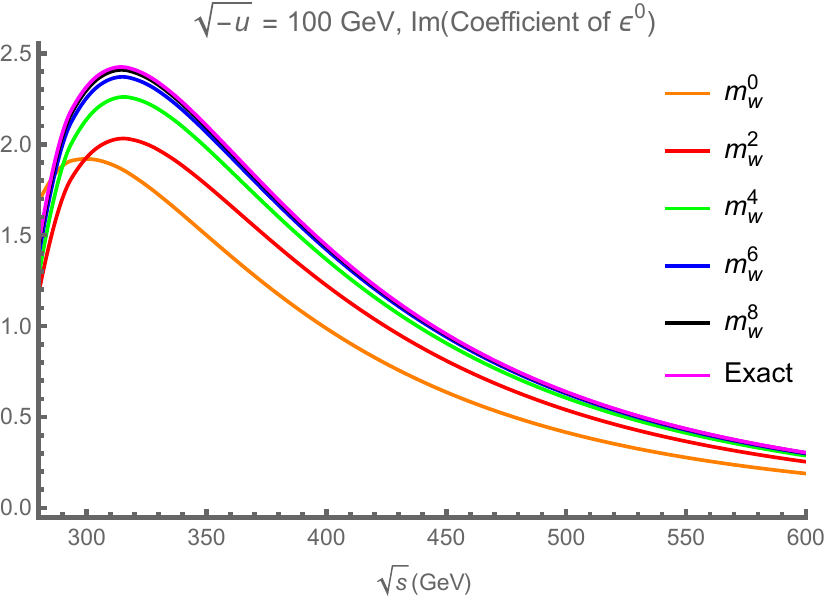}
	\end{minipage}
	\begin{minipage}{0.45\linewidth}
		\centering
		\includegraphics[width=1.0\linewidth]{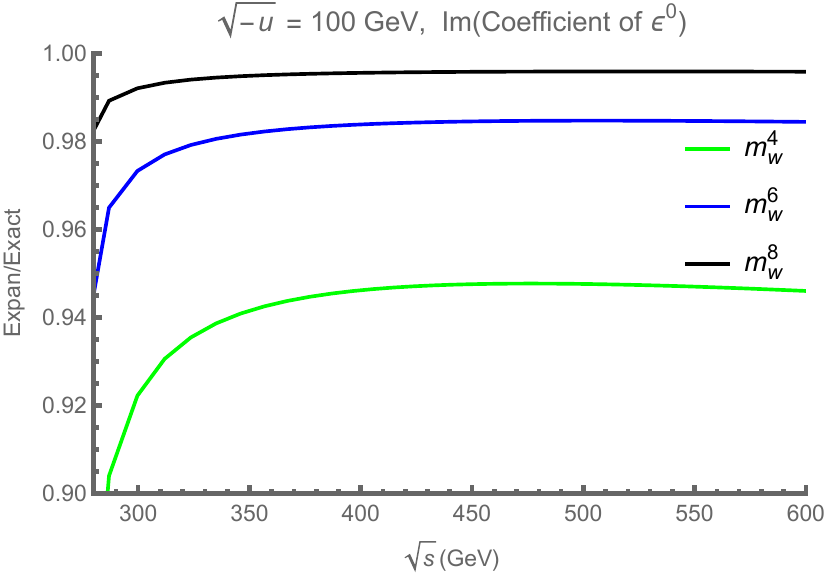}
	\end{minipage}
\caption{The real (upper plots) and imaginary (lower plots) parts of the $\epsilon^0$ coefficient of $(1/m_t^6)I^{\text{NP2}}_{1,1,1,1,1,1,1,0,0}(m_W^2)$ as a function of $\sqrt{s}$ with $\sqrt{-u}$ = 100 GeV.}
\label{fig:u100-expansion}
\end{figure}

Here we present the result of the integral
$
(1/m_t^6)I^{\text{NP2}}_{1,1,1,1,1,1,1,0,0}(m_W^2)
$
as an example. 
At the moment, we do not have the full analytical expression  at $m_W \neq0$, and take the numerical result  using the {\tt AMFlow} package  as the exact value.
In figure \ref{fig:u100-expansion}, we compare the expanded result for the coefficient of $\epsilon^{0}$ with the exact one as a function of $\sqrt{s}$.
One can see that the expansions converge rapidly to the exact results.
The difference between the expansion results up to ${\cal O}({m_W^8})$ and exact values is about ${\cal O}(10^{-3})$.
One would expect that the expansion series has a better convergence behaviour at larger $\sqrt{s}$.
This is indeed the case for the imaginary part of the integral, as shown in the figure.
However, the real part does not seem to follow this pattern. 
The reason is that the absolute values of the expansions and the exact results are almost vanishing.

\begin{figure}[ht]
	\centering
	\begin{minipage}{0.45\linewidth}
		\centering
		\includegraphics[width=1.0\linewidth]{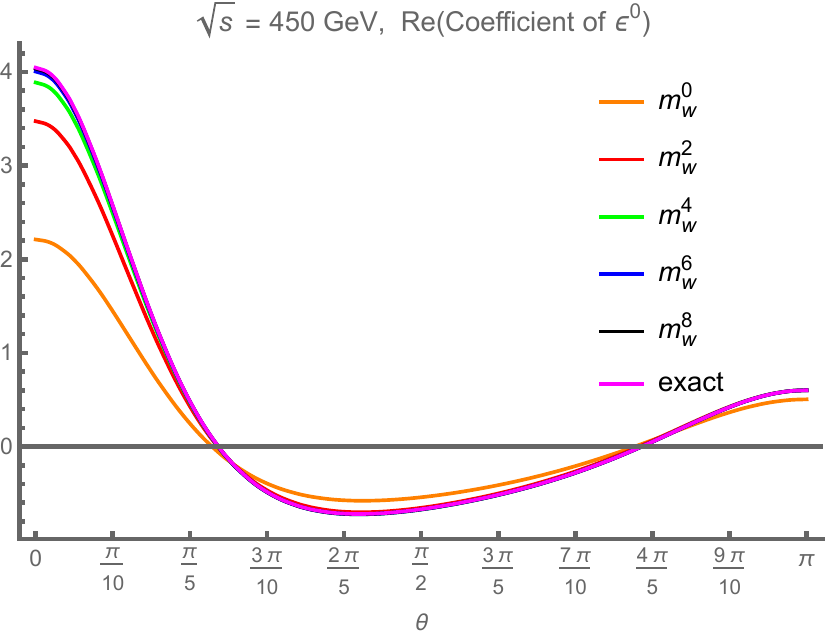}
	\end{minipage}
	\begin{minipage}{0.45\linewidth}
		\centering
		\includegraphics[width=1.0\linewidth]{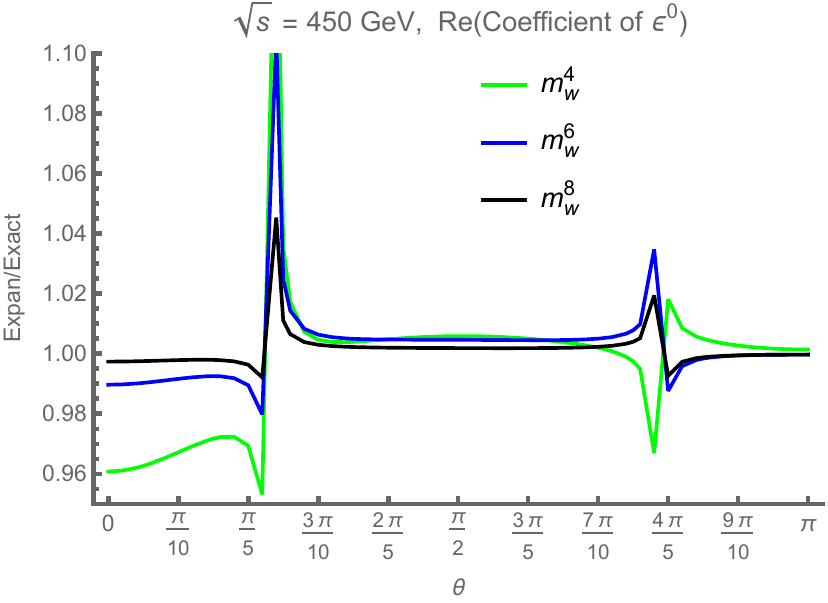}
	\end{minipage}
	\begin{minipage}{0.45\linewidth}
		\centering
		\includegraphics[width=1.0\linewidth]{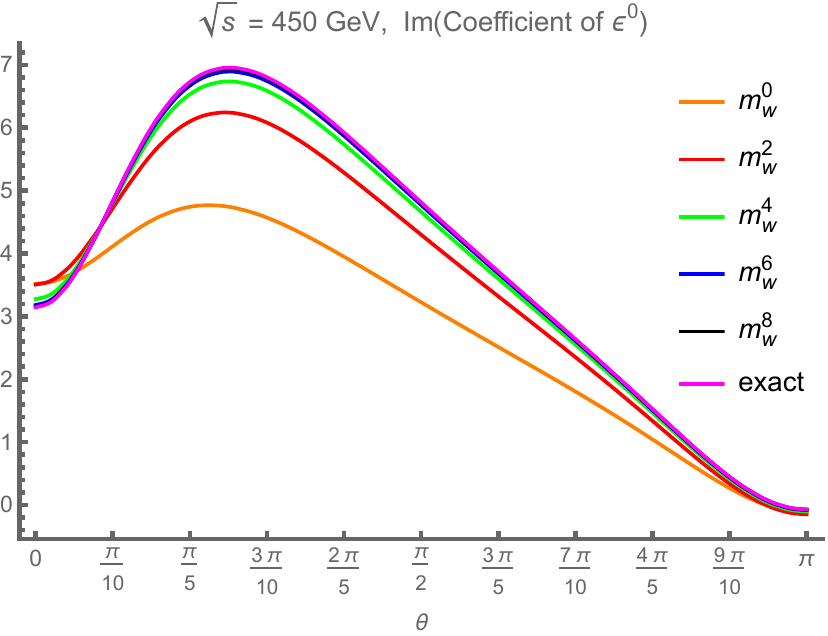}
	\end{minipage}
	\begin{minipage}{0.45\linewidth}
		\centering
		\includegraphics[width=1.0\linewidth]{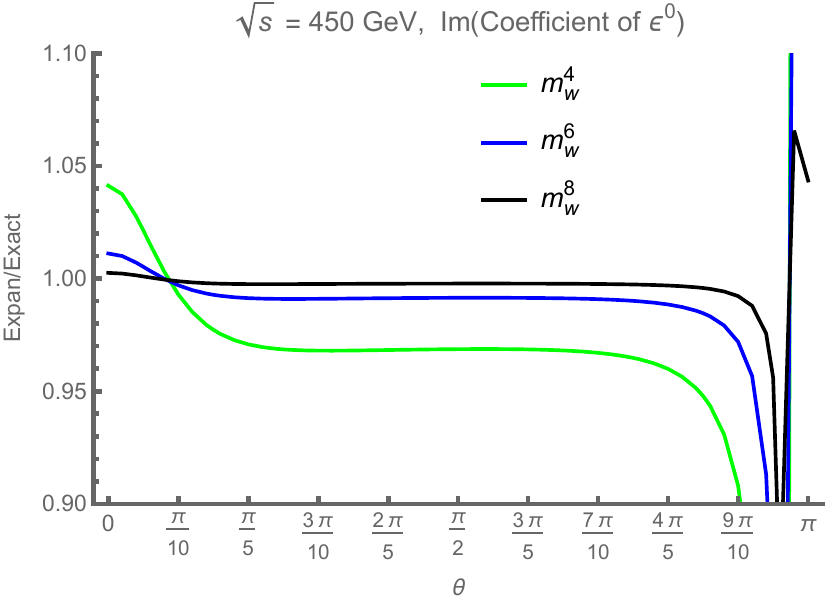}
	\end{minipage}
	\caption{The real (upper plots) and  imaginary (lower plots) parts of the $\epsilon^0$ coefficient of $(1/m_t^6)I^{\text{NP2}}_{1,1,1,1,1,1,1,0,0}(m_W^2)$ as a function of $\theta$ with $\sqrt{s}$ = 450 GeV.}
	\label{fig:s450-expansion}
\end{figure}

To have a global picture, we also show the comparison as a function of $\sqrt{-u}$ in figure \ref{fig:s450-expansion}.
In the plot we use the angle $\theta$ between the final-state $W$ boson and the initial-state gluon as the parameter, and $u$ is related to $\theta$ by
\begin{align}
u = \frac{m_t^2+m_W^2-s-r^\text{P2}_1 \cos\theta }{2}.
\end{align}
Again, we find that the differences between the expansions up to ${\cal O}({m_W^8})$ and exact values are at the level of ${\cal O}(10^{-3})$.
This accuracy can be further improved by taking expansion to higher orders,
which is not difficult to achieve since the MIs are the same.

\section{Conclusions}
\label{sec:conclusions}

The associated $tW$ production is an important process at hadron colliders.
Precision prediction of the cross section of this process requires the calculation of relevant two-loop Feynman integrals.
We present the analytical results for the two-loop master integrals that contain two massive top-quark propagators, following our previous work on integrals of one massive propagator.
There are four integral families in this category.
For the planar integral family, we construct the integral basis whose differential equations are in the canonical form so that the final results can be written in terms of multiple polylogarithms.
To simplify the calculation of the non-planar integral families, we adopt the method of expansion in the $W$ boson mass $m_W^2$.
Under the condition $m_W^2=0$, 
 the numbers of both the master integrals and the square roots in the differential equations are reduced.
In the NP2 family, only one square root is involved and can be rationalized.
However, there are still four square roots in the NP3 family, which can not be rationalized simultaneously. 
We observe that the differential equations of a proper linear combination of the basis exhibit such a dependence on the square roots that all the basis integrals can be solved in terms of multiple polylogarithms up to weight three.
Meanwhile, the weight-four results can be obtained as single integrals, whose numerical evaluation can be performed simply using the classical Newton-Cotes formulas.
With the same technique, all the integrals in the NP4 family are expressible in terms of multiple polylogarithms up to weight four. 
Our analytical results for the master integrals have been used in the computation of the leading color two-loop amplitudes in \cite{Chen:2022yni},
and could also contribute to the full two-loop amplitudes in future.

\section*{Acknowledgements}
We would like to thank Jiaqi Chen, Long-Bin Chen, Hai Tao Li and Zhao Li for helpful discussions. This work was supported in part by the National Science Foundation of China (grant Nos. 12005117 and 12147154) and the Taishan Scholar Foundation of Shandong province (tsqn201909011).
The topology diagrams in this paper were drawn using the TikZ-Feynman package \cite{Ellis:2016jkw}.

\appendix
\section{Topologies of the master integrals }

The topology diagrams of the master integrals in the P2, NP2, NP3 and NP4 families are displayed below.
\begin{figure}[H]
	\centering
	\begin{minipage}{0.15\linewidth}
		\centering
		\includegraphics[width=0.7\linewidth]{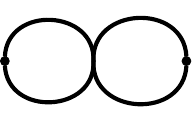}
		\caption*{$\M^\text{P2}_{1}$}
	\end{minipage}
	\begin{minipage}{0.15\linewidth}
		\centering
		\includegraphics[width=0.8\linewidth]{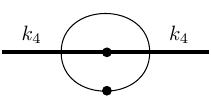}
		\caption*{$\M^\text{P2}_{2}$}
	\end{minipage}
	\begin{minipage}{0.15\linewidth}
		\centering
		\includegraphics[width=0.8\linewidth]{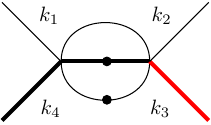}
		\caption*{$\M^\text{P2}_{3}$}
	\end{minipage}
	\begin{minipage}{0.15\linewidth}
		\centering
		\includegraphics[width=0.8\linewidth]{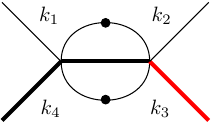}
		\caption*{$\M^\text{P2}_{4}$}
	\end{minipage}
	\begin{minipage}{0.15\linewidth}
		\centering
		\includegraphics[width=0.8\linewidth]{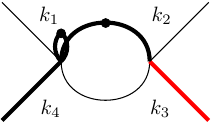}
		\caption*{$\M^\text{P2}_{5}$}
	\end{minipage}
	\begin{minipage}{0.15\linewidth}
		\centering
		\includegraphics[width=0.8\linewidth]{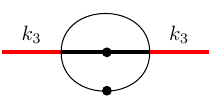}
		\caption*{$\M^\text{P2}_{6}$}
	\end{minipage}
	\begin{minipage}{0.15\linewidth}
		\centering
		\includegraphics[width=0.8\linewidth]{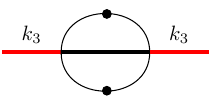}
		\caption*{$\M^\text{P2}_{7}$}
	\end{minipage}
	\begin{minipage}{0.15\linewidth}
		\centering
		\includegraphics[width=0.5\linewidth]{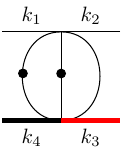}
		\caption*{$\M^\text{P2}_{8}$}
	\end{minipage}
	\begin{minipage}{0.15\linewidth}
		\centering
		\includegraphics[width=0.8\linewidth]{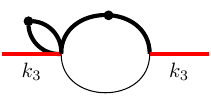}
		\caption*{$\M^\text{P2}_{9}$}
	\end{minipage}
	\begin{minipage}{0.15\linewidth}
		\centering
		\includegraphics[width=0.8\linewidth]{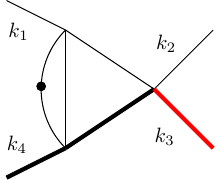}
		\caption*{$\M^\text{P2}_{10}$}
	\end{minipage}
	\begin{minipage}{0.15\linewidth}
		\centering
		\includegraphics[width=1.0\linewidth]{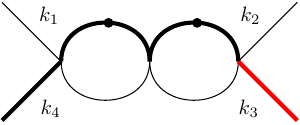}
		\caption*{$\M^\text{P2}_{11}$}
	\end{minipage}
	\begin{minipage}{0.15\linewidth}
		\centering
		\includegraphics[width=0.8\linewidth]{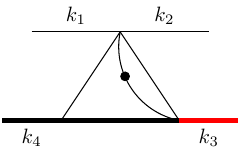}
		\caption*{$\M^\text{P2}_{12}$}
	\end{minipage}
	\begin{minipage}{0.15\linewidth}
		\centering
		\includegraphics[width=0.8\linewidth]{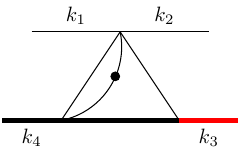}
		\caption*{$\M^\text{P2}_{13}$}
	\end{minipage}
	\begin{minipage}{0.15\linewidth}
		\centering
		\includegraphics[width=0.8\linewidth]{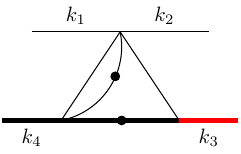}
		\caption*{$\M^\text{P2}_{14}$}
	\end{minipage}
	\begin{minipage}{0.15\linewidth}
		\centering
		\includegraphics[width=0.8\linewidth]{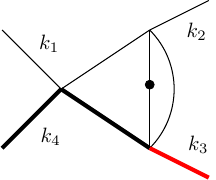}
		\caption*{$\M^\text{P2}_{15}$}
	\end{minipage}
	\begin{minipage}{0.15\linewidth}
		\centering
		\includegraphics[width=1.0\linewidth]{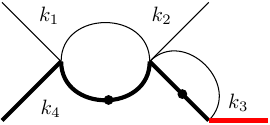}
		\caption*{$\M^\text{P2}_{16}$}
	\end{minipage}
	\begin{minipage}{0.15\linewidth}
		\centering
		\includegraphics[width=0.8\linewidth]{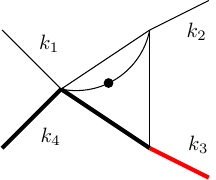}
		\caption*{$\M^\text{P2}_{17}$}
	\end{minipage}
	\begin{minipage}{0.15\linewidth}
		\centering
		\includegraphics[width=0.8\linewidth]{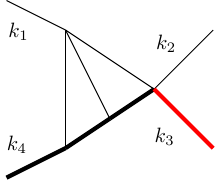}
		\caption*{$\M^\text{P2}_{18}$}
	\end{minipage}
	\begin{minipage}{0.15\linewidth}
		\centering
		\includegraphics[width=0.8\linewidth]{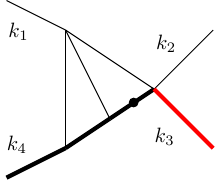}
		\caption*{$\M^\text{P2}_{19}$}
	\end{minipage}
	\begin{minipage}{0.15\linewidth}
		\centering
		\includegraphics[width=0.8\linewidth]{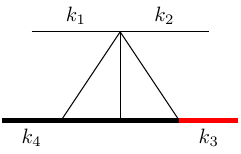}
		\caption*{$\M^\text{P2}_{20}$}
	\end{minipage}
	\begin{minipage}{0.15\linewidth}
		\centering
		\includegraphics[width=0.8\linewidth]{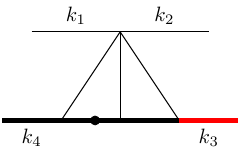}
		\caption*{$\M^\text{P2}_{21}$}
	\end{minipage}
	\begin{minipage}{0.15\linewidth}
		\centering
		\includegraphics[width=0.8\linewidth]{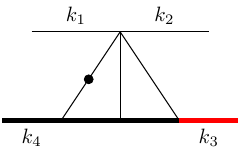}
		\caption*{$\M^\text{P2}_{22}$}
	\end{minipage}
	\begin{minipage}{0.15\linewidth}
		\centering
		\includegraphics[width=0.8\linewidth]{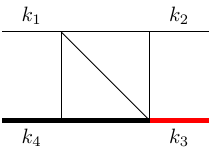}
		\caption*{$\M^\text{P2}_{23}$}
	\end{minipage}
	\begin{minipage}{0.15\linewidth}
		\centering
		\includegraphics[width=0.8\linewidth]{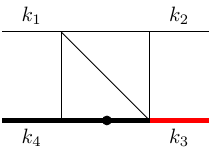}
		\caption*{$\M^\text{P2}_{24}$}
	\end{minipage}
	\begin{minipage}{0.15\linewidth}
		\centering
		\includegraphics[width=0.8\linewidth]{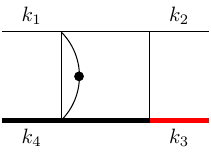}
		\caption*{$\M^\text{P2}_{25}$}
	\end{minipage}
	\begin{minipage}{0.15\linewidth}
		\centering
		\includegraphics[width=0.8\linewidth]{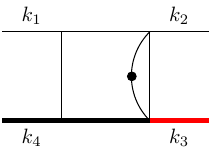}
		\caption*{$\M^\text{P2}_{26}$}
	\end{minipage}
	\begin{minipage}{0.15\linewidth}
		\centering
		\includegraphics[width=0.8\linewidth]{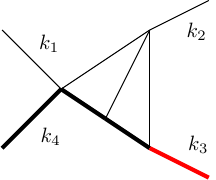}
		\caption*{$\M^\text{P2}_{27}$}
	\end{minipage}
	\begin{minipage}{0.15\linewidth}
		\centering
		\includegraphics[width=0.8\linewidth]{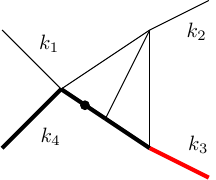}
		\caption*{$\M^\text{P2}_{28}$}
	\end{minipage}
	\begin{minipage}{0.15\linewidth}
		\centering
		\includegraphics[width=0.8\linewidth]{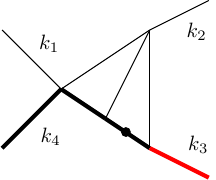}
		\caption*{$\M^\text{P2}_{29}$}
	\end{minipage}
	\begin{minipage}{0.15\linewidth}
		\centering
		\includegraphics[width=0.8\linewidth]{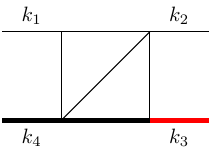}
		\caption*{$\M^\text{P2}_{30}$}
	\end{minipage}
	\begin{minipage}{0.15\linewidth}
		\centering
		\includegraphics[width=0.8\linewidth]{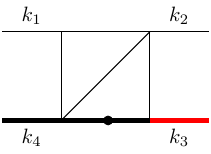}
		\caption*{$\M^\text{P2}_{31}$}
	\end{minipage}
	\begin{minipage}{0.15\linewidth}
		\centering
		\includegraphics[width=0.8\linewidth]{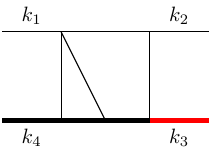}
		\caption*{$\M^\text{P2}_{32},\M^\text{P2}_{33}$}
	\end{minipage}
	\begin{minipage}{0.15\linewidth}
		\centering
		\includegraphics[width=0.8\linewidth]{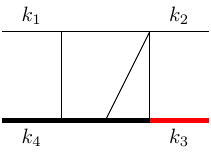}
		\caption*{$\M^\text{P2}_{34}-\M^\text{P2}_{36}$}
	\end{minipage}
	\begin{minipage}{0.15\linewidth}
		\centering
		\includegraphics[width=0.9\linewidth]{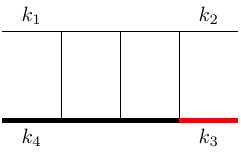}
		\caption*{$\M^\text{P2}_{37},\M^\text{P2}_{38}$}
	\end{minipage}
	\caption{Master integrals in the P2 topology. The thick  black and red lines stand for the top quark and $W$ boson, respectively.
	One black dot indicates one additional power of the corresponding propagator. The numerators can be found in the text.}
	\label{fig:P2_MIs}	
\end{figure}
\begin{figure}[H]
	\centering
	\begin{minipage}{0.15\linewidth}
		\centering
		\includegraphics[width=0.7\linewidth]{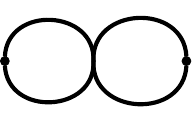}
		\caption*{$\M^\text{NP2}_{1}$}
	\end{minipage}
	\begin{minipage}{0.15\linewidth}
		\centering
		\includegraphics[width=0.5\linewidth]{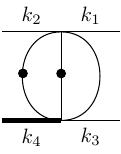}
		\caption*{$\M^\text{NP2}_{2}$}
	\end{minipage}
	\begin{minipage}{0.15\linewidth}
		\centering
		\includegraphics[width=0.8\linewidth]{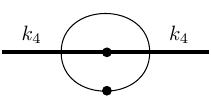}
		\caption*{$\M^\text{NP2}_{3}$}
	\end{minipage}
	\begin{minipage}{0.15\linewidth}
		\centering
		\includegraphics[width=0.8\linewidth]{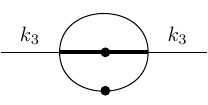}
		\caption*{$\M^\text{NP2}_{4}$}
	\end{minipage}
	\begin{minipage}{0.15\linewidth}
		\centering
		\includegraphics[width=0.5\linewidth]{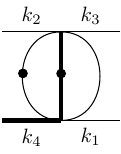}
		\caption*{$\M^\text{NP2}_{5}$}
	\end{minipage}
	\begin{minipage}{0.15\linewidth}
		\centering
		\includegraphics[width=0.5\linewidth]{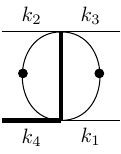}
		\caption*{$\M^\text{NP2}_{6}$}
	\end{minipage}
	\begin{minipage}{0.15\linewidth}
		\centering
		\includegraphics[width=0.8\linewidth]{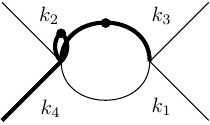}
		\caption*{$\M^\text{NP2}_{7}$}
	\end{minipage}
	\begin{minipage}{0.15\linewidth}
		\centering
		\includegraphics[width=0.8\linewidth]{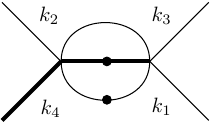}
		\caption*{$\M^\text{NP2}_{8}$}
	\end{minipage}
	\begin{minipage}{0.15\linewidth}
		\centering
		\includegraphics[width=0.8\linewidth]{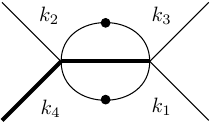}
		\caption*{$\M^\text{NP2}_{9}$}
	\end{minipage}
	\begin{minipage}{0.15\linewidth}
		\centering
		\includegraphics[width=0.8\linewidth]{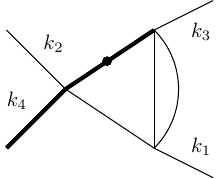}
		\caption*{$\M^\text{NP2}_{10}$}
	\end{minipage}
	\begin{minipage}{0.15\linewidth}
		\centering
		\includegraphics[width=0.9\linewidth]{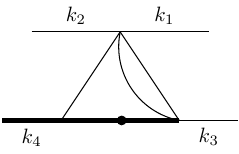}
		\caption*{$\M^\text{NP2}_{11}$}
	\end{minipage}
	\begin{minipage}{0.15\linewidth}
		\centering
		\includegraphics[width=0.9\linewidth]{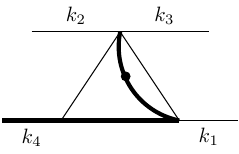}
		\caption*{$\M^\text{NP2}_{12}$}
	\end{minipage}
	\begin{minipage}{0.15\linewidth}
		\centering
		\includegraphics[width=0.9\linewidth]{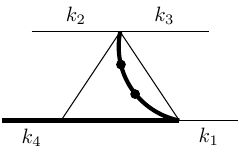}
		\caption*{$\M^\text{NP2}_{13}$}
	\end{minipage}
	\begin{minipage}{0.15\linewidth}
		\centering
		\includegraphics[width=0.9\linewidth]{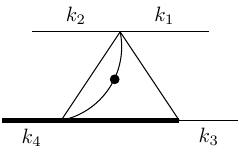}
		\caption*{$\M^\text{NP2}_{14}$}
	\end{minipage}
	\begin{minipage}{0.15\linewidth}
		\centering
		\includegraphics[width=0.8\linewidth]{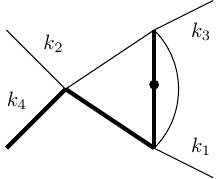}
		\caption*{$\M^\text{NP2}_{15}$}
	\end{minipage}
	\begin{minipage}{0.15\linewidth}
		\centering
		\includegraphics[width=0.8\linewidth]{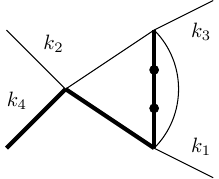}
		\caption*{$\M^\text{NP2}_{16}$}
	\end{minipage}
	\begin{minipage}{0.15\linewidth}
		\centering
		\includegraphics[width=0.8\linewidth]{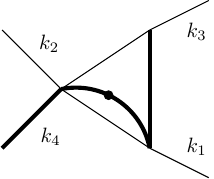}
		\caption*{$\M^\text{NP2}_{17}$}
	\end{minipage}
	\begin{minipage}{0.15\linewidth}
		\centering
		\includegraphics[width=0.8\linewidth]{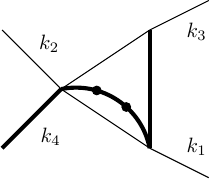}
		\caption*{$\M^\text{NP2}_{18}$}
	\end{minipage}
	\begin{minipage}{0.15\linewidth}
		\centering
		\includegraphics[width=0.8\linewidth]{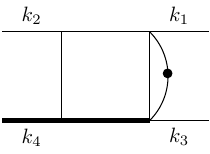}
		\caption*{$\M^\text{NP2}_{19}$}
	\end{minipage}
	\begin{minipage}{0.15\linewidth}
		\centering
		\includegraphics[width=0.8\linewidth]{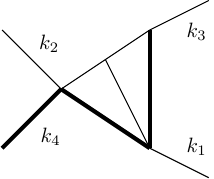}
		\caption*{$\M^\text{NP2}_{20}$}
	\end{minipage}
	\begin{minipage}{0.15\linewidth}
		\centering
		\includegraphics[width=0.8\linewidth]{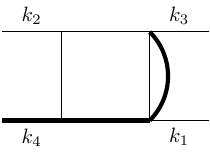}
		\caption*{$\M^\text{NP2}_{21}$}
	\end{minipage}
	\begin{minipage}{0.15\linewidth}
		\centering
		\includegraphics[width=0.8\linewidth]{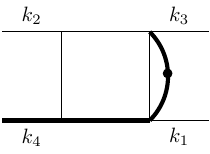}
		\caption*{$\M^\text{NP2}_{22}$}
	\end{minipage}
	\begin{minipage}{0.15\linewidth}
		\centering
		\includegraphics[width=0.8\linewidth]{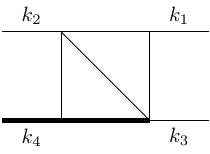}
		\caption*{$\M^\text{NP2}_{23}$}
	\end{minipage}
	\begin{minipage}{0.15\linewidth}
		\centering
		\includegraphics[width=0.8\linewidth]{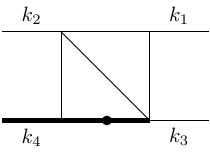}
		\caption*{$\M^\text{NP2}_{24}$}
	\end{minipage}
	\begin{minipage}{0.15\linewidth}
		\centering
		\includegraphics[width=0.8\linewidth]{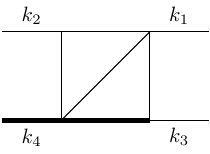}
		\caption*{$\M^\text{NP2}_{25}$}
	\end{minipage}
	\begin{minipage}{0.15\linewidth}
		\centering
		\includegraphics[width=0.8\linewidth]{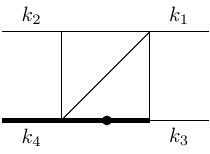}
		\caption*{$\M^\text{NP2}_{26}$}
	\end{minipage}
	\begin{minipage}{0.15\linewidth}
		\centering
		\includegraphics[width=0.8\linewidth]{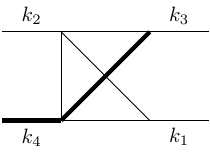}
		\caption*{$\M^\text{NP2}_{27}$}
	\end{minipage}
	\begin{minipage}{0.15\linewidth}
		\centering
		\includegraphics[width=0.8\linewidth]{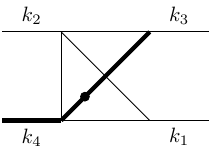}
		\caption*{$\M^\text{NP2}_{28}$}
	\end{minipage}
	\begin{minipage}{0.15\linewidth}
		\centering
		\includegraphics[width=0.8\linewidth]{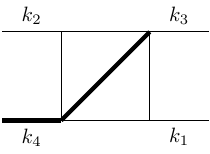}
		\caption*{$\M^\text{NP2}_{29}$}
	\end{minipage}
	\begin{minipage}{0.15\linewidth}
		\centering
		\includegraphics[width=0.8\linewidth]{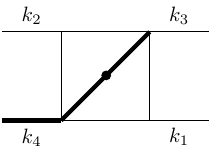}
		\caption*{$\M^\text{NP2}_{30}$}
	\end{minipage}
	\begin{minipage}{0.15\linewidth}
		\centering
		\includegraphics[width=0.8\linewidth]{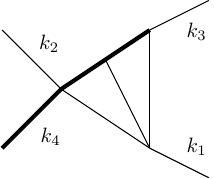}
		\caption*{$\M^\text{NP2}_{31}$}
	\end{minipage}
	\begin{minipage}{0.15\linewidth}
		\centering
		\includegraphics[width=0.8\linewidth]{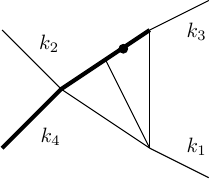}
		\caption*{$\M^\text{NP2}_{32}$}
	\end{minipage}
	\begin{minipage}{0.15\linewidth}
		\centering
		\includegraphics[width=0.8\linewidth]{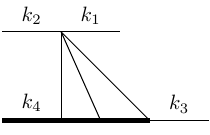}
		\caption*{$\M^\text{NP2}_{33}$}
	\end{minipage}
	\begin{minipage}{0.15\linewidth}
		\centering
		\includegraphics[width=0.8\linewidth]{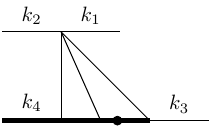}
		\caption*{$\M^\text{NP2}_{34}$}
	\end{minipage}
	\begin{minipage}{0.15\linewidth}
		\centering
		\includegraphics[width=0.8\linewidth]{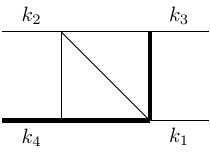}
		\caption*{$\M^\text{NP2}_{35}$}
	\end{minipage}
	\begin{minipage}{0.15\linewidth}
		\centering
		\includegraphics[width=0.8\linewidth]{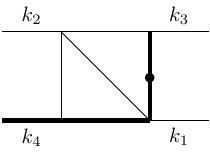}
		\caption*{$\M^\text{NP2}_{36}$}
	\end{minipage}
	\begin{minipage}{0.15\linewidth}
		\centering
		\includegraphics[width=0.8\linewidth]{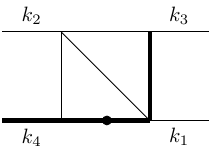}
		\caption*{$\M^\text{NP2}_{37}$}
	\end{minipage}
	\begin{minipage}{0.15\linewidth}
		\centering
		\includegraphics[width=0.8\linewidth]{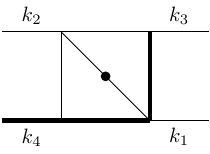}
		\caption*{$\M^\text{NP2}_{38}$}
	\end{minipage}
	\begin{minipage}{0.15\linewidth}
		\centering
		\includegraphics[width=0.8\linewidth]{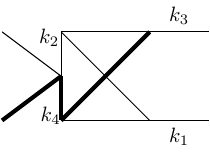}
		\caption*{$\M^\text{NP2}_{39}$}
	\end{minipage}
	\begin{minipage}{0.15\linewidth}
		\centering
		\includegraphics[width=0.8\linewidth]{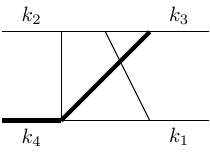}
		\caption*{$\M^\text{NP2}_{40}$}
	\end{minipage}
	\begin{minipage}{0.15\linewidth}
		\centering
		\includegraphics[width=0.8\linewidth]{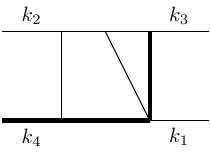}
		\caption*{$\M^\text{NP2}_{41},\M^\text{NP2}_{42}$}
	\end{minipage}
	\begin{minipage}{0.15\linewidth}
		\centering
		\includegraphics[width=0.8\linewidth]{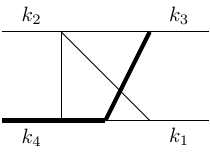}
		\caption*{$\M^\text{NP2}_{43}-\M^\text{NP2}_{46}$}
	\end{minipage}
	\begin{minipage}{0.15\linewidth}
		\centering
		\includegraphics[width=1.05\linewidth]{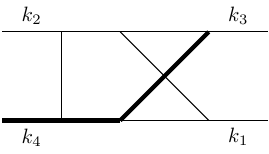}
		\caption*{$\M^\text{NP2}_{47}-\M^\text{NP2}_{49}$}
	\end{minipage}
	\caption{Master integrals in the NP2 topology.  The thick lines stand for massive top quarks while the others represent massless particles.  One black dot indicates one additional power of the corresponding propagator. The numerators can be found in the text.}
	\label{fig:NP2_MIs}
\end{figure}
\begin{figure}[H]
	\centering
	\begin{minipage}{0.15\linewidth}
		\centering
		\includegraphics[width=0.7\linewidth]{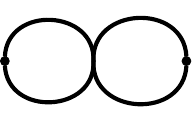}
		\caption*{$\M^\text{NP3}_{1}$}
	\end{minipage}
	\begin{minipage}{0.15\linewidth}
		\centering
		\includegraphics[width=0.8\linewidth]{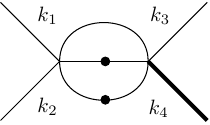}
		\caption*{$\M^\text{NP3}_{2}$}
	\end{minipage}
	\begin{minipage}{0.15\linewidth}
		\centering
		\includegraphics[width=0.8\linewidth]{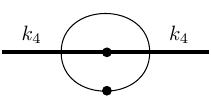}
		\caption*{$\M^\text{NP3}_{3}$}
	\end{minipage}
	\begin{minipage}{0.15\linewidth}
		\centering
		\includegraphics[width=0.8\linewidth]{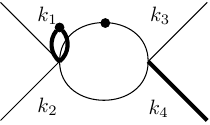}
		\caption*{$\M^\text{NP3}_{4}$}
	\end{minipage}
	\begin{minipage}{0.15\linewidth}
		\centering
		\includegraphics[width=0.8\linewidth]{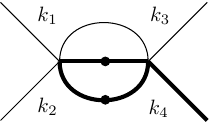}
		\caption*{$\M^\text{NP3}_{5}$}
	\end{minipage}
	\begin{minipage}{0.15\linewidth}
		\centering
		\includegraphics[width=0.8\linewidth]{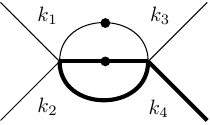}
		\caption*{$\M^\text{NP3}_{6}$}
	\end{minipage}
	\begin{minipage}{0.15\linewidth}
		\centering
		\includegraphics[width=0.8\linewidth]{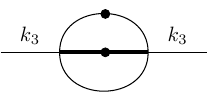}
		\caption*{$\M^\text{NP3}_{7}$}
	\end{minipage}
	\begin{minipage}{0.15\linewidth}
		\centering
		\includegraphics[width=0.5\linewidth]{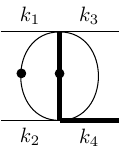}
		\caption*{$\M^\text{NP3}_{8}$}
	\end{minipage}
	\begin{minipage}{0.15\linewidth}
		\centering
		\includegraphics[width=0.5\linewidth]{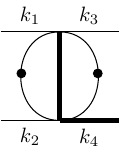}
		\caption*{$\M^\text{NP3}_{9}$}
	\end{minipage}
	\begin{minipage}{0.15\linewidth}
		\centering
		\includegraphics[width=0.5\linewidth]{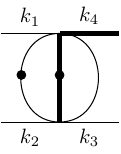}
		\caption*{$\M^\text{NP3}_{10}$}
	\end{minipage}
	\begin{minipage}{0.15\linewidth}
		\centering
		\includegraphics[width=0.5\linewidth]{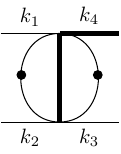}
		\caption*{$\M^\text{NP3}_{11}$}
	\end{minipage}
	\begin{minipage}{0.15\linewidth}
		\centering
		\includegraphics[width=0.8\linewidth]{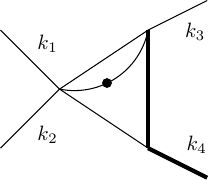}
		\caption*{$\M^\text{NP3}_{12}$}
	\end{minipage}
	\begin{minipage}{0.15\linewidth}
		\centering
		\includegraphics[width=0.8\linewidth]{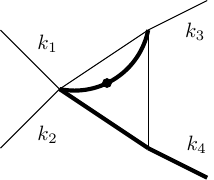}
		\caption*{$\M^\text{NP3}_{13}$}
	\end{minipage}
	\begin{minipage}{0.15\linewidth}
		\centering
		\includegraphics[width=0.8\linewidth]{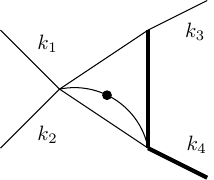}
		\caption*{$\M^\text{NP3}_{14}$}
	\end{minipage}
	\begin{minipage}{0.15\linewidth}
		\centering
		\includegraphics[width=0.9\linewidth]{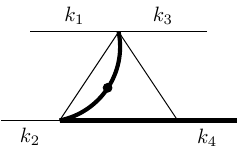}
		\caption*{$\M^\text{NP3}_{15}$}
	\end{minipage}
	\begin{minipage}{0.15\linewidth}
		\centering
		\includegraphics[width=0.9\linewidth]{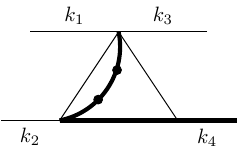}
		\caption*{$\M^\text{NP3}_{16}$}
	\end{minipage}
	\begin{minipage}{0.15\linewidth}
		\centering
		\includegraphics[width=0.8\linewidth]{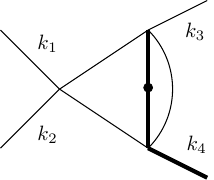}
		\caption*{$\M^\text{NP3}_{17}$}
	\end{minipage}
	\begin{minipage}{0.15\linewidth}
		\centering
		\includegraphics[width=0.8\linewidth]{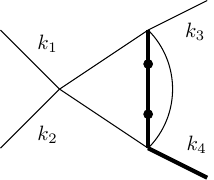}
		\caption*{$\M^\text{NP3}_{18}$}
	\end{minipage}
	\begin{minipage}{0.15\linewidth}
		\centering
		\includegraphics[width=0.8\linewidth]{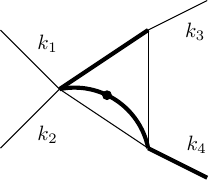}
		\caption*{$\M^\text{NP3}_{19}$}
	\end{minipage}
	\begin{minipage}{0.15\linewidth}
		\centering
		\includegraphics[width=0.8\linewidth]{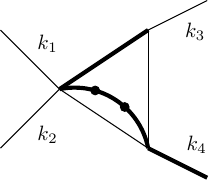}
		\caption*{$\M^\text{NP3}_{20}$}
	\end{minipage}
	\begin{minipage}{0.15\linewidth}
		\centering
		\includegraphics[width=0.9\linewidth]{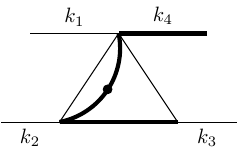}
		\caption*{$\M^\text{NP3}_{21}$}
	\end{minipage}
	\begin{minipage}{0.15\linewidth}
		\centering
		\includegraphics[width=0.9\linewidth]{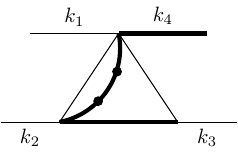}
		\caption*{$\M^\text{NP3}_{22}$}
	\end{minipage}
	\begin{minipage}{0.15\linewidth}
		\centering
		\includegraphics[width=0.8\linewidth]{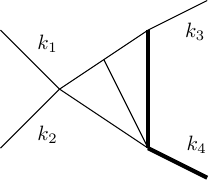}
		\caption*{$\M^\text{NP3}_{23}$}
	\end{minipage}
	\begin{minipage}{0.15\linewidth}
		\centering
		\includegraphics[width=0.8\linewidth]{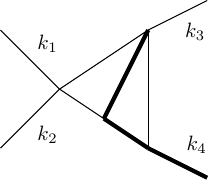}
		\caption*{$\M^\text{NP3}_{24}$}
	\end{minipage}
	\begin{minipage}{0.15\linewidth}
		\centering
		\includegraphics[width=0.8\linewidth]{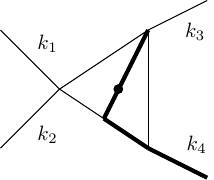}
		\caption*{$\M^\text{NP3}_{25}$}
	\end{minipage}
	\begin{minipage}{0.15\linewidth}
		\centering
		\includegraphics[width=0.8\linewidth]{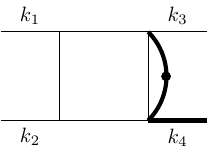}
		\caption*{$\M^\text{NP3}_{26}$}
	\end{minipage}
	\begin{minipage}{0.15\linewidth}
		\centering
		\includegraphics[width=0.8\linewidth]{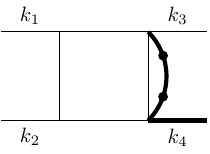}
		\caption*{$\M^\text{NP3}_{27}$}
	\end{minipage}
	\begin{minipage}{0.15\linewidth}
		\centering
		\includegraphics[width=0.8\linewidth]{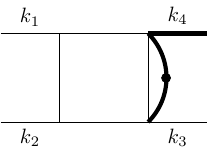}
		\caption*{$\M^\text{NP3}_{28}$}
	\end{minipage}
	\begin{minipage}{0.15\linewidth}
		\centering
		\includegraphics[width=0.8\linewidth]{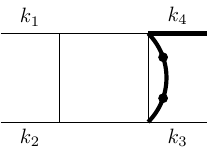}
		\caption*{$\M^\text{NP3}_{29}$}
	\end{minipage}
	\begin{minipage}{0.15\linewidth}
		\centering
		\includegraphics[width=0.8\linewidth]{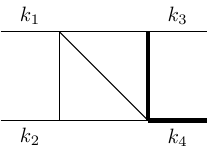}
		\caption*{$\M^\text{NP3}_{30}$}
	\end{minipage}
	\begin{minipage}{0.15\linewidth}
		\centering
		\includegraphics[width=0.8\linewidth]{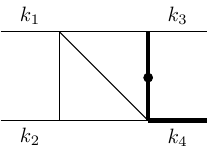}
		\caption*{$\M^\text{NP3}_{31}$}
	\end{minipage}
	\begin{minipage}{0.15\linewidth}
		\centering
		\includegraphics[width=0.8\linewidth]{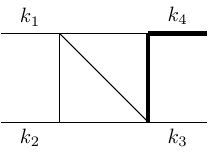}
		\caption*{$\M^\text{NP3}_{32}$}
	\end{minipage}
	\begin{minipage}{0.15\linewidth}
		\centering
		\includegraphics[width=0.8\linewidth]{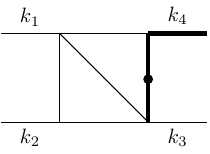}
		\caption*{$\M^\text{NP3}_{33}$}
	\end{minipage}
	\begin{minipage}{0.15\linewidth}
		\centering
		\includegraphics[width=0.8\linewidth]{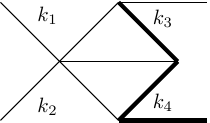}
		\caption*{$\M^\text{NP3}_{34}$}
	\end{minipage}
	\begin{minipage}{0.15\linewidth}
		\centering
		\includegraphics[width=0.8\linewidth]{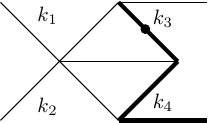}
		\caption*{$\M^\text{NP3}_{35}$}
	\end{minipage}
	\begin{minipage}{0.15\linewidth}
		\centering
		\includegraphics[width=0.8\linewidth]{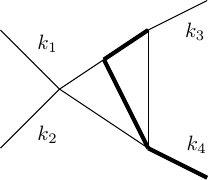}
		\caption*{$\M^\text{NP3}_{36}$}
	\end{minipage}
	\begin{minipage}{0.15\linewidth}
		\centering
		\includegraphics[width=0.8\linewidth]{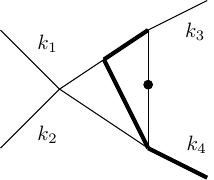}
		\caption*{$\M^\text{NP3}_{37}$}
	\end{minipage}
	\begin{minipage}{0.15\linewidth}
		\centering
		\includegraphics[width=0.8\linewidth]{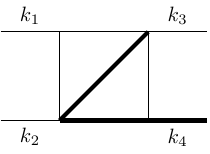}
		\caption*{$\M^\text{NP3}_{38}$}
	\end{minipage}
	\begin{minipage}{0.15\linewidth}
		\centering
		\includegraphics[width=0.8\linewidth]{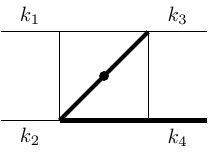}
		\caption*{$\M^\text{NP3}_{39}$}
	\end{minipage}
	\begin{minipage}{0.15\linewidth}
		\centering
		\includegraphics[width=0.8\linewidth]{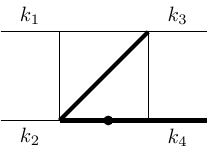}
		\caption*{$\M^\text{NP3}_{40}$}
	\end{minipage}
	\begin{minipage}{0.15\linewidth}
		\centering
		\includegraphics[width=0.8\linewidth]{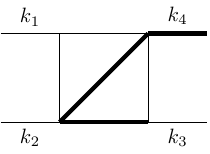}
		\caption*{$\M^\text{NP3}_{41}$}
	\end{minipage}
	\begin{minipage}{0.15\linewidth}
		\centering
		\includegraphics[width=0.8\linewidth]{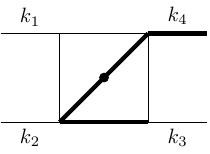}
		\caption*{$\M^\text{NP3}_{42}$}
	\end{minipage}
	\begin{minipage}{0.15\linewidth}
		\centering
		\includegraphics[width=0.8\linewidth]{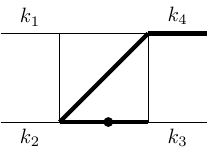}
		\caption*{$\M^\text{NP3}_{43}$}
	\end{minipage}
	\begin{minipage}{0.15\linewidth}
		\centering
		\includegraphics[width=0.8\linewidth]{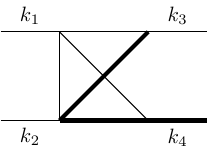}
		\caption*{$\M^\text{NP3}_{44}$}
	\end{minipage}
	\begin{minipage}{0.15\linewidth}
		\centering
		\includegraphics[width=0.8\linewidth]{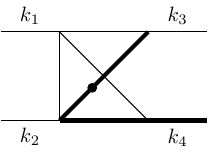}
		\caption*{$\M^\text{NP3}_{45}$}
	\end{minipage}
	\begin{minipage}{0.15\linewidth}
		\centering
		\includegraphics[width=0.8\linewidth]{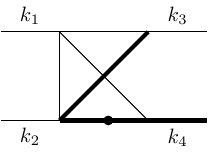}
		\caption*{$\M^\text{NP3}_{46}$}
	\end{minipage}
	\begin{minipage}{0.15\linewidth}
		\centering
		\includegraphics[width=0.8\linewidth]{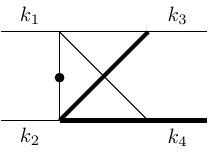}
		\caption*{$\M^\text{NP3}_{47}$}
	\end{minipage}
	\begin{minipage}{0.15\linewidth}
		\centering
		\includegraphics[width=0.8\linewidth]{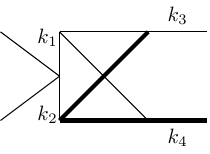}
		\caption*{$\M^\text{NP3}_{48}$}
	\end{minipage}
	\begin{minipage}{0.15\linewidth}
		\centering
		\includegraphics[width=0.8\linewidth]{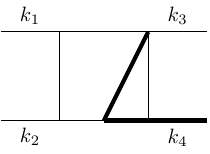}
		\caption*{$\M^\text{NP3}_{49}$}
	\end{minipage}
	\begin{minipage}{0.15\linewidth}
		\centering
		\includegraphics[width=0.8\linewidth]{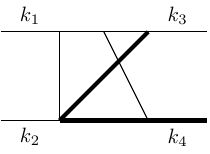}
		\caption*{$\M^\text{NP3}_{50}$, $\M^\text{NP3}_{51}$}
	\end{minipage}
	\begin{minipage}{0.15\linewidth}
		\centering
		\includegraphics[width=0.8\linewidth]{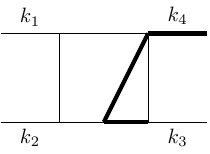}
		\caption*{$\M^\text{NP3}_{52}$}
	\end{minipage}
	\begin{minipage}{0.15\linewidth}
		\centering
		\includegraphics[width=0.8\linewidth]{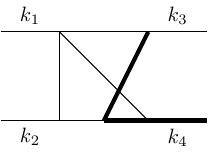}
		\caption*{$\M^\text{NP3}_{53}-\M^\text{NP3}_{56}$}
	\end{minipage}
	\begin{minipage}{0.15\linewidth}
		\centering
		\includegraphics[width=1.05\linewidth]{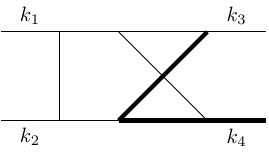}
		\caption*{$\M^\text{NP3}_{57}-\M^\text{NP3}_{60}$}
	\end{minipage}
	\caption{Master integrals in the NP3 topology. The thick lines stand for massive top quarks while the others represent massless particles.  One black dot indicates one additional power of the corresponding propagator. The numerators can be found in the text.}
	\label{fig:NP3_MIs}
\end{figure}
\begin{figure}[H]
	\centering
	\begin{minipage}{0.15\linewidth}
		\centering
		\includegraphics[width=0.7\linewidth]{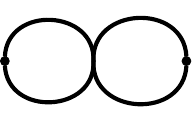}
		\caption*{$\M^\text{NP4}_{1}$}
	\end{minipage}
	\begin{minipage}{0.15\linewidth}
		\centering
		\includegraphics[width=0.8\linewidth]{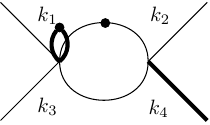}
		\caption*{$\M^\text{NP4}_{2}$}
	\end{minipage}
	\begin{minipage}{0.15\linewidth}
		\centering
		\includegraphics[width=0.5\linewidth]{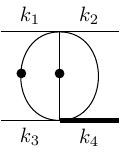}
		\caption*{$\M^\text{NP4}_{3}$}
	\end{minipage}
	\begin{minipage}{0.15\linewidth}
		\centering
		\includegraphics[width=0.8\linewidth]{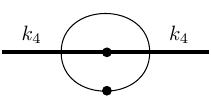}
		\caption*{$\M^\text{NP4}_{4}$}
	\end{minipage}
	\begin{minipage}{0.15\linewidth}
		\centering
		\includegraphics[width=0.8\linewidth]{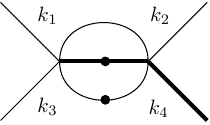}
		\caption*{$\M^\text{NP4}_{5}$}
	\end{minipage}
	\begin{minipage}{0.15\linewidth}
		\centering
		\includegraphics[width=0.8\linewidth]{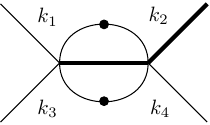}
		\caption*{$\M^\text{NP4}_{6}$}
	\end{minipage}
	\begin{minipage}{0.15\linewidth}
		\centering
		\includegraphics[width=0.8\linewidth]{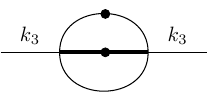}
		\caption*{$\M^\text{NP4}_{7}$}
	\end{minipage}
	\begin{minipage}{0.15\linewidth}
		\centering
		\includegraphics[width=0.5\linewidth]{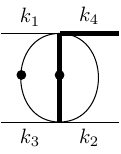}
		\caption*{$\M^\text{NP4}_{8}$}
	\end{minipage}
	\begin{minipage}{0.15\linewidth}
		\centering
		\includegraphics[width=0.5\linewidth]{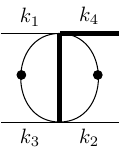}
		\caption*{$\M^\text{NP4}_{9}$}
	\end{minipage}
	\begin{minipage}{0.15\linewidth}
		\centering
		\includegraphics[width=0.8\linewidth]{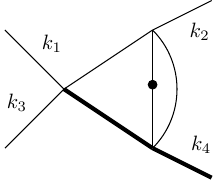}
		\caption*{$\M^\text{NP4}_{10}$}
	\end{minipage}
	\begin{minipage}{0.15\linewidth}
		\centering
		\includegraphics[width=0.8\linewidth]{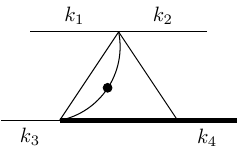}
		\caption*{$\M^\text{NP4}_{11}$}
	\end{minipage}
	\begin{minipage}{0.15\linewidth}
		\centering
		\includegraphics[width=0.8\linewidth]{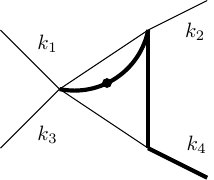}
		\caption*{$\M^\text{NP4}_{12}$}
	\end{minipage}
	\begin{minipage}{0.15\linewidth}
		\centering
		\includegraphics[width=0.8\linewidth]{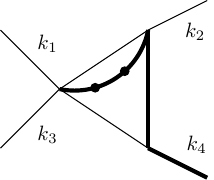}
		\caption*{$\M^\text{NP4}_{13}$}
	\end{minipage}
	\begin{minipage}{0.15\linewidth}
		\centering
		\includegraphics[width=0.8\linewidth]{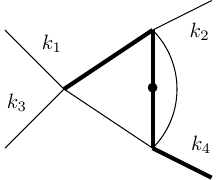}
		\caption*{$\M^\text{NP4}_{14}$}
	\end{minipage}
	\begin{minipage}{0.15\linewidth}
		\centering
		\includegraphics[width=0.8\linewidth]{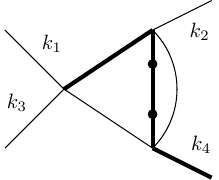}
		\caption*{$\M^\text{NP4}_{15}$}
	\end{minipage}
	\begin{minipage}{0.15\linewidth}
		\centering
		\includegraphics[width=0.8\linewidth]{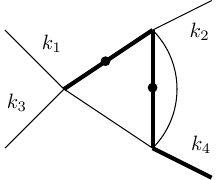}
		\caption*{$\M^\text{NP4}_{16}$}
	\end{minipage}
	\begin{minipage}{0.15\linewidth}
		\centering
		\includegraphics[width=0.9\linewidth]{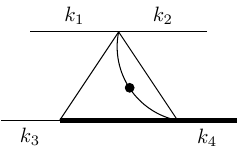}
		\caption*{$\M^\text{NP4}_{17}$}
	\end{minipage}
	\begin{minipage}{0.15\linewidth}
		\centering
		\includegraphics[width=0.9\linewidth]{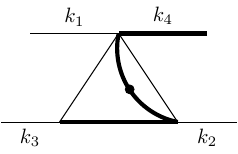}
		\caption*{$\M^\text{NP4}_{18}$}
	\end{minipage}
	\begin{minipage}{0.15\linewidth}
		\centering
		\includegraphics[width=0.9\linewidth]{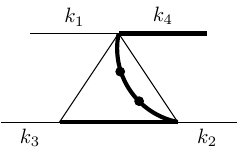}
		\caption*{$\M^\text{NP4}_{19}$}
	\end{minipage}
	\begin{minipage}{0.15\linewidth}
		\centering
		\includegraphics[width=0.8\linewidth]{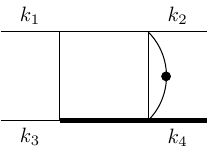}
		\caption*{$\M^\text{NP4}_{20}$}
	\end{minipage}
	\begin{minipage}{0.15\linewidth}
		\centering
		\includegraphics[width=0.8\linewidth]{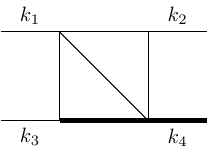}
		\caption*{$\M^\text{NP4}_{21}$}
	\end{minipage}
	\begin{minipage}{0.15\linewidth}
		\centering
		\includegraphics[width=0.8\linewidth]{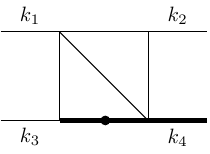}
		\caption*{$\M^\text{NP4}_{22}$}
	\end{minipage}
	\begin{minipage}{0.15\linewidth}
		\centering
		\includegraphics[width=0.8\linewidth]{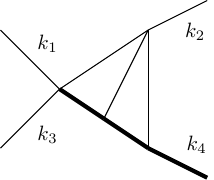}
		\caption*{$\M^\text{NP4}_{23}$}
	\end{minipage}
	\begin{minipage}{0.15\linewidth}
		\centering
		\includegraphics[width=0.8\linewidth]{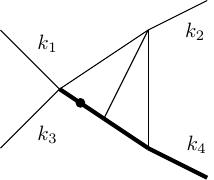}
		\caption*{$\M^\text{NP4}_{24}$}
	\end{minipage}
	\begin{minipage}{0.15\linewidth}
		\centering
		\includegraphics[width=0.8\linewidth]{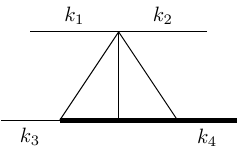}
		\caption*{$\M^\text{NP4}_{25}$}
	\end{minipage}
	\begin{minipage}{0.15\linewidth}
		\centering
		\includegraphics[width=0.8\linewidth]{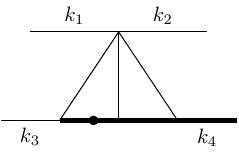}
		\caption*{$\M^\text{NP4}_{26}$}
	\end{minipage}
	\begin{minipage}{0.15\linewidth}
		\centering
		\includegraphics[width=0.8\linewidth]{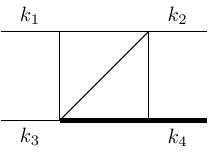}
		\caption*{$\M^\text{NP4}_{27}$}
	\end{minipage}
	\begin{minipage}{0.15\linewidth}
		\centering
		\includegraphics[width=0.8\linewidth]{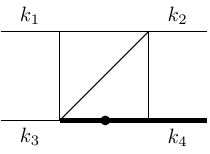}
		\caption*{$\M^\text{NP4}_{28}$}
	\end{minipage}
	\begin{minipage}{0.15\linewidth}
		\centering
		\includegraphics[width=0.8\linewidth]{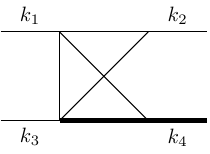}
		\caption*{$\M^\text{NP4}_{29}$}
	\end{minipage}
	\begin{minipage}{0.15\linewidth}
		\centering
		\includegraphics[width=0.8\linewidth]{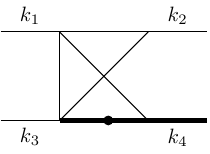}
		\caption*{$\M^\text{NP4}_{30}$}
	\end{minipage}
	\begin{minipage}{0.15\linewidth}
		\centering
		\includegraphics[width=0.8\linewidth]{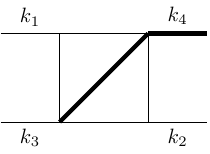}
		\caption*{$\M^\text{NP4}_{31}$}
	\end{minipage}
	\begin{minipage}{0.15\linewidth}
		\centering
		\includegraphics[width=0.8\linewidth]{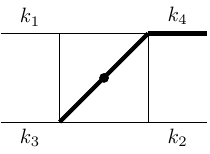}
		\caption*{$\M^\text{NP4}_{32}$}
	\end{minipage}
	\begin{minipage}{0.15\linewidth}
		\centering
		\includegraphics[width=0.8\linewidth]{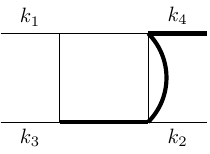}
		\caption*{$\M^\text{NP4}_{33}$}
	\end{minipage}
	\begin{minipage}{0.15\linewidth}
		\centering
		\includegraphics[width=0.8\linewidth]{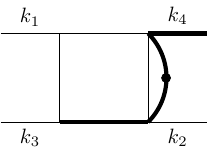}
		\caption*{$\M^\text{NP4}_{34}$}
	\end{minipage}
	\begin{minipage}{0.15\linewidth}
		\centering
		\includegraphics[width=0.8\linewidth]{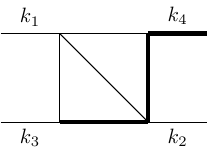}
		\caption*{$\M^\text{NP4}_{35}$}
	\end{minipage}
	\begin{minipage}{0.15\linewidth}
		\centering
		\includegraphics[width=0.8\linewidth]{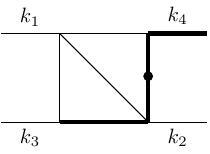}
		\caption*{$\M^\text{NP4}_{36}$}
	\end{minipage}
	\begin{minipage}{0.15\linewidth}
		\centering
		\includegraphics[width=0.8\linewidth]{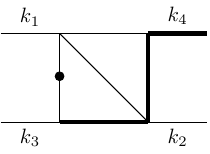}
		\caption*{$\M^\text{NP4}_{37}$}
	\end{minipage}
	\begin{minipage}{0.15\linewidth}
		\centering
		\includegraphics[width=0.8\linewidth]{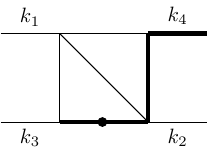}
		\caption*{$\M^\text{NP4}_{38}$}
	\end{minipage}
	\begin{minipage}{0.15\linewidth}
		\centering
		\includegraphics[width=0.8\linewidth]{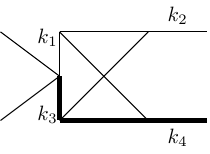}
		\caption*{$\M^\text{NP4}_{39}$}
	\end{minipage}
	\begin{minipage}{0.15\linewidth}
		\centering
		\includegraphics[width=0.8\linewidth]{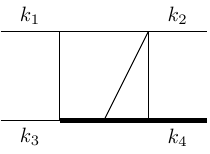}
		\caption*{$\M^\text{NP4}_{40},\M^\text{NP4}_{41}$}
	\end{minipage}
	\begin{minipage}{0.15\linewidth}
		\centering
		\includegraphics[width=0.8\linewidth]{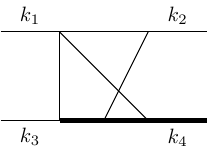}
		\caption*{$\M^\text{NP4}_{42}-\M^\text{NP4}_{45}$}
	\end{minipage}
	\begin{minipage}{0.15\linewidth}
		\centering
		\includegraphics[width=1.05\linewidth]{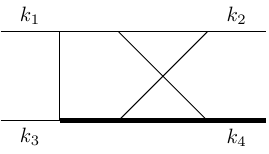}
		\caption*{$\M^\text{NP4}_{46}-\M^\text{NP4}_{48}$}
	\end{minipage}
	\caption{Master integrals in the NP4 topology. The thick lines stand for massive top quarks while the others represent massless particles.  One black dot indicates one additional power of the corresponding propagator. The numerators can be found in the text.}
	\label{fig:NP4_MIs}
\end{figure}

\bibliography{NP2_NP3_NP4_draft}
\bibliographystyle{JHEP}

\end{document}